%% file: paper.tex
\documentclass[usenatbib]{mn2e}
\usepackage[dvips]{graphicx}
\usepackage{amssymb}
\usepackage{natbib}
\usepackage{float}
\usepackage{latexsym}
\usepackage{amsfonts}
\usepackage{graphicx}
\usepackage{booktabs}
\usepackage{color}

\input{mydefs}

\begin{document}

\title[Gas distribution in Local Group galaxies]{The distribution of gas in the Local Group from constrained cosmological simulations: the case for Andromeda and the Milky Way galaxies}

\author[Nuza et al.]{Sebasti\'an E. Nuza$^{1}$\thanks{snuza@aip.de}, 
  Florencia Parisi$^{1,2}$,  
  Cecilia Scannapieco$^{1}$, Philipp Richter$^{1,3}$, \newauthor 
  Stefan Gottl\"ober$^{1}$ and Matthias Steinmetz$^{1}$\\
$^1$ Leibniz-Institut f\"ur Astropysik Potsdam (AIP), An der Sternwarte 16, D-14482, Potsdam, Germany\\
$^2$ Consejo Nacional de Investigaciones Cient\'{\i}ficas y T\'ecnicas (CONICET), Argentina\\
$^3$ Institut f\"ur Physik und Astronomie, Universit\"at Potsdam, Haus 28, Karl-Liebknecht-Str. 24/25, 14476 Golm (Potsdam), Germany\\}

   \maketitle

   \begin{abstract}
 We study the gas distribution in the Milky Way and Andromeda using a constrained cosmological simulation of the 
 Local Group (LG) within the context of the CLUES (Constrained Local UniversE Simulations) project. We analyse the properties of gas 
 in the simulated galaxies at $z=0$ for three different phases:  `cold', `hot' and  H\,{\sc i}, and compare our results with observations. 
 The amount of material in the hot halo ($M_{\rm hot}\approx 4-5\times10^{10}\,\Msun$), and 
 the cold ($M_{\rm cold}(r\lesssim10\,{\rm kpc})\approx10^{8}\,\Msun$) and 
 H\,{\sc i} ($M_{\rm HI}(r\lesssim50\,{\rm kpc})\approx 3-4\times10^8\,\Msun$) components display a reasonable agreement 
 with observations. We also compute the accretion/ejection rates together with the H\,{\sc i} (radial and all-sky) covering 
 fractions. The integrated H\,{\sc i} accretion rate within $r=50\,$kpc gives $\sim$$0.2-0.3\,\Msun\,{\rm yr}^{-1}$, i.e. close to 
 that obtained from high-velocity clouds in the Milky Way. We find that the global accretion rate is dominated by hot material, although 
 ionized gas with $T\lesssim10^5\,$K can contribute significantly too. The {\it net} accretion rates of {\it all} material at 
 the virial radii are $6-8\,\Msun\,{\rm yr}^{-1}$.  At $z=0$, we find a significant gas excess between the two galaxies, 
 as compared to any other direction, resulting from the overlap of their gaseous haloes. In our simulation, the gas excess 
 first occurs at $z\sim1$, as a consequence of the kinematical evolution of the LG.
  \end{abstract}

\begin{keywords}  
methods: numerical -- Galaxy: halo -- intergalactic medium -- Local Group -- large-scale structure of Universe.
\end{keywords}

\section{Introduction}

\begin{figure*}
\begin{center}                        
{\includegraphics[width=55mm]{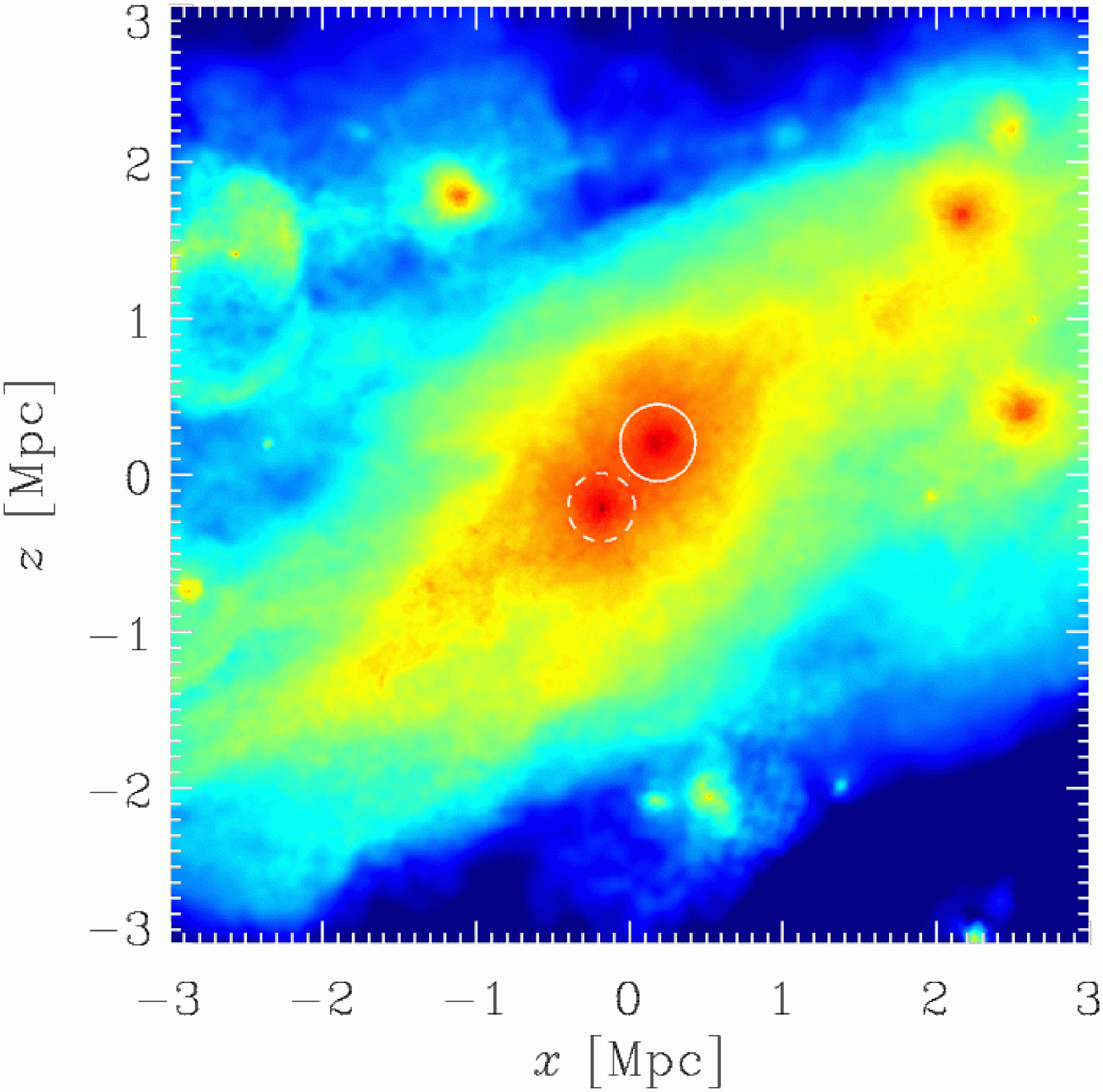}\includegraphics[width=55mm]{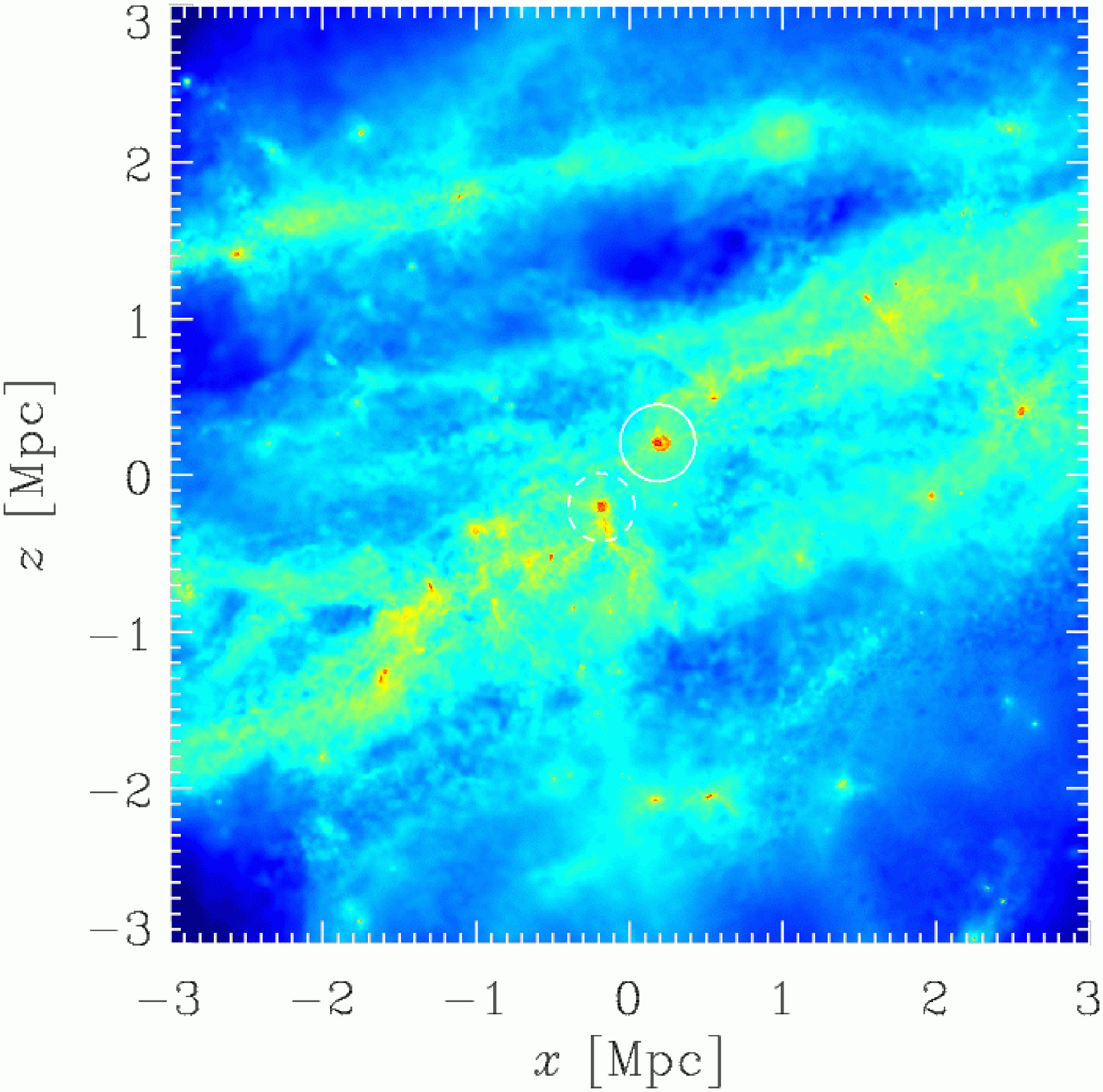}\includegraphics[width=55mm]{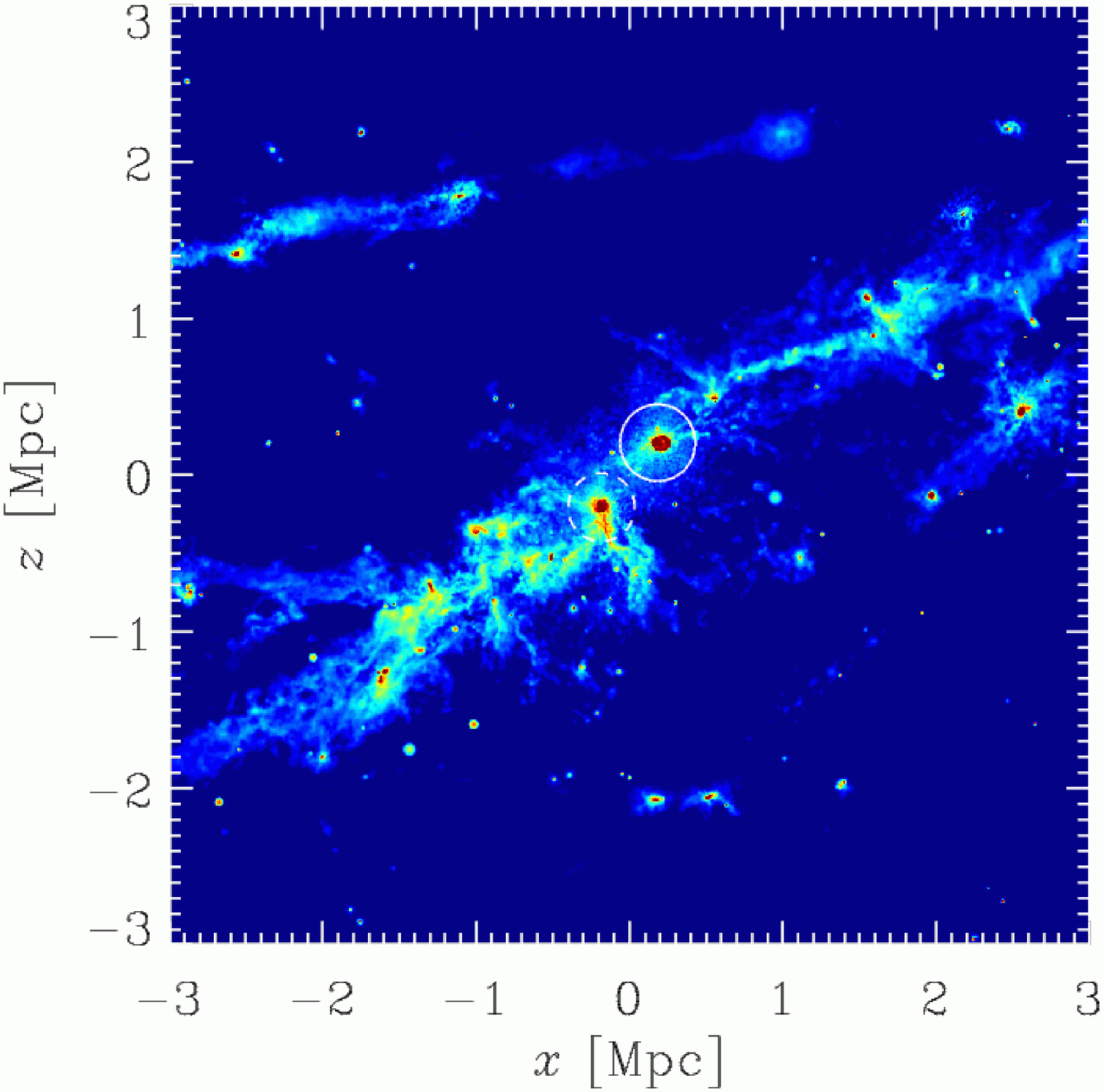}}
\caption{Gas density maps of the simulated LG at $z=0$ for the hot ($T\geq10^{5}\,$K; left-hand panel), cold ($T<10^{5}\,$K; middle panel) 
 and H\,{\sc i} (right-hand panel)  components. The plots are centred in the MW$^{\rm c}$/M31$^{\rm c}$ system. The virial radii of our Milky Way and Andromeda candidates are shown as dashed and solid lines respectively.
In order to highlight the differences in the distribution of the hot, cold and neutral gas 
components, each plot shows the projected density in a colour scale covering four orders of 
magnitude.}
\label{large_scale_env}
\end{center}
\end{figure*}

The distribution of gas in galaxies, their haloes and their adjacent regions is 
a key aspect of galaxy formation and evolution and can give us important clues on the 
current evolutionary state of the galactic systems, as well as on their recent past history. 
For instance, the presence of cold, inflowing gas, may point towards an immediate 
future of active star formation, whereas the existence of outflows may indicate that 
star formation and feedback has taken place in the recent past, in a way that these 
outflows could develop.

During the last two decades, significant progress was achieved in our
understanding of the distribution and physical properties of circumgalactic and 
intergalactic gas in the Local Group \citep[hereafter LG; e.g.,][for reviews on this topic]{Wakker97,Richter06,Putman12}.
Recent observational campaigns including $21\,$cm radio observations of neutral
hydrogen \citep[e.g.,][]{Thilker04,Kalberla06,Kerp11}, and 
absorption-line measurements of neutral and ionized gas 
in the optical and ultraviolet (UV; e.g., Sembach et al. 2003; Collins et
al. 2005, 2009; Ben Bekhti et al. 2008, 2012; Richter et al. 2009, 2012
[hereafter R12]; Herenz et al. 2013; Rao et al. 2013) indicate the extreme multi-phase character of 
circumgalactic gas within the virial radii of the Milky Way and Andromeda galaxies. 
One major difficulty in disentangling the various gaseous components in the
LG and constraining their origin and fate lies in the fact that the 
three-dimensional distribution of the gas is very difficult to establish
from our perspective within the Milky Way. This is because direct
distance estimates of circumgalactic gas complexes are very difficult to
obtain, as they require a large observational effort \citep[][]{Wakker07,Wakker08}.

Our current understanding is that the Milky Way and Andromeda, the two major spiral galaxies 
in the LG, are both surrounded by multi-phase gas haloes, whose
properties reflect the various processes that circulate gaseous material 
in their circumgalactic environments (i.e., outflows, inflows and tidal interactions). 
Observations imply that the cold ($T<10^4\,$K), and predominantly neutral,
circumgalactic gas that gives rise to detectable H\,{\sc i} $21\,$cm emission 
appears to be concentrated within a distance of $\sim$$50$ kpc from 
the Milky Way and Andromeda galaxies \citep[][]{Thilker04,Wakker07,Wakker08,Richter12}. 
At larger distances, and even beyond the virial radii of the
two galaxies, the LG is possibly filled with highly-ionized, hot ($T\gtrsim10^5$ K) gas 
that gives rise to absorption of highly ionized oxygen in the UV 
\citep[O\,{\sc vi}][]{Sembach03} and in the X-ray band \citep[O\,{\sc vii} and O\,{\sc viii};][]{Gupta12} 
in the spectra of distant quasars, blazars and other
UV and X-ray bright background sources. It is therefore possible that this warm-hot circumgalactic 
medium harbours a large fraction of the baryons present in the LG, which could explain the 
apparent discrepancy between the baryon content found in galaxy groups and the mean 
cosmological value \citep[see also][]{Richter08}. This line of evidence is strengthened by the 
recent discovery of X-ray haloes surrounding regular, although massive, spiral galaxies that are not 
related to merger interactions, starbursts and/or emission from compact objects 
(e.g., \citealt{Anderson11,Dai12,Bogdan13a,Bogdan13b}). 

According to the standard picture of galaxy formation, two main accretion modes are responsible 
of feeding galaxies with a fresh supply of cold gas that can be further processed into stars: 
a `hot'-mode, originated as high-density gas surrounding the galaxy cools radiatively 
and collapses towards the centre \citep[e.g.,][]{White78}, and a `cold'-mode, driven by cold gas 
filaments penetrating well inside the virial radius of the halo 
\citep[e.g.,][]{NavarroSteinmetz97,Keres05}.

Recently, several authors have studied the distribution of gas, and their associated accretion modes, 
for isolated galaxy haloes using cosmological simulations. 
For instance, \citet{Fernandez12} focused on the cold accretion mode in 
an isolated Milky Way-mass galaxy with low velocity dispersion 
under the condition that the mean environmental density at $z=0$ is close to that of the actual LG. 
Similarly, \citet{Joung12} extended this work by considering 
the accretion of warm-hot material on to the halo. Their results indicate that the 
accretion rate in the system is dominated by ionized warm-hot gas over a large range of 
distances extending up to the virial radius, whereas cooler gas tends to occupy the central 
regions.

In this work, we study the properties of gas in a constrained simulation of the LG of galaxies, 
therefore including the interaction between their members, and make detailed comparisons 
with available observations. 
Our simulation reproduces the local environment of the LG, going beyond the
simple criteria adopted by \citet{Fernandez12}, while including the 
main observed large scale structures that surround the LG 
(e.g., the Virgo cluster among other mass aggregations). 
The simulated LG contains two main galaxies that, at $z=0$, approximately resemble 
the Milky Way and Andromeda galaxies in terms of their masses, separation, relative velocity 
and orientation of the discs. 
Our aim is to characterize the spatial distribution and physical conditions of the different 
gas phases in the simulated LG, as well as to compare these results with recent multi-wavelength 
observations of gas in the circumgalactic environment of the Milky Way and
Andromeda. In a companion paper (Scannapieco et al., in preparation) we  study 
the formation of the  Milky Way and Andromeda candidates in the same
simulation, but focusing on the evolution of stellar discs and their 
merger histories in relation to the environment.

Our paper is organized as follows: in Section\,2 we describe
the main aspects of the simulation and simulated LG;
in Section\,3 we analyse the properties of the gas at the largest 
scales of the LG, as well as within the simulated Milky Way and Andromeda haloes. 
In Section\,4 we compare the distribution of hot, cold and neutral 
gas belonging to the galaxies with observations and discuss
how well the covering fractions of neutral Hydrogen compare
with observational results in the Milky Way and Andromeda. 
In Section\,5, we discuss accretion and ejection rates of gas 
on to the main LG haloes, and in Section\,6 we analyse the gas distribution 
around the simulated Milky Way/Andromeda system. 
In Section\,7, we compare some of our findings with 
previous work and in Section\,8 we present our conclusions.

\section{The Constrained Cosmological Simulation}
\label{sec:sims}

This study is carried out within the context of the Constrained Local UniversE Simulations (CLUES) project 
(www.clues-project.org) which aims at simulating a realistic LG including the effects 
of the local environment as well as of the most prominent surrounding structures.

The initial conditions (ICs) are constructed using the Hoffman-Ribak algorithm \citep[][]{Hoffman91} as a way to 
impose a set of Gaussian constraints at high redshift using a number of observational constraints at $z\approx0$. 
The initial matter distribution consists of a cubic box of $64\,\Mpch$ side length which contains 
a high-resolution spherical region of $2\,\Mpch$ radius located at its centre.

The simulation contains structures resembling the Virgo, Great Attractor, Fornax and 
Perseus clusters \citep[see e.g.,][]{Klypin03}. Since typical constraints are of the order of a few Mpc, 
structure formation at 
the smallest scales is essentially random. For this reason, a series of realizations were run in order to get 
the best possible match of our Milky Way and Andromeda galaxy candidates with the available estimates of masses 
and relative radial velocity of the LG galaxies. Further details on the generation of ICs can be 
found in \citet{CLUES10}.

The simulation was started at a redshift of $z=50$ assuming a $\Lambda$ cold dark matter cosmology 
({\it Wilkinson Microwave Anisotropy Probe} 5). 
For the cosmological parameters we adopted a matter density of $\Omega_{\rm m}=0.279$, a baryon density 
of $\Omega_{\rm b}=0.046$, a cosmological constant density of $\Omega_{\Lambda}=0.721$, a Hubble constant of 
$100h$ km s$^{-1}$ Mpc$^{-1}$ with $h=0.7$, and a power spectrum normalization of $\sigma_8=0.8$. 
The masses of the gas and dark matter particles inside the high-resolution sphere are 
$M_{\rm g}=3.89\times10^5\,\Msunh$ and $M_{\rm dm}=1.97\times10^6\,\Msunh$ respectively. 
The interested reader is referred to Scannapieco et al. (in preparation) for 
further details on this simulation.

To identify (sub)structures in the simulation, 
we use the halo finder {\sc subfind} \citep[][]{Springel01,Dolag09}. 
This code allows us to link each particle in the simulation (gas, stars
and dark matter) to a given substructure (containing a minimum of 32 particles)
based on its binding energy. In this way, it is possible to identify 
the main haloes within the simulation at each redshift as well as their associated 
satellite systems.

\subsection{The Simulation Code}

For this study, we use the smoothed particle hydrodynamics (SPH) simulation code {\sc gadget3} \citep{Springel08}
with the extensions of \citet{Scann05, Scann06}. The code includes metal-dependent
cooling (above $10^4$ K; \citealt{S&D93}) and chemical enrichment, a multiphase model for the gas component 
and supernova (thermal) feedback. We refer the interested reader
to \citet{Scann05, Scann06} for details on the implementation of these processes.

The multiphase and feedback models naturally generate galactic winds
after significant star formation bursts \citep{Scann06, Scann08}. This feature allows
to study in a more realistic manner the distribution of gas in and around
haloes, as well as in the intergalactic medium, since there is no a priori
assumption on the occurrence and/or strength of the winds.
Moreover, since the code follows the enrichment of the interstellar medium
from supernova Type Ia and Type II explosions, it is possible to investigate
the chemical properties of the gas component in the simulation and, in
particular, the distribution of the H\,{\sc i} gas.

The Scannapieco et al. model has been successfully applied to the study
of Milky Way-mass galaxies \citep{Scann09,Scann10,Scann11} and of dwarf spheroidal galaxies
\citep{Sawala11,Sawala12}. Particularly relevant for this study is that
the same input parameters are used in the aforementioned studies, indicating
that the code can properly follow the formation of systems of different mass.
This is possible because the implementation of supernova feedback, whose effects
strongly depend on the mass of a system, has been specifically designed to avoid
the introduction of scale-dependent parameters (see \citealt{Scann06, Scann08}).
The ability to simulate different mass systems with the same
choice of input parameters  is of relevance in cosmological simulations, where systems
of all masses are forming simultaneously at any time.

Recently, there has been considerable discussion on the ability of the
SPH technique to capture some small-scale processes, particularly fluid mixing and
jumps in physical properties (see.
e.g. \citealt{Agertz07,Read12,Hopkins13} and references therein). 
We do not expect these  problems to be particularly severe in our simulation.
On one hand, SPH gives accurate gas properties in the intergalactic medium,
where the gas is at relatively low densities and there is no significant
spatial coexistence of different phases. Within the haloes, the gas
has a complex multi-phase structure, which standard SPH might not be 
able to accurately resolve.
However, our multi-phase model has proven efficient at describing
the coexistence of gas at different phases \citep{Scann06}, and therefore
we do not expect our results to be strongly affected by numerical
problems connected to the SPH technique.

We note that our simulation does not include the effects of radiation 
pressure (RP) provided by massive stars prior to supernova explosions, 
a process that has been included only very recently in cosmological 
simulations (e.g., \citealt{Aumer13, Stinson13}). 
These works found that RP efficiently regulates star formation,
particularly at high redshift, producing  halo-to-stellar mass relations
in better agreement with the predictions of abundance matching models.
RP affects mainly the central regions of haloes, but should not
strongly affect the global properties of the gaseous haloes or intergalactic
medium.

\subsection{The simulated LG}
\label{sim_LG}

\begin{figure}
\begin{center}
\includegraphics[width=87mm]{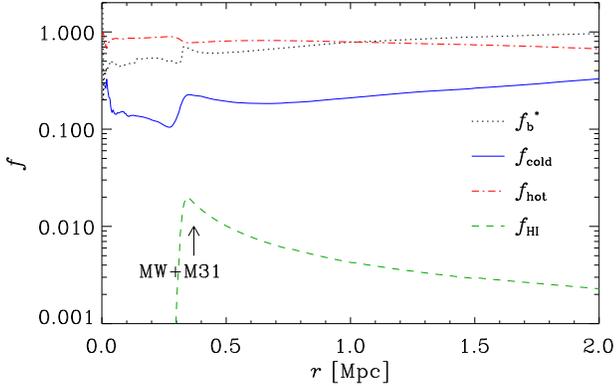}
\caption{Fraction of cold gas ($f_{\rm cold}$; solid line), hot gas ($f_{\rm hot}$; dot-dashed line), H\,{\sc i} gas ($f_{\rm HI}$; dashed line) 
 and baryons ($f^*_{\rm b}$, normalized to the universal baryon fraction; dotted line) for increasingly larger 
 spheres centred in the MW$^{\rm c}$/M31$^{\rm c}$ system. The arrow indicates the position 
 of the MW$^{\rm c}$ and M31$^{\rm c}$ galaxies.} 
\label{LG_gas_fractions}
\end{center}
\end{figure}

The simulated LG at the present time includes two main massive haloes 
within the $2\,\Mpch\,$high-resolution region that can be associated with 
the Milky Way and Andromeda galaxies.  
In what follows, we will refer to our simulated galaxy candidates using a `c' 
superscript, i.e. MW$^{\rm c}$ and M31$^{\rm c}$ respectively, while keeping the names 
Milky Way and Andromeda for reference to the actual systems and 
related observational quantities. Note that, as stated above, the simulation 
does not allow us to directly constrain the properties at galaxy-size scales, and thus 
the two simulated galaxies have dynamical and kinematical properties that can 
only approximate those of the actual Milky Way and Andromeda.

As in the real LG, M31$^{\rm c}$ and MW$^{\rm c}$ are approaching each other. 
The relative radial velocity between the mass centres of the simulated galaxies is
$-138$ ${\rm km\,s}^{-1}$, a value which is close to the recent observational
estimate of $-109.3\pm4.4$ ${\rm km\,s}^{-1}$ for the Andromeda and Milky Way
system presented by \citet{vandermarel12}. 
On the other hand, the distance between M31$^{\rm c}$ and MW$^{\rm c}$ is 
$652$ kpc, in comparison to the observed value of $770\pm40$ kpc \citep[e.g.,][]{Ribas05}.

\begin{figure*}
\begin{center}
{\includegraphics[width=56mm]{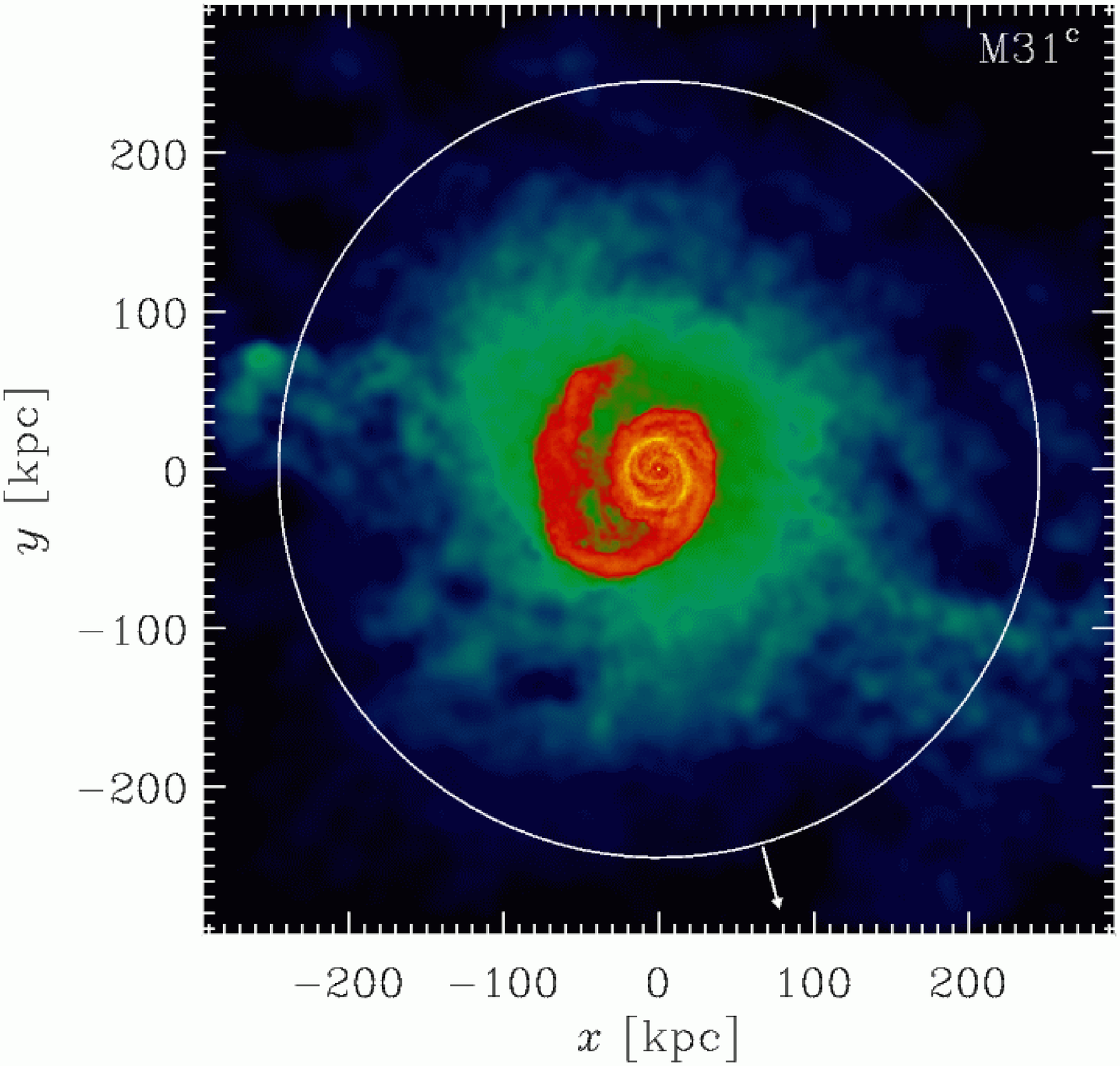}
\includegraphics[width=56mm]{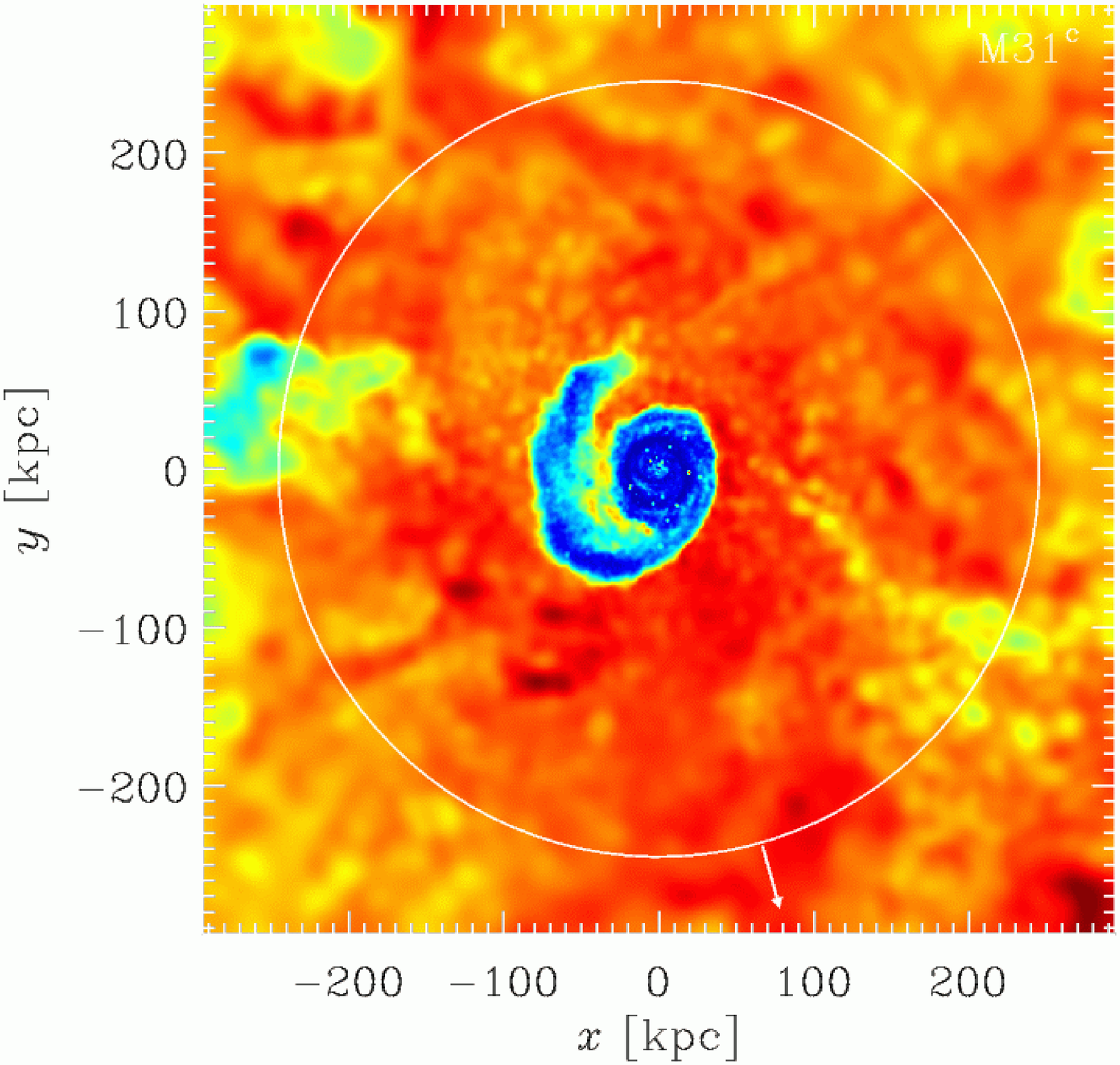}
\includegraphics[width=56mm]{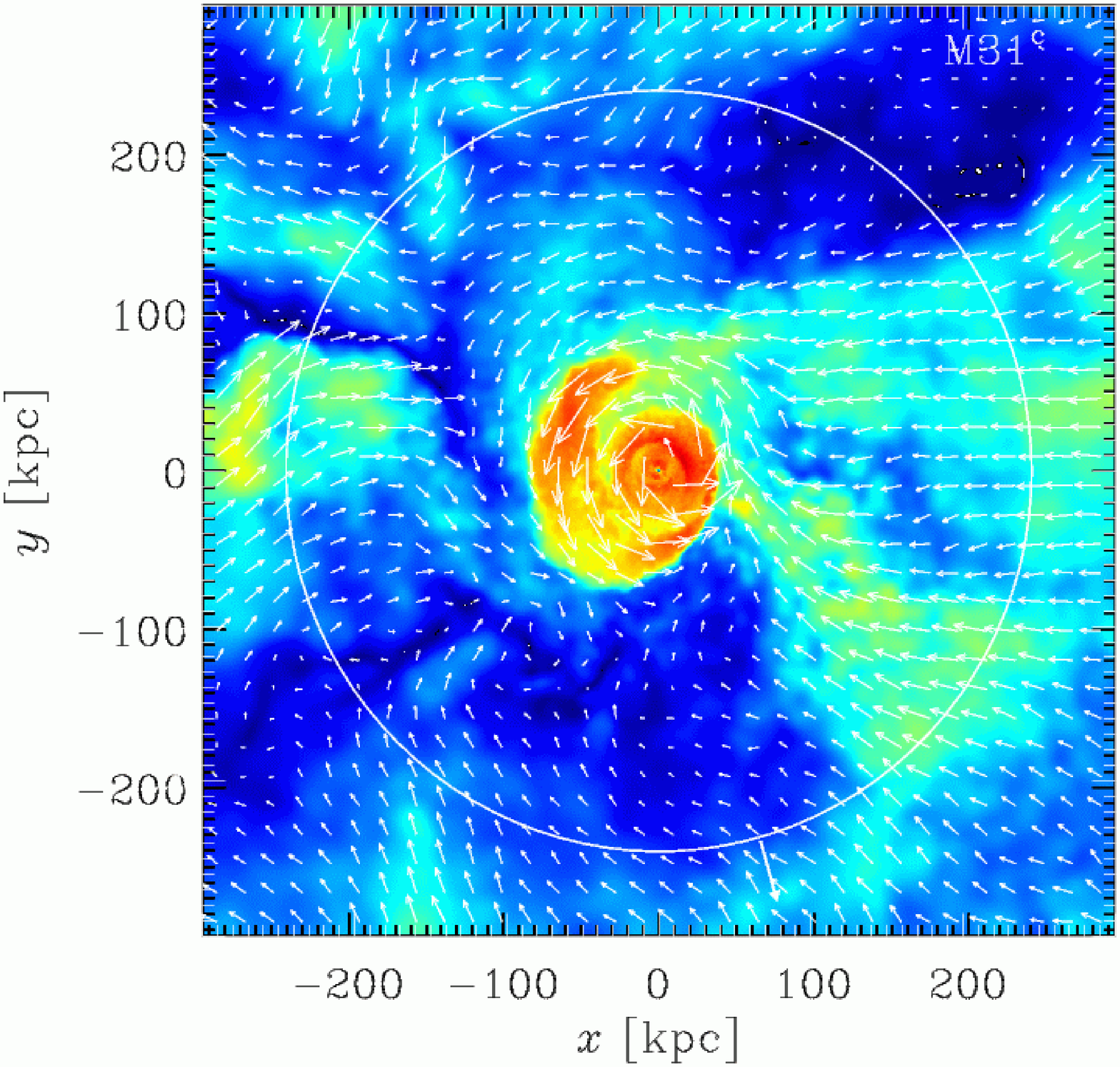}}

{\includegraphics[width=56mm]{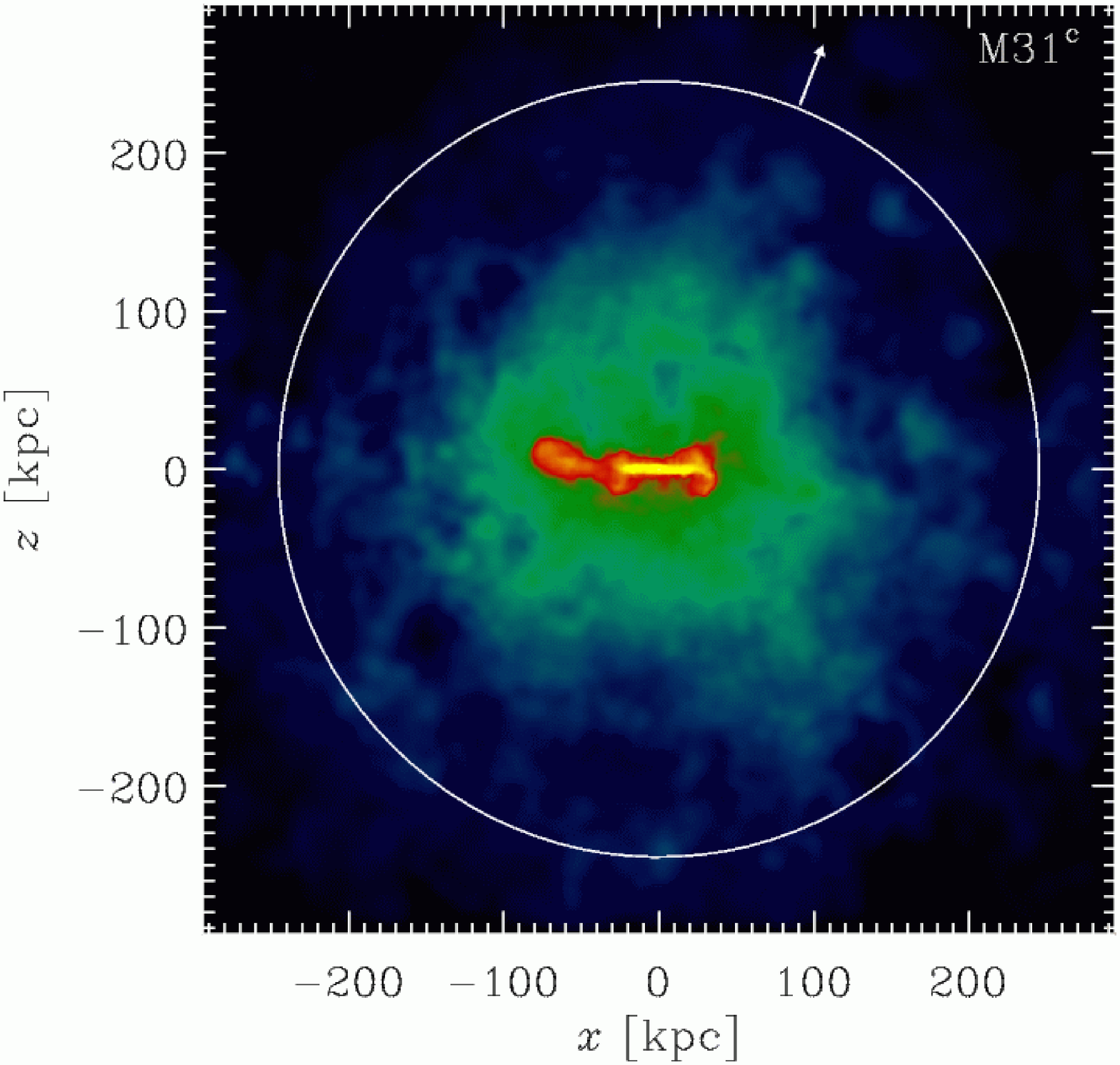}
\includegraphics[width=56mm]{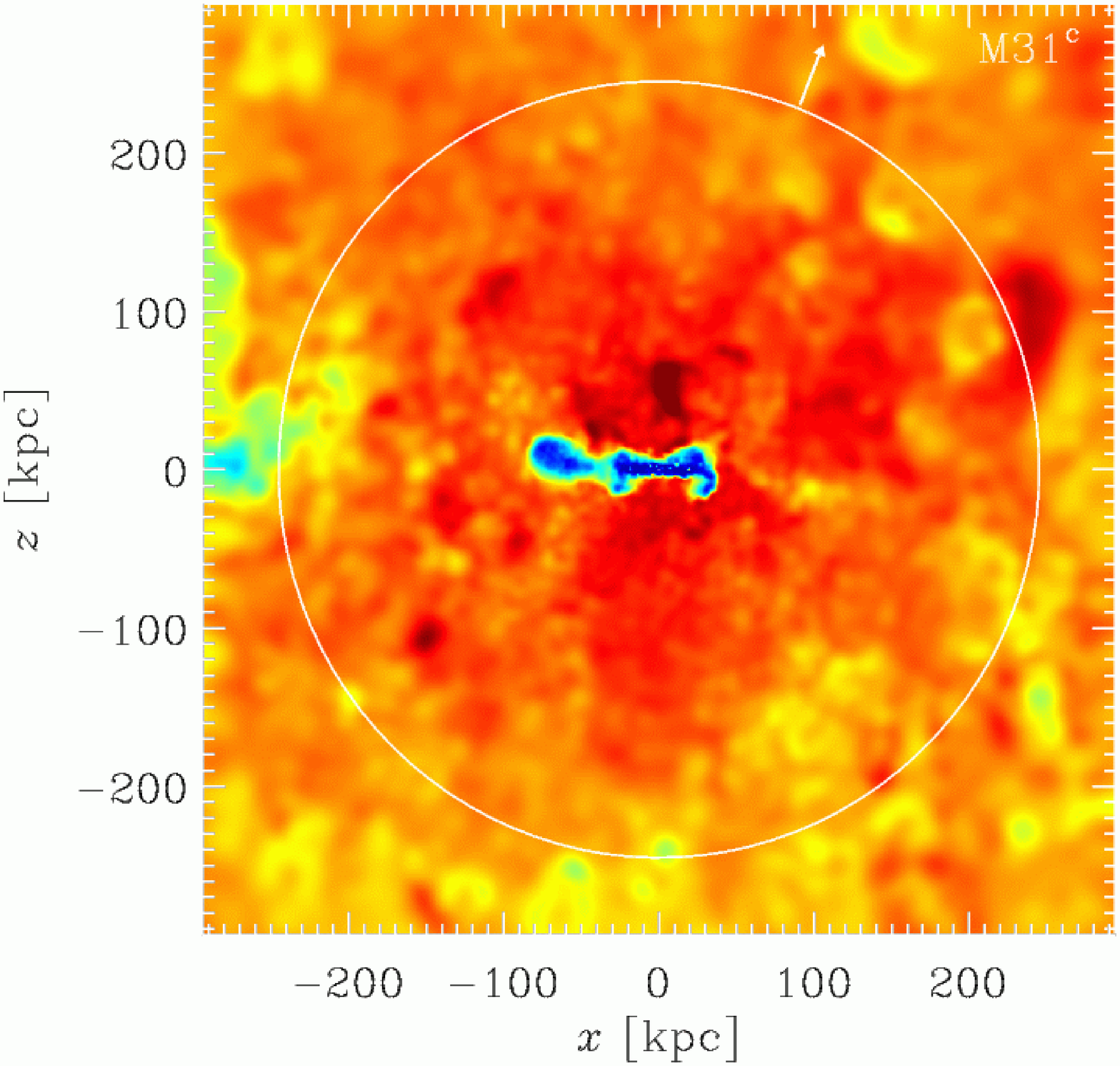}
\includegraphics[width=56mm]{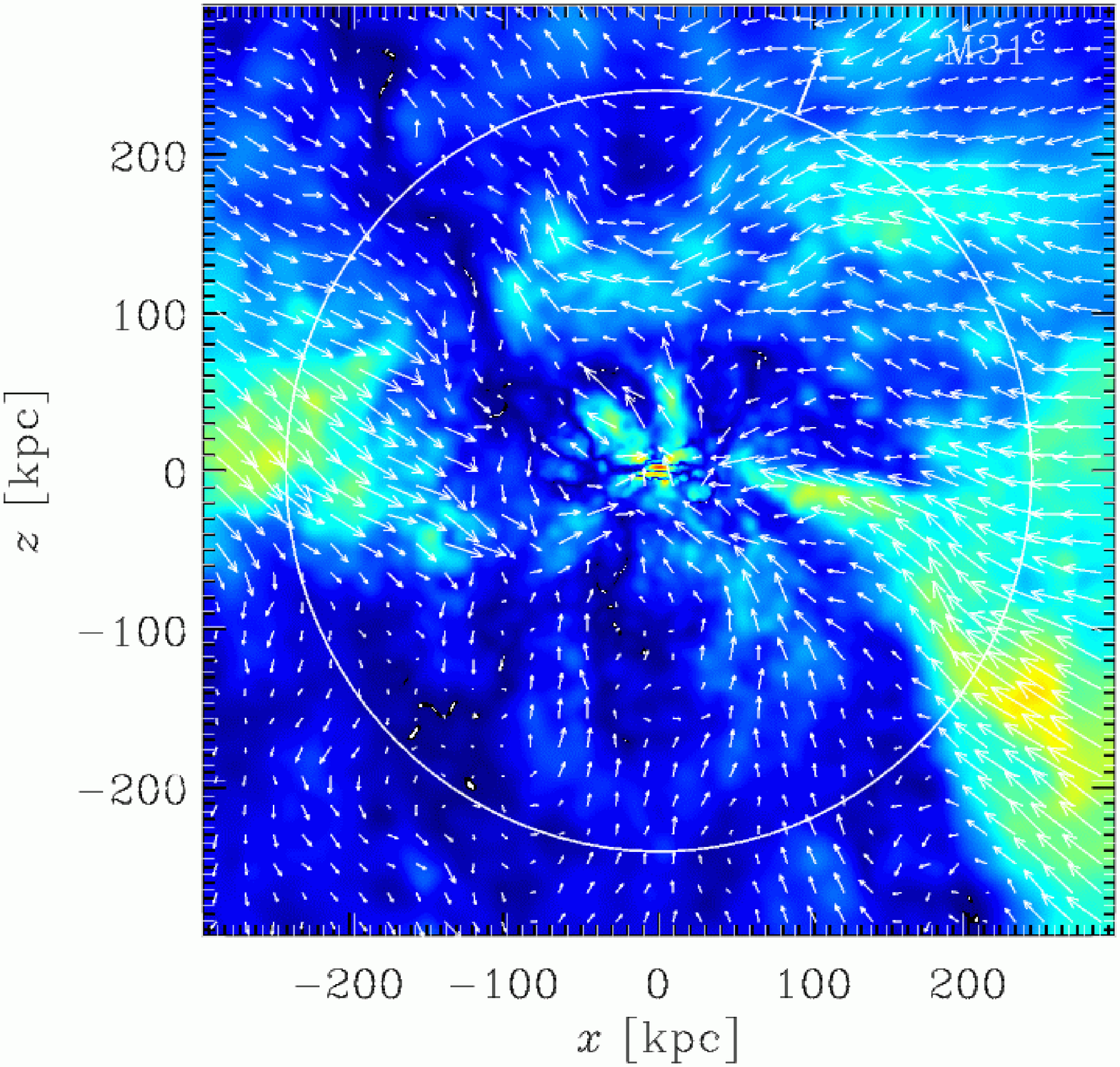}}

{\includegraphics[width=56mm]{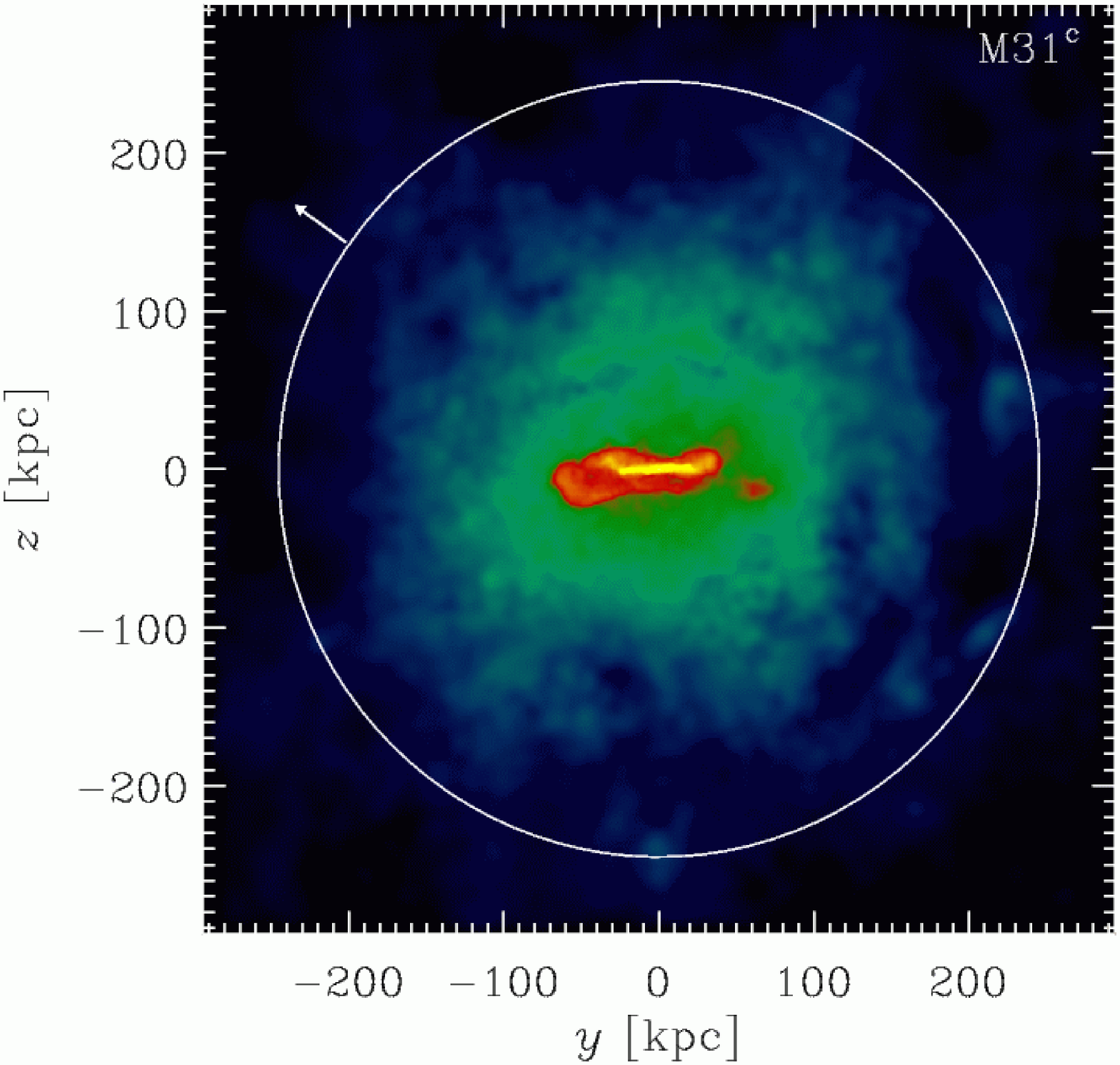}
\includegraphics[width=56mm]{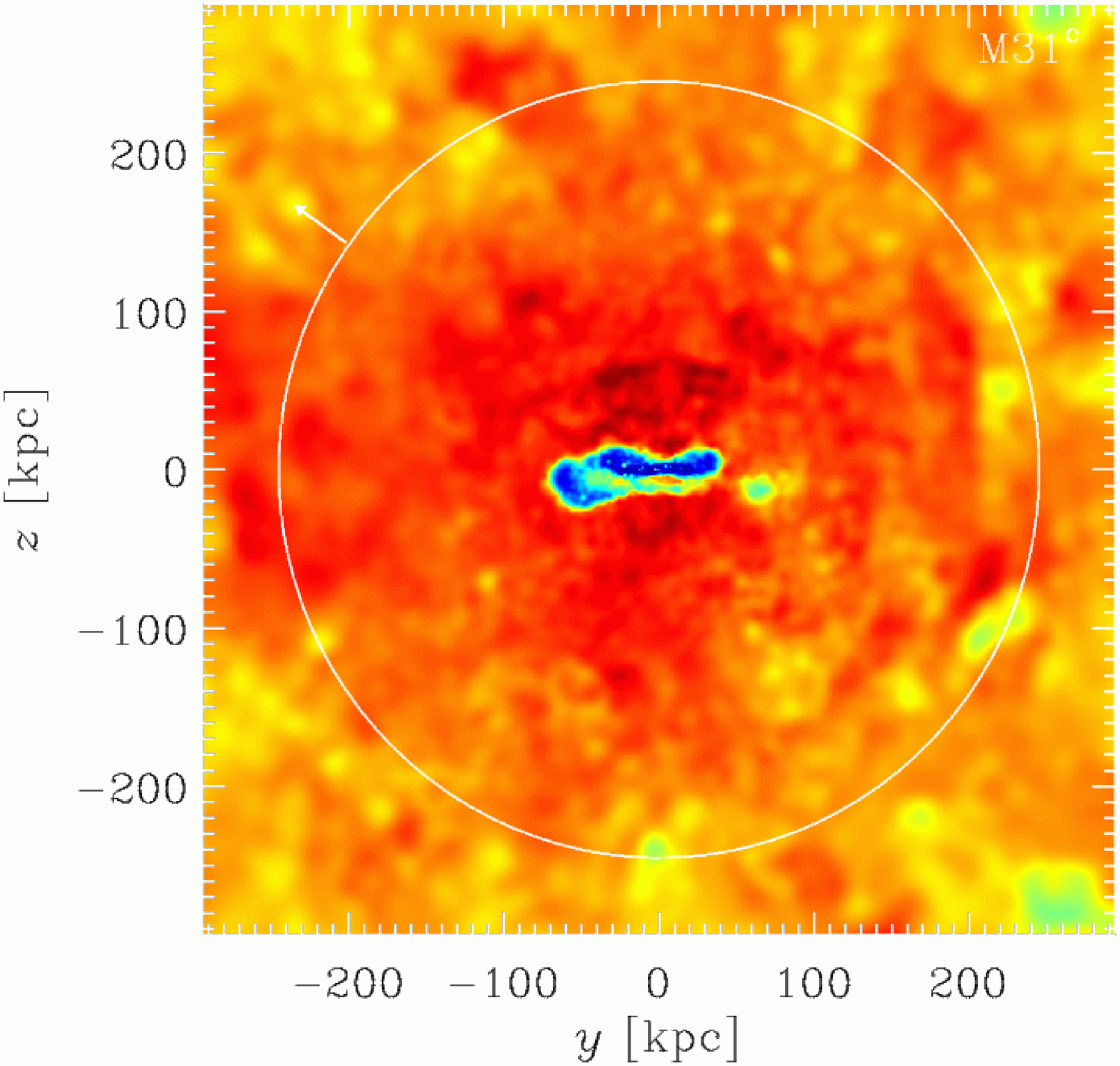}
\includegraphics[width=56mm]{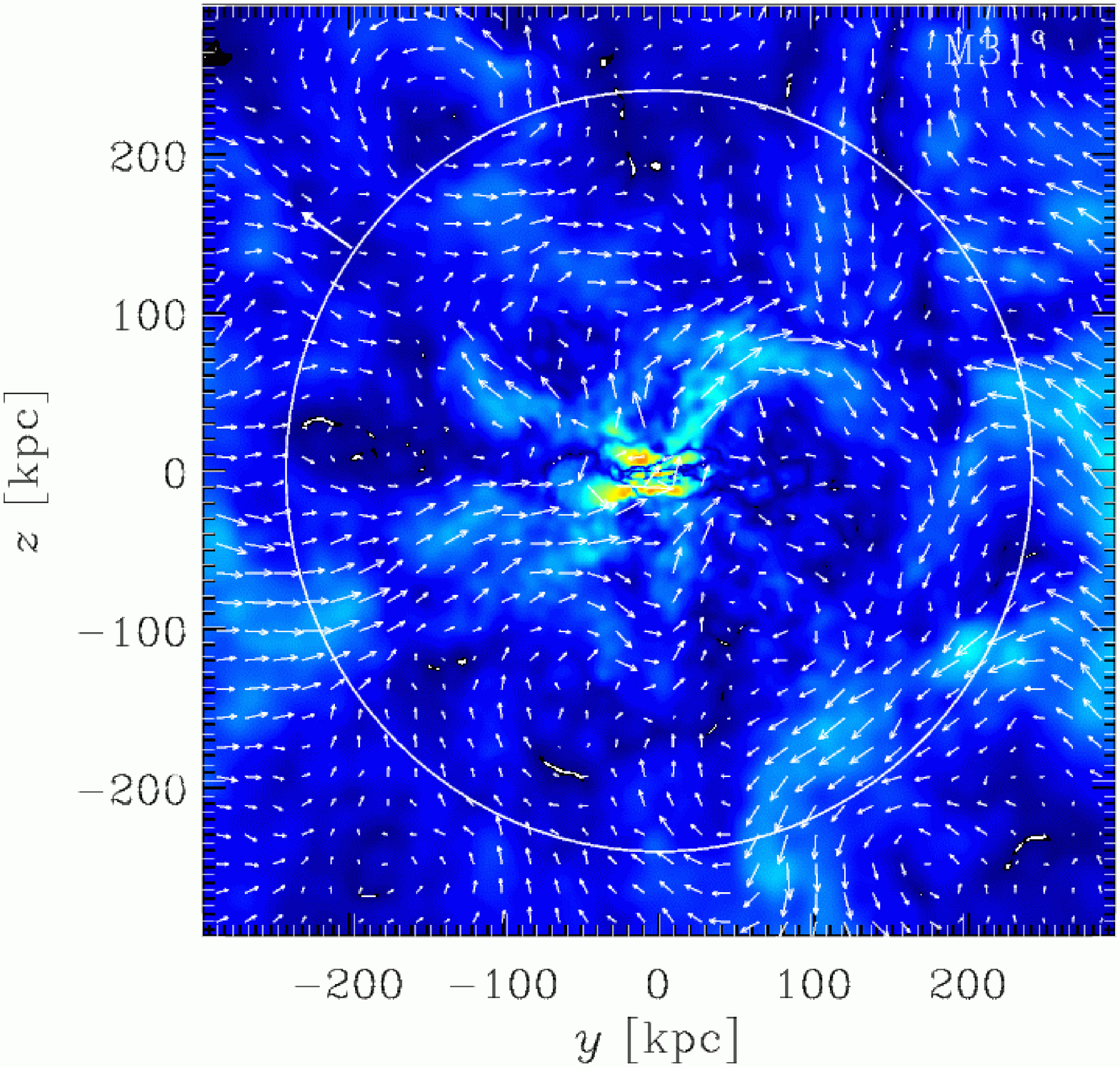}}

{\includegraphics[width=56mm]{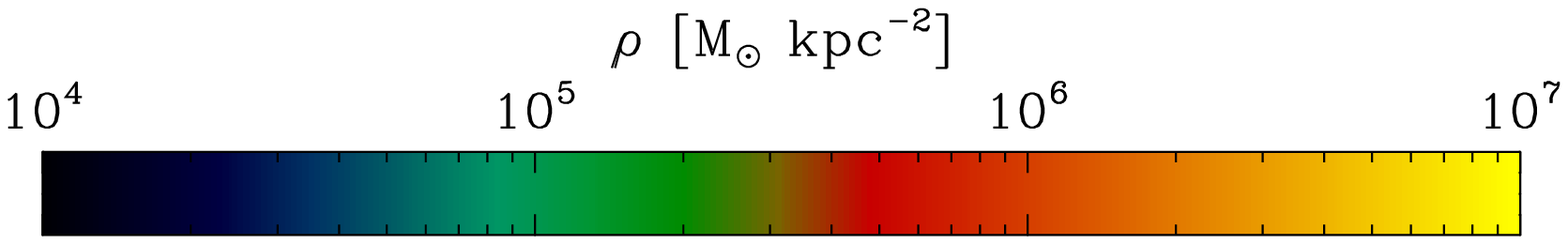}\hspace{1mm}
\includegraphics[width=56mm]{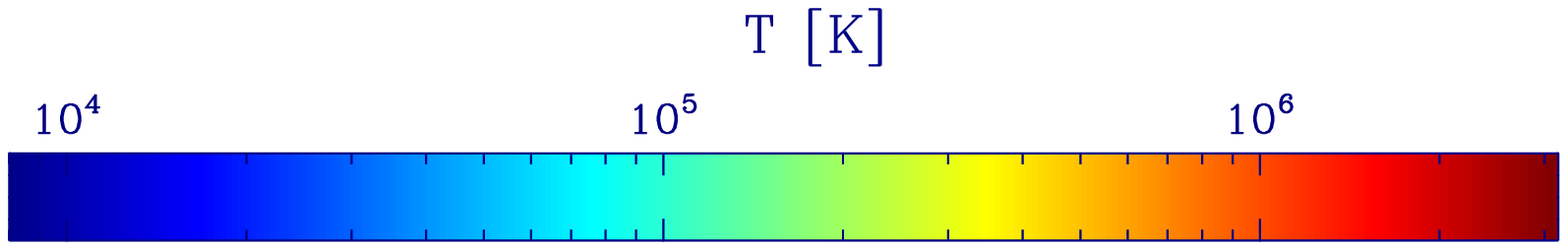}\hspace{1mm}
\includegraphics[width=56mm]{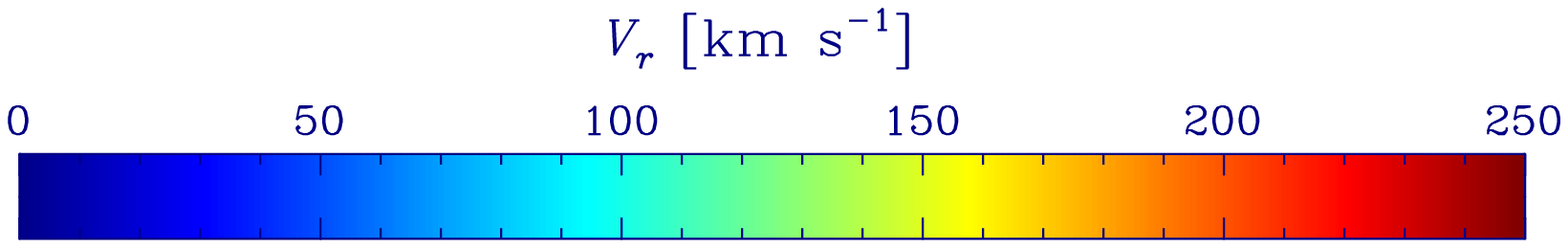}}

\caption{Face-on (upper panels) and edge-on (middle and lower panels) maps of gas density, 
temperature and projected velocity
with respect to centre of mass of the simulated M31$^{\rm c}$. For the latter, 
the colours represent the absolute magnitude of the corresponding
velocity and the arrows show the direction of the velocity field at
each position. For the gas
density, we include all mass within a cubic box of $1.2\times R_{\rm vir}$
on a side. In the case of temperatures and velocities, the plots
are also  $1.2\times R_{\rm vir}$
on a side, but  include only mass within thin slices $0.24\times R_{\rm vir}$ wide.
The circles
indicate the location of the virial radius and the arrows point towards the
position of the  MW$^{\rm c}$.} 
\label{M31}
\end{center}
\end{figure*}

\begin{figure*}
\begin{center}
{\includegraphics[width=56mm]{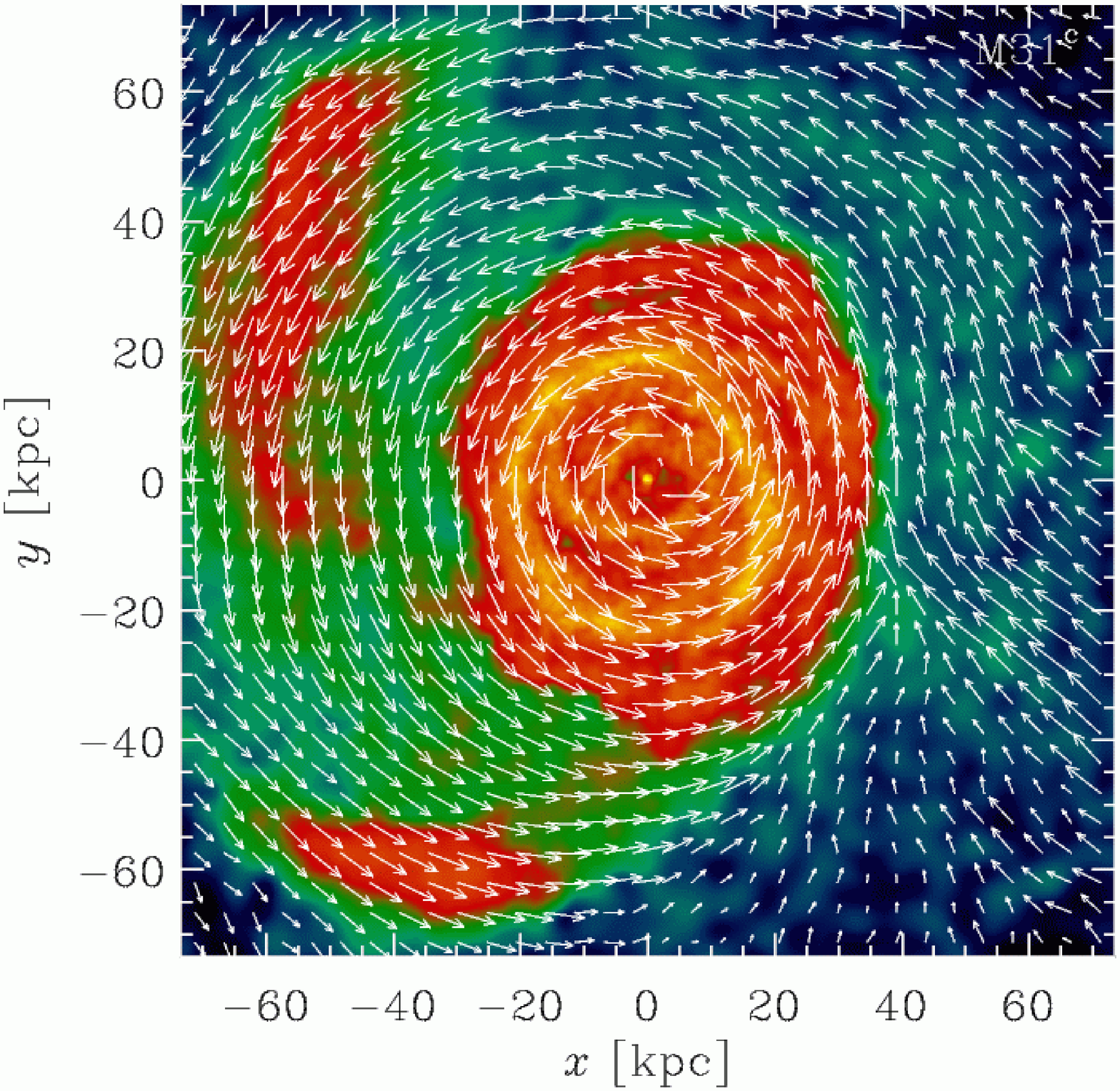}
\includegraphics[width=56mm]{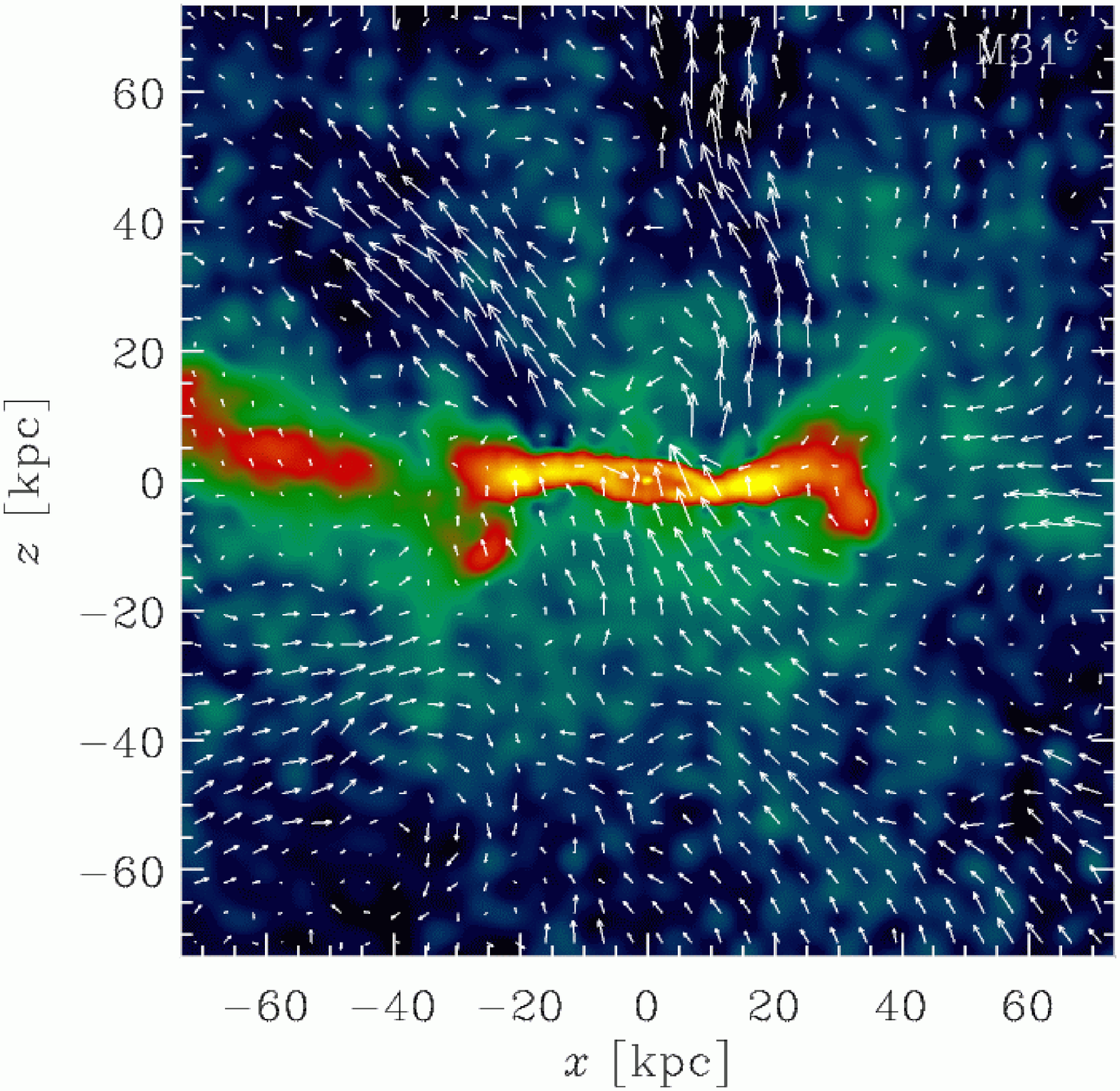}
\includegraphics[width=56mm]{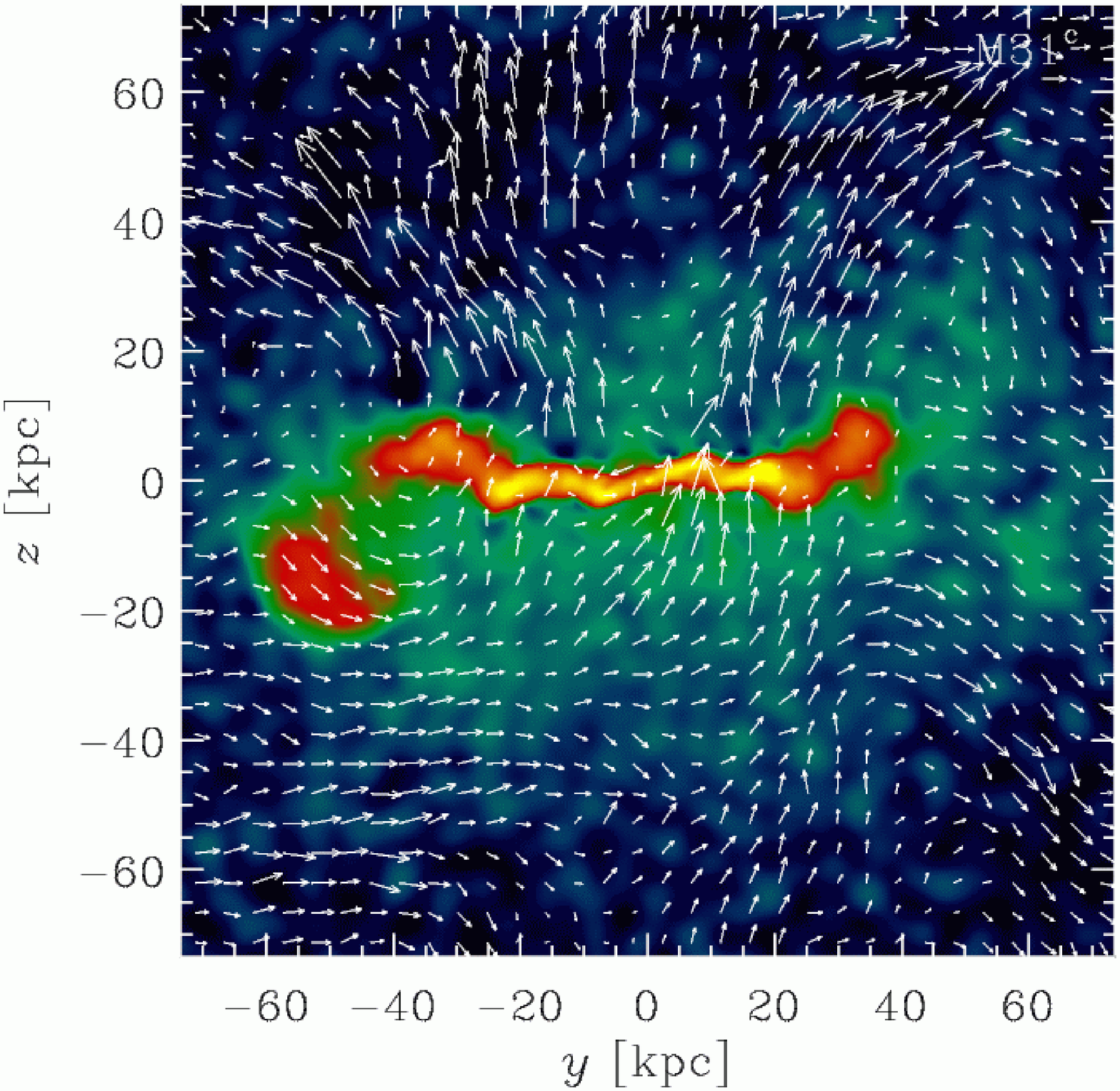}}

\caption{Zoom in of the projected gas density for M31$^{\rm c}$, within the inner
$0.3 \times R_{\rm vir}$, and corresponding velocity field (arrows). 
The colour scale is the same as that of Fig.~\ref{M31}.}
\label{M31_central}
\end{center}
\end{figure*}

\begin{figure*}
\begin{center}
{\includegraphics[width=58mm]{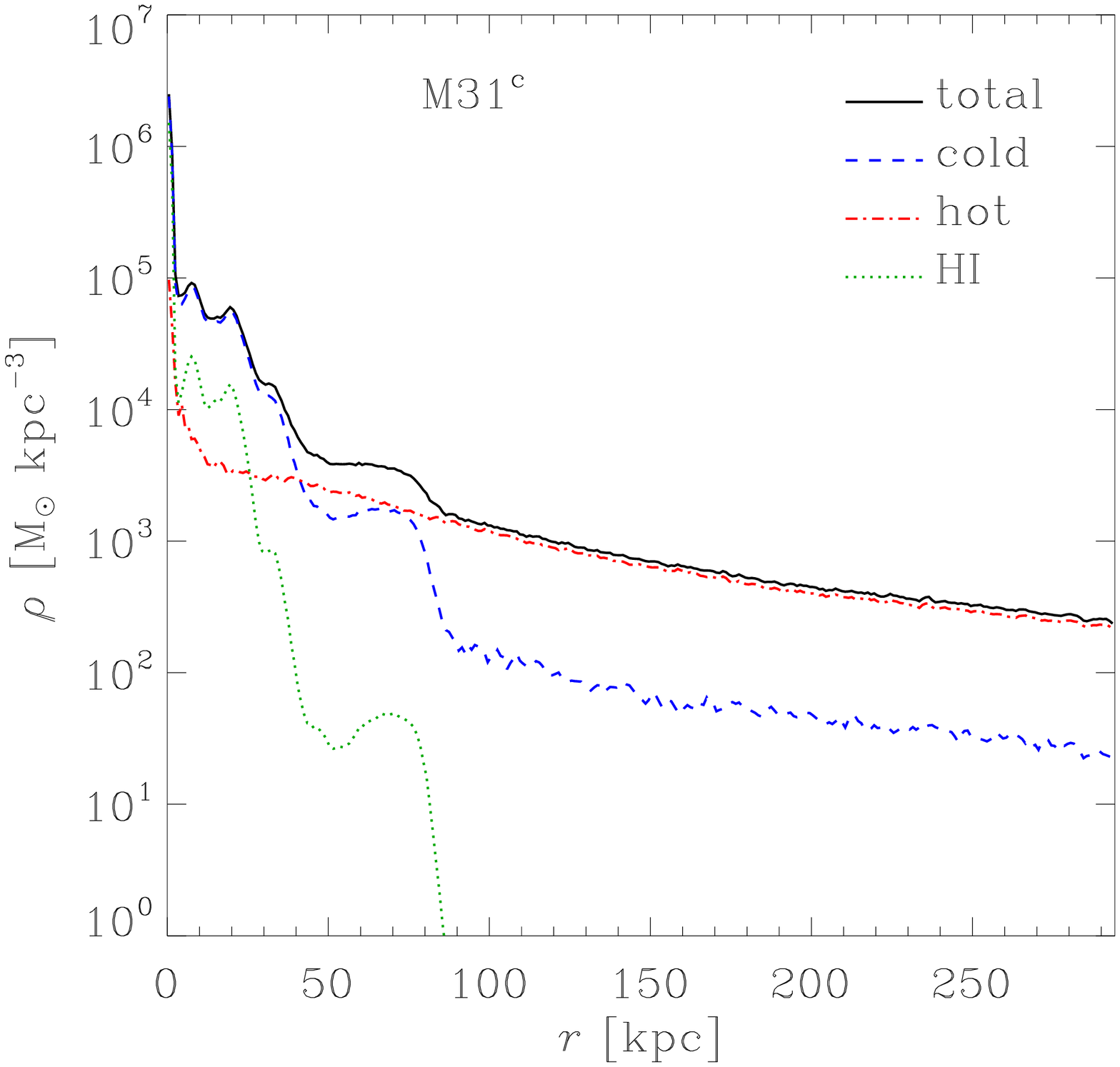}
\includegraphics[width=58mm]{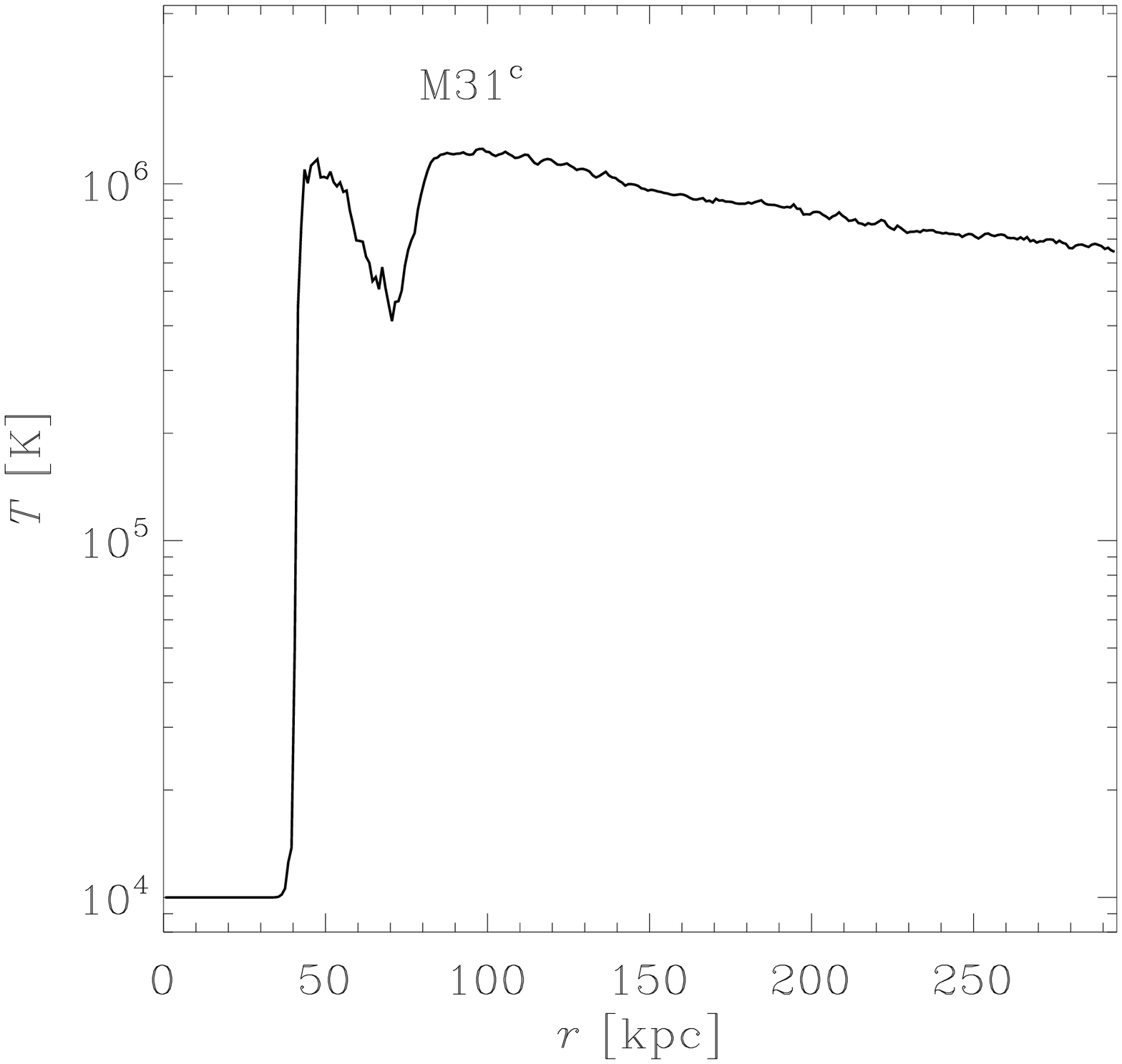}
\includegraphics[width=58mm]{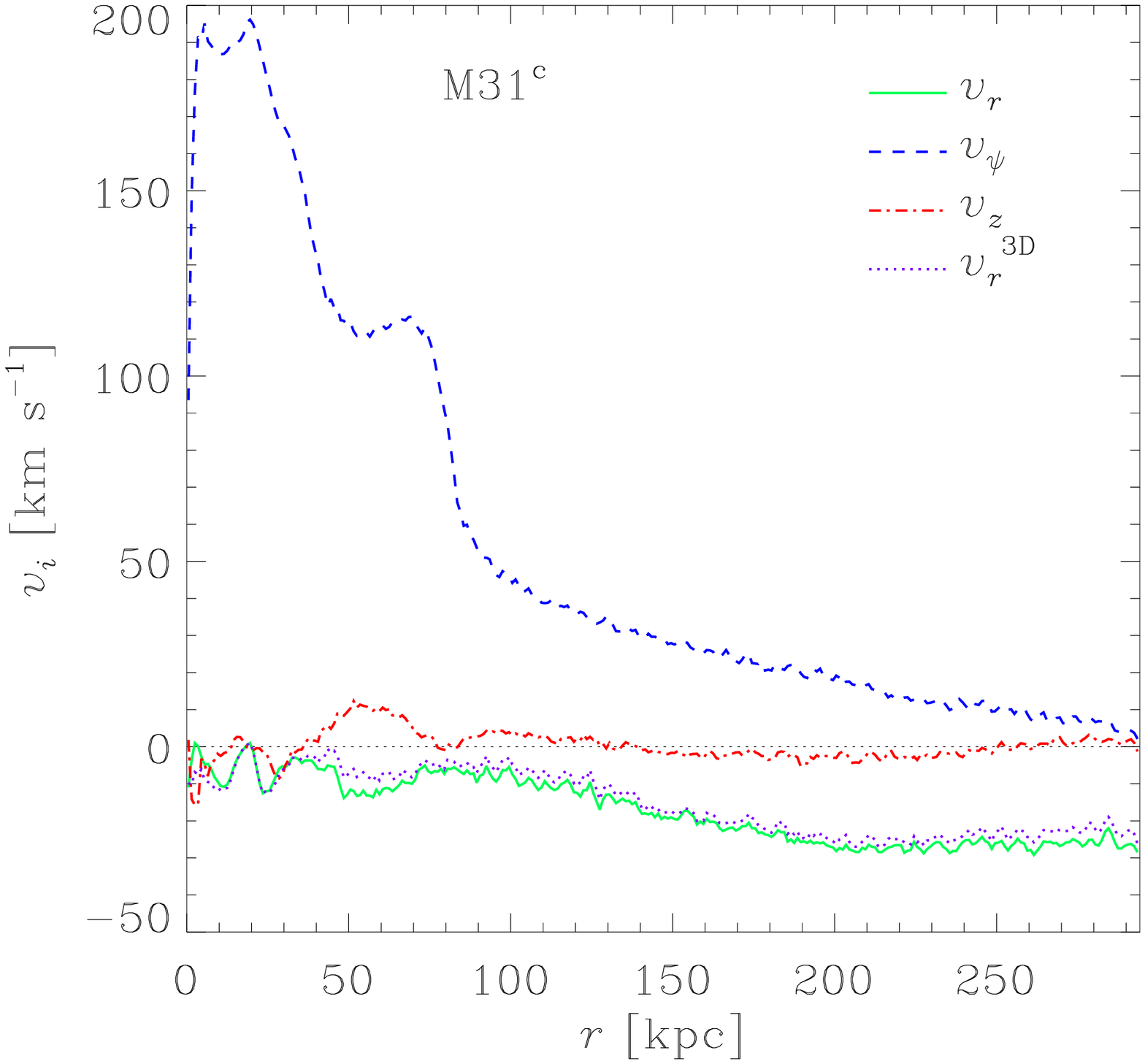}}

\caption{Density, temperature and velocity profiles for M31$^{\rm c}$, up to 
$1.2\times R_{\rm vir}$. For the density, we show also the profiles
of the hot ($T\geq10^5$ K, dotted-dashed line), cold ($T<10^5$ K, dashed line) and H\,{\sc i} (dotted line) gas separately. 
In the case of the velocities, we show the profiles of radial
(solid green line), 
tangential (dashed blue line) and vertical (dot-dashed red line)  
velocity in a cylindrical coordinate
system. Furthermore, for the vertical velocity, 
we show $V_z^*\equiv v_z$sign$(z)$ in order
to easily differentiate between inflows ($V_z^*\leq0$) and outflows ($V_z^*>0$).
}
\label{profiles_M31}
\end{center}
\end{figure*}

Observational estimates of both the Milky Way and Andromeda virial masses range from about 
$10^{12}\,\Msun$ to $2.5 \times 10^{12}\,\Msun$  
\citep[][and references therein]{vandermarel12}.
To compute the virial mass of the simulated galaxies we rely on the 
definition of the virial radius, noted $R_{\rm vir}$, as the one comprising a mass density 200 times 
that of the critical density of the universe. As a result, the simulated M31$^{\rm c}$ and MW$^{\rm c}$ systems have virial 
masses of $1.68\times 10^{12}$ M$_\odot$ and $1.23\times 10^{12}$ M$_\odot$ respectively, where we have decided 
to take the less massive system as the Milky Way candidate.

The uniqueness of our constrained simulation is that it comprises two massive Milky Way-mass galaxies 
located in a large-scale environment that reproduces the distribution of the most massive
structures of the local Universe. As an additional characteristic, the resulting orientations of
the gaseous galaxy discs at $z=0$ resemble those of the real Milky Way and
Andromeda galaxies as it will be shown in Section~\ref{HI_CovFracs}.

In Table~\ref{Table_global} we present some of the global properties of the simulated LG 
massive galaxies including virial mass, virial radius, virial velocity and total dark matter and gas masses 
within the virial radius.

\begin{figure*}
\begin{center}
{\includegraphics[width=56mm]{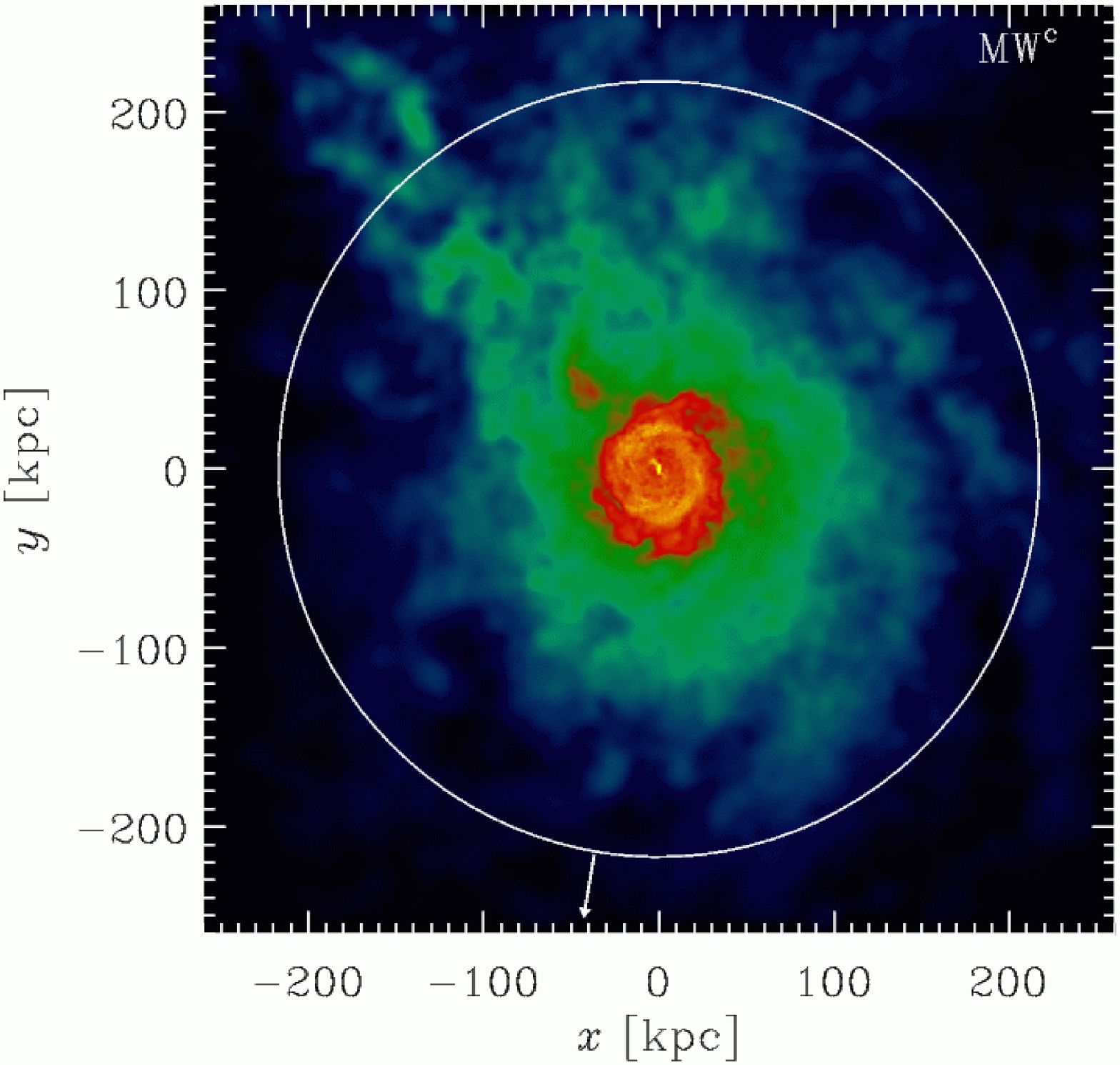}
\includegraphics[width=56mm]{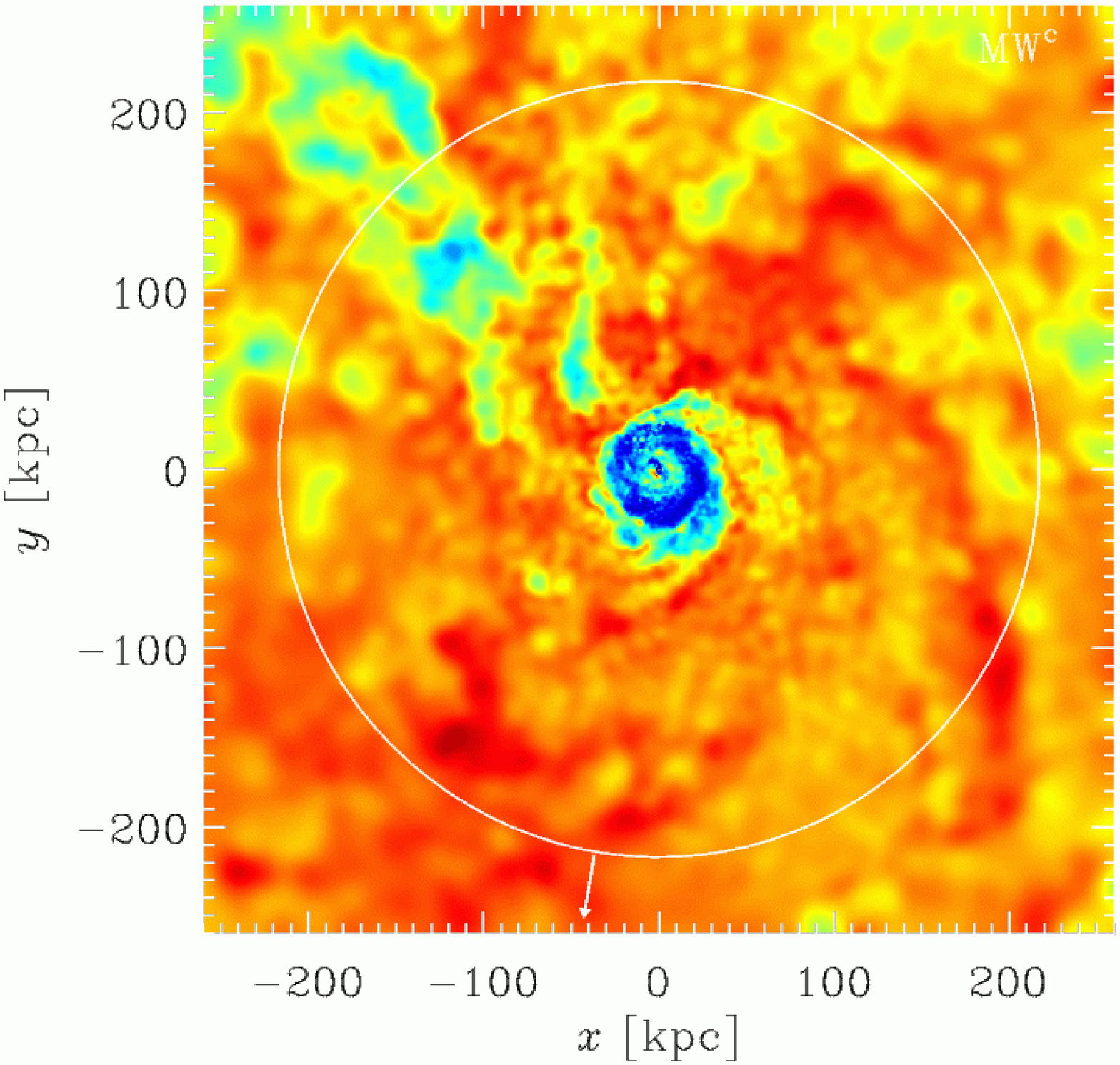}
\includegraphics[width=56mm]{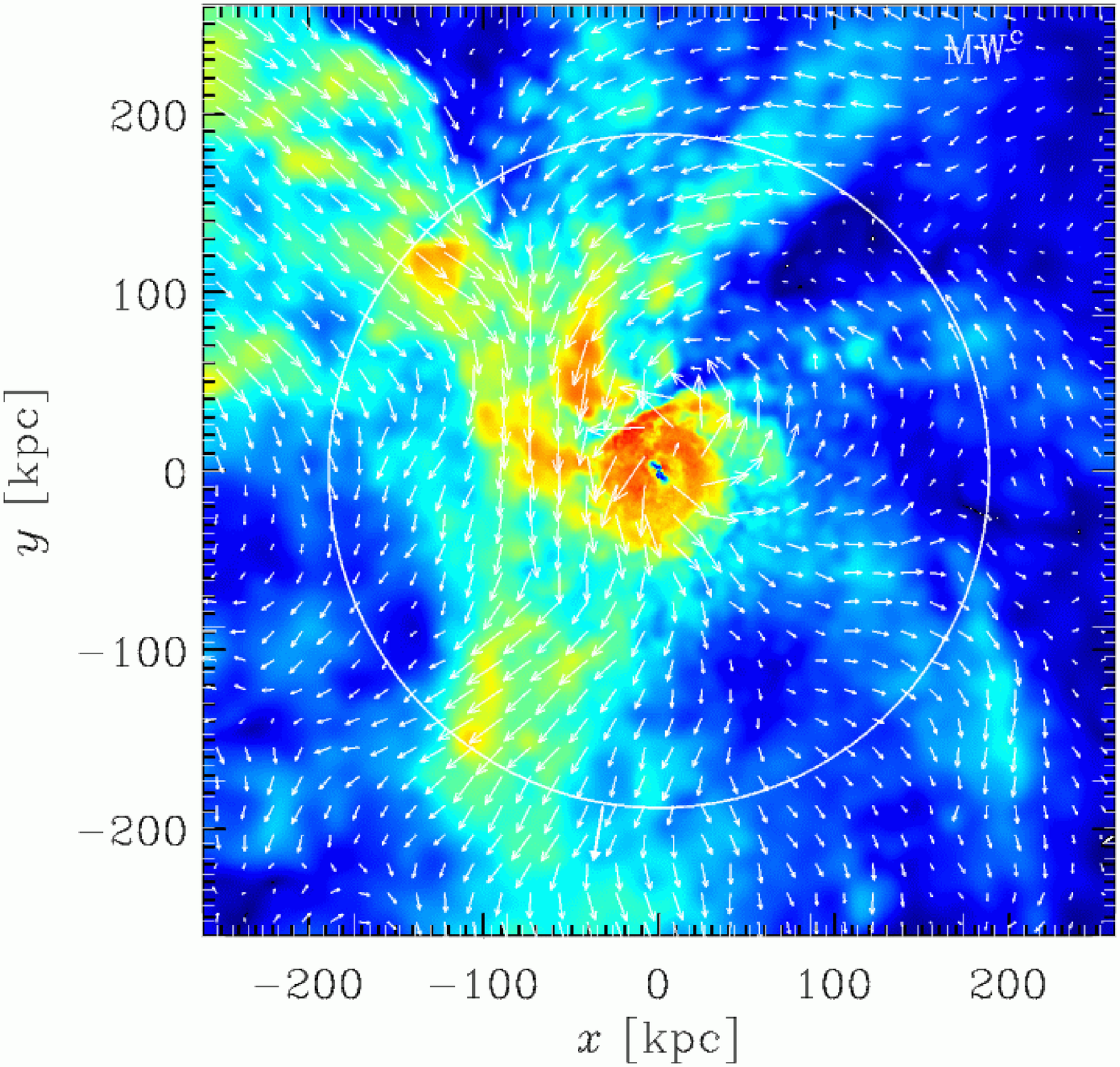}}

{\includegraphics[width=56mm]{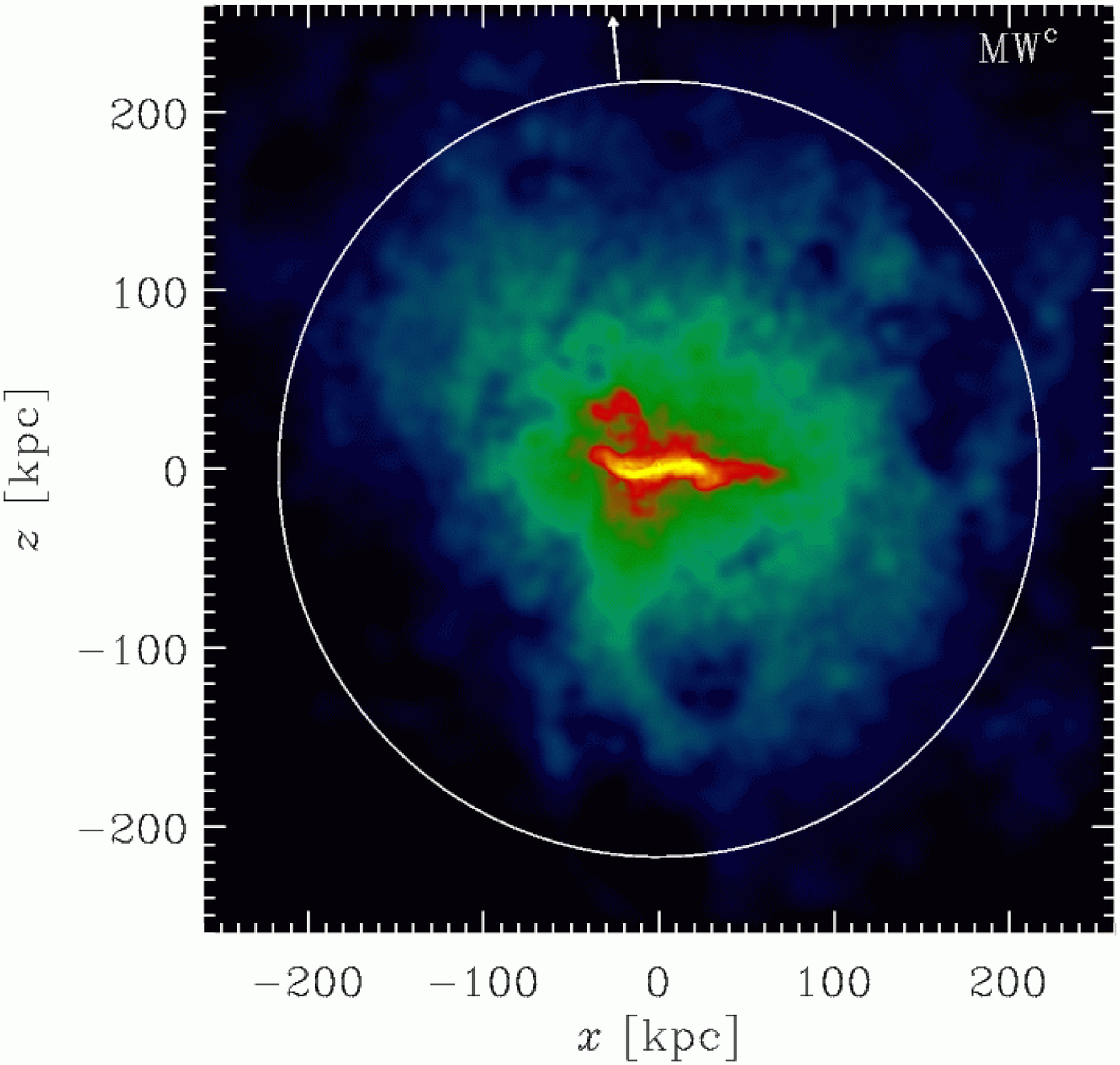}
\includegraphics[width=56mm]{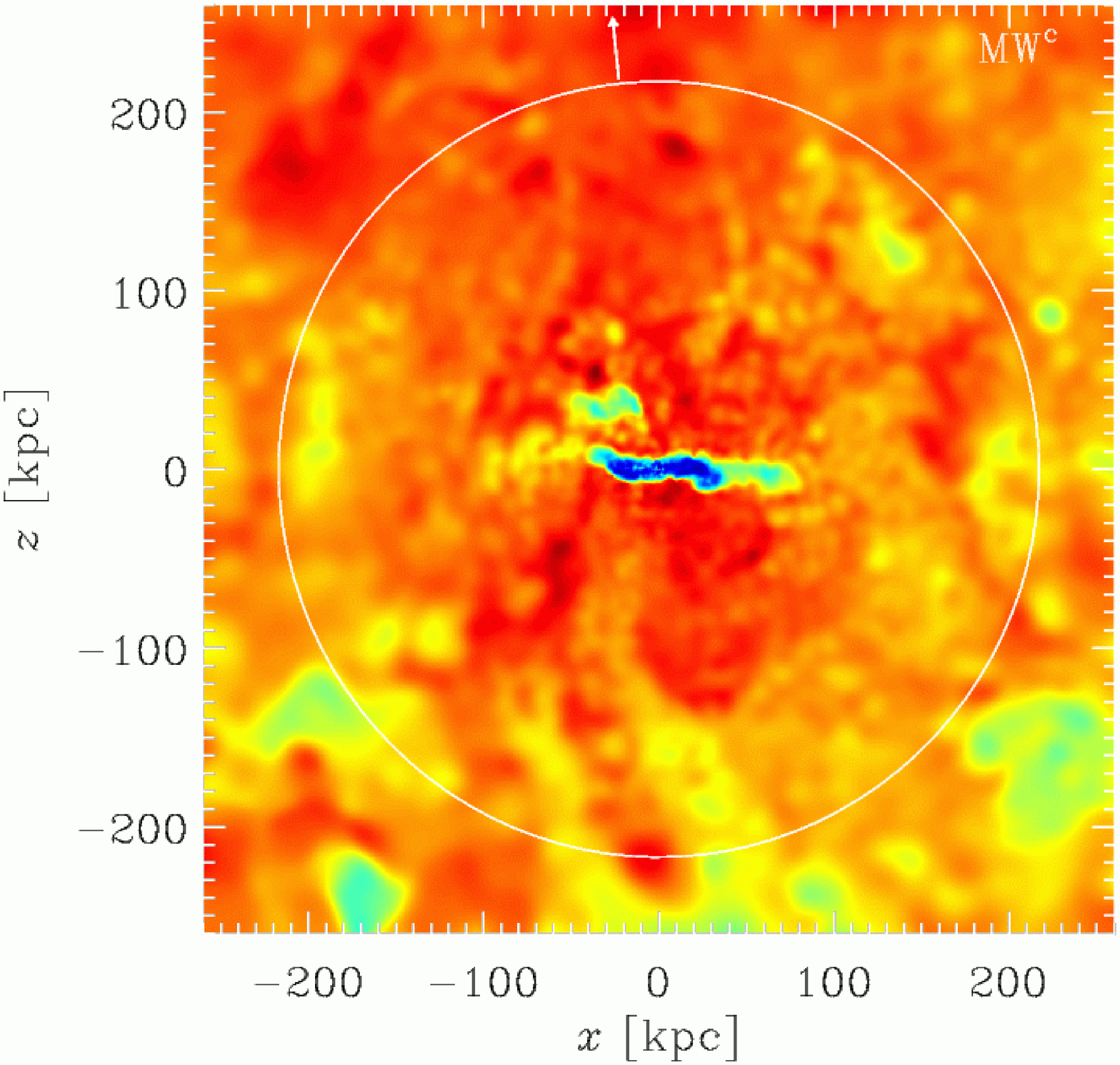}
\includegraphics[width=56mm]{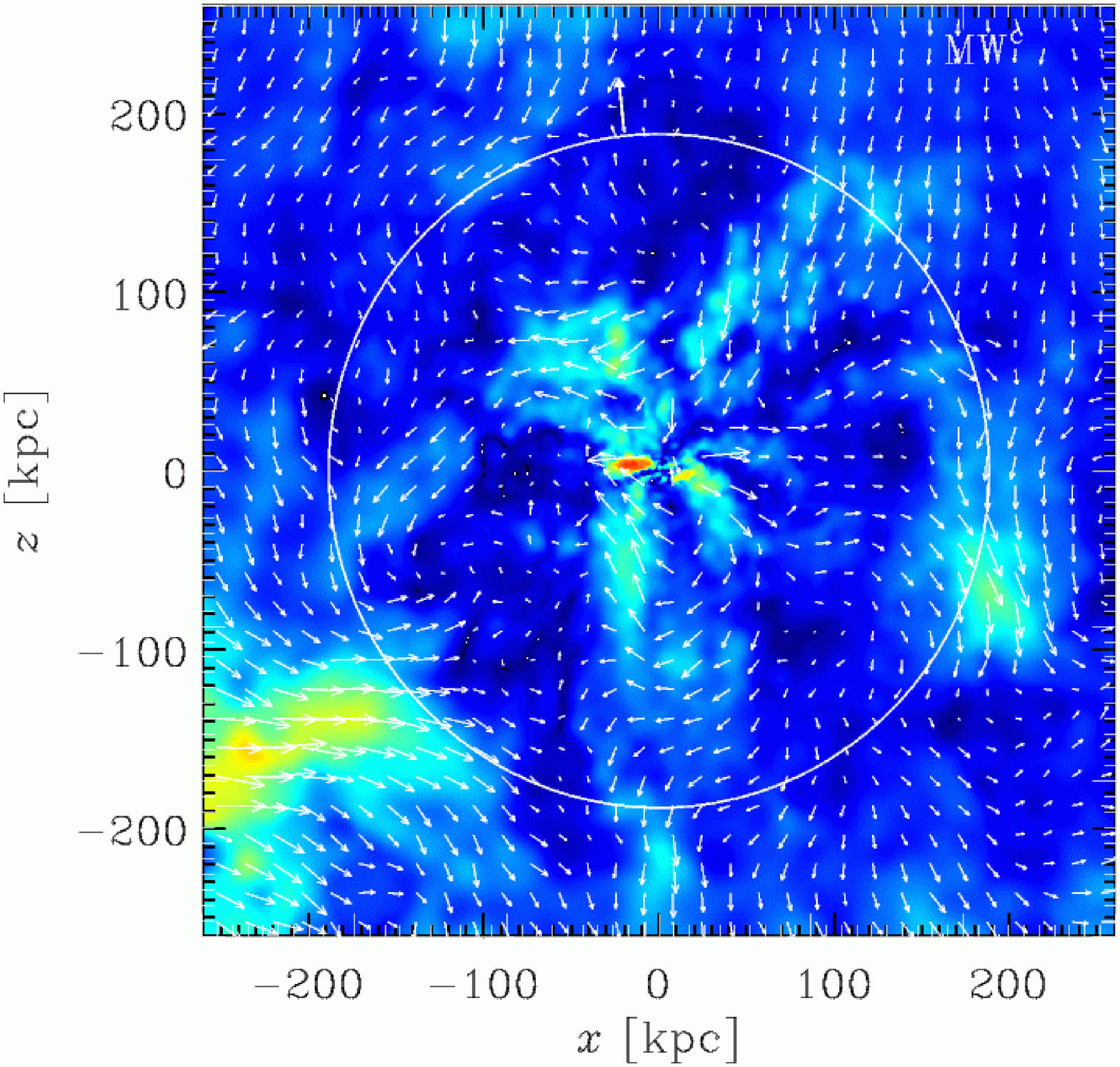}}

{\includegraphics[width=56mm]{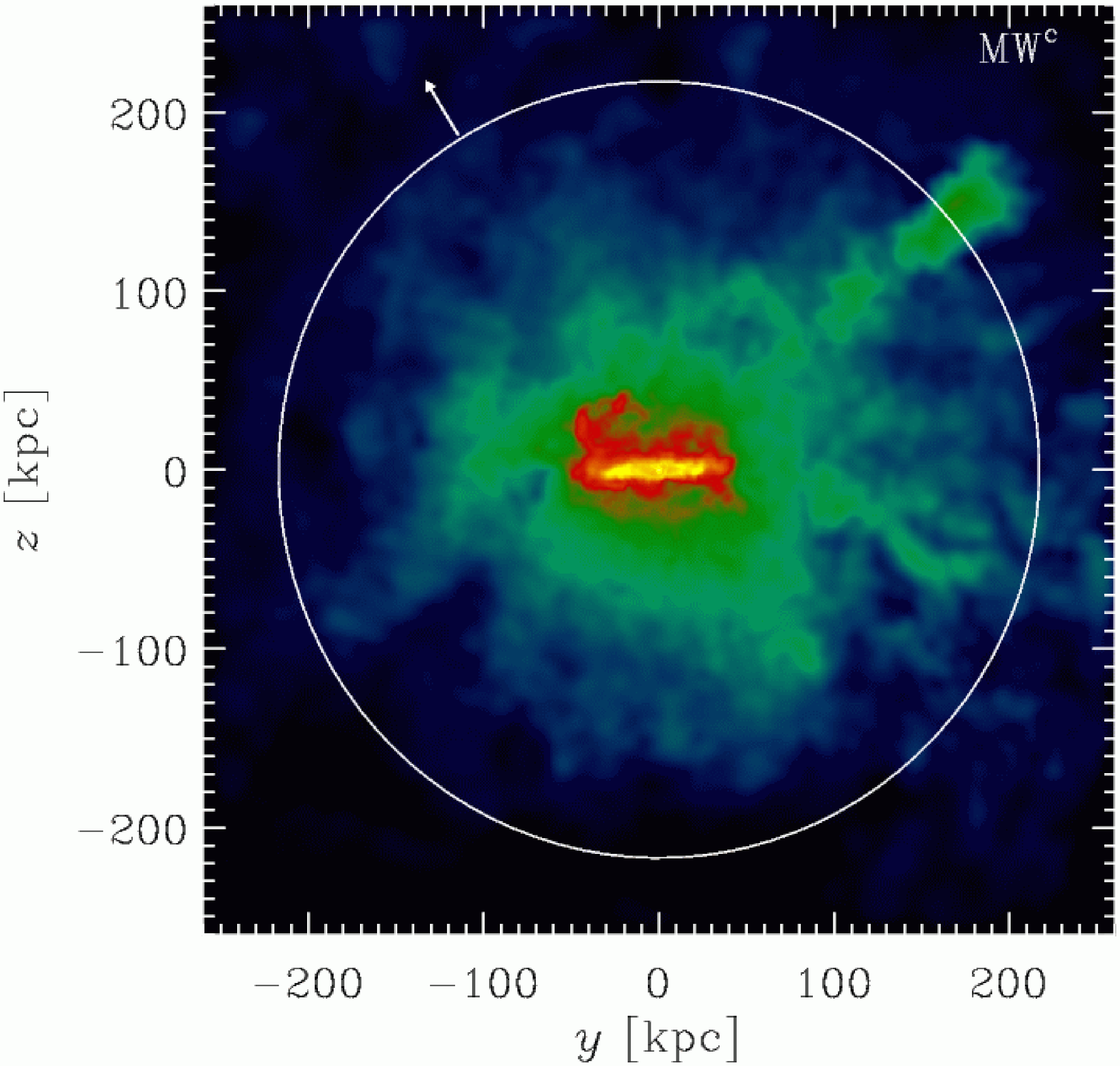}
\includegraphics[width=56mm]{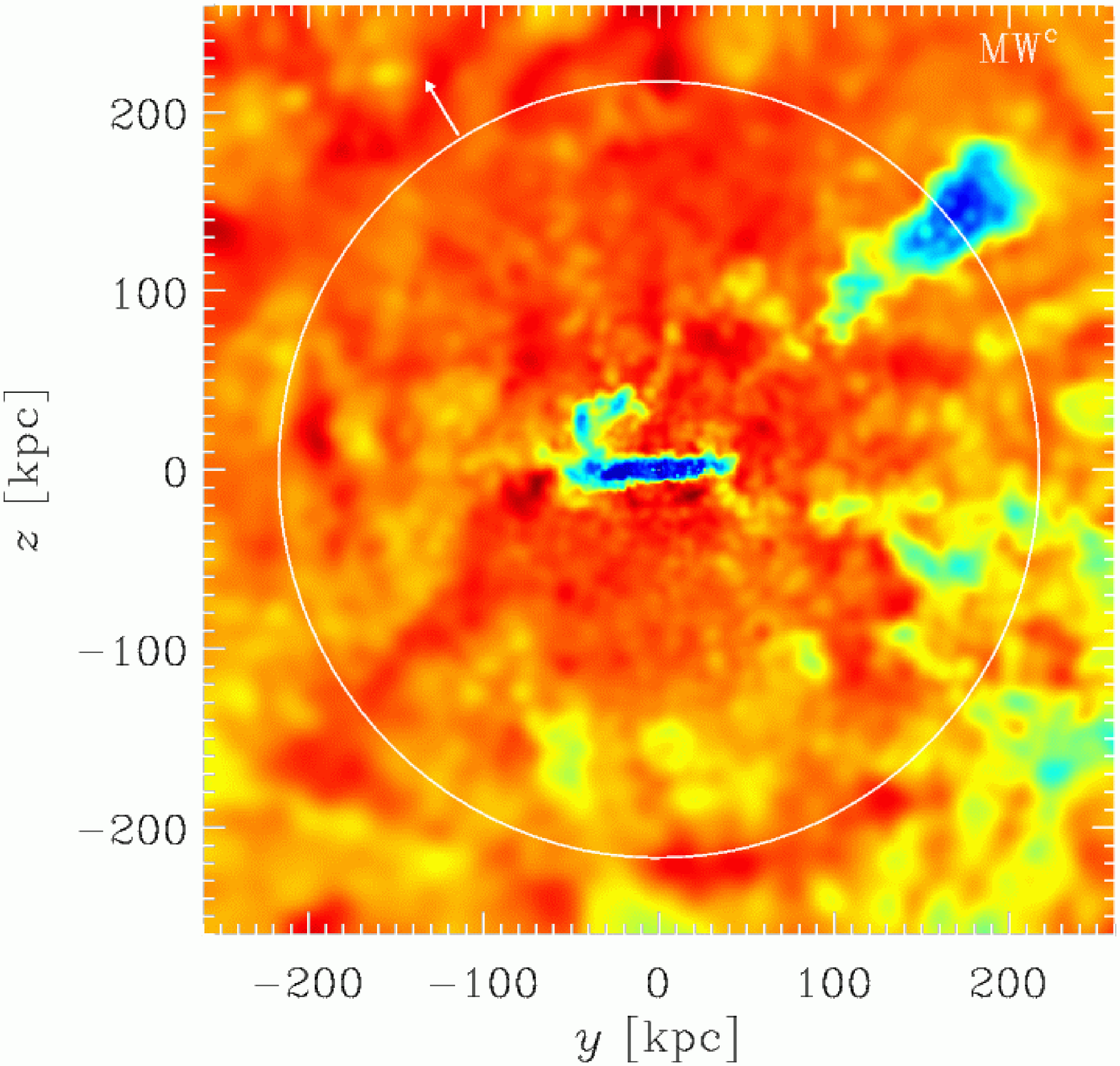}
\includegraphics[width=56mm]{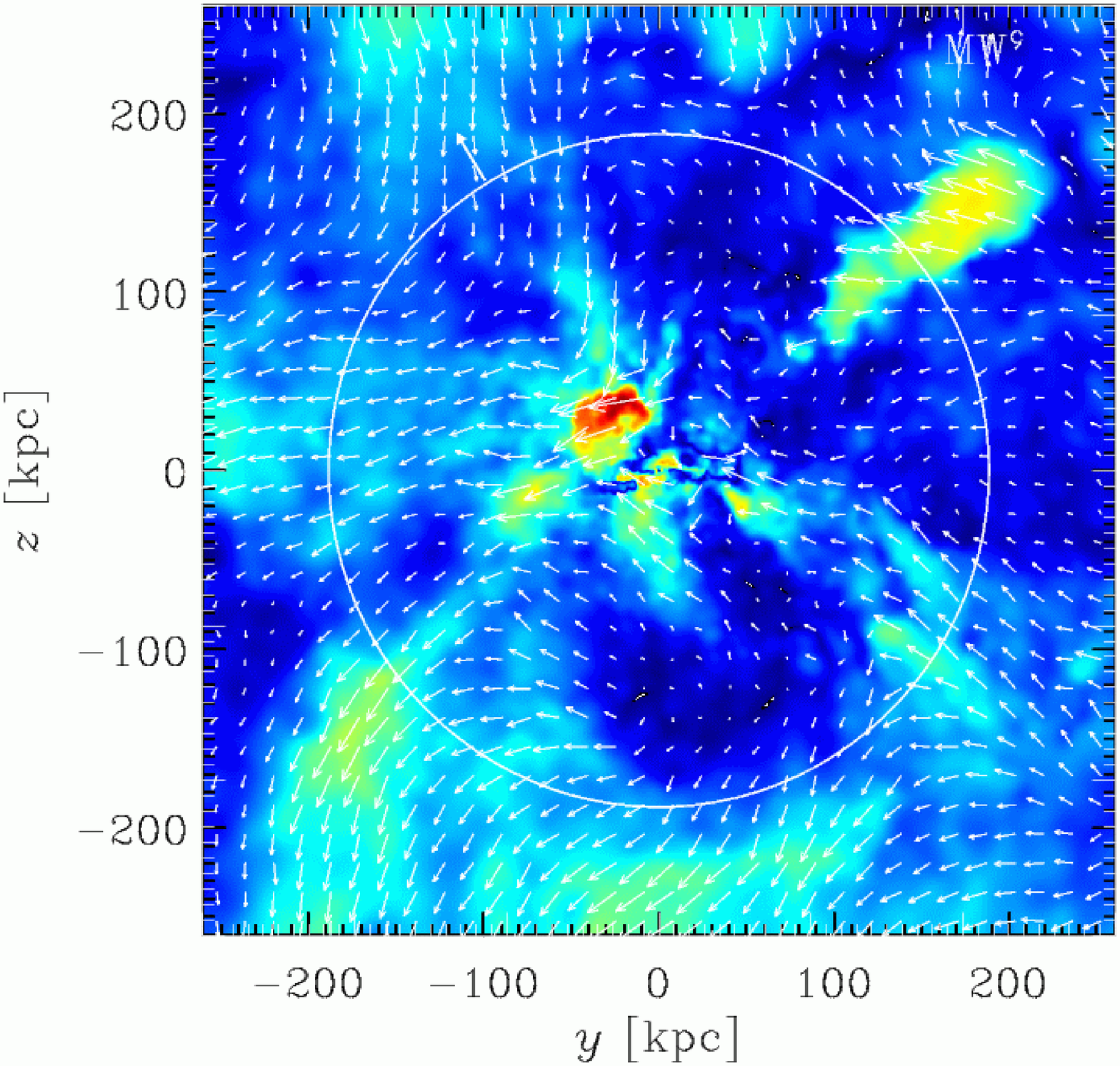}}

{\includegraphics[width=56mm]{Figures/ColorBar_density.ps}\hspace{1mm}
\includegraphics[width=56mm]{Figures/ColorBar_temp.ps}\hspace{1mm}
\includegraphics[width=56mm]{Figures/ColorBar_vr.ps}}

\caption{Face-on (upper panels) and edge-on (middle and lower panels) maps of gas density, 
temperature and projected velocity with respect to centre of mass of the simulated 
 MW$^{\rm c}$  (idem as Fig.~\ref{M31} but for the  MW$^{\rm c}$  case). The circles indicate the location 
of the virial radius and the arrows point towards the position of M31$^{\rm c}$.} 
\label{MW}
\end{center}
\end{figure*}

\begin{figure*}
\begin{center}
{\includegraphics[width=56mm]{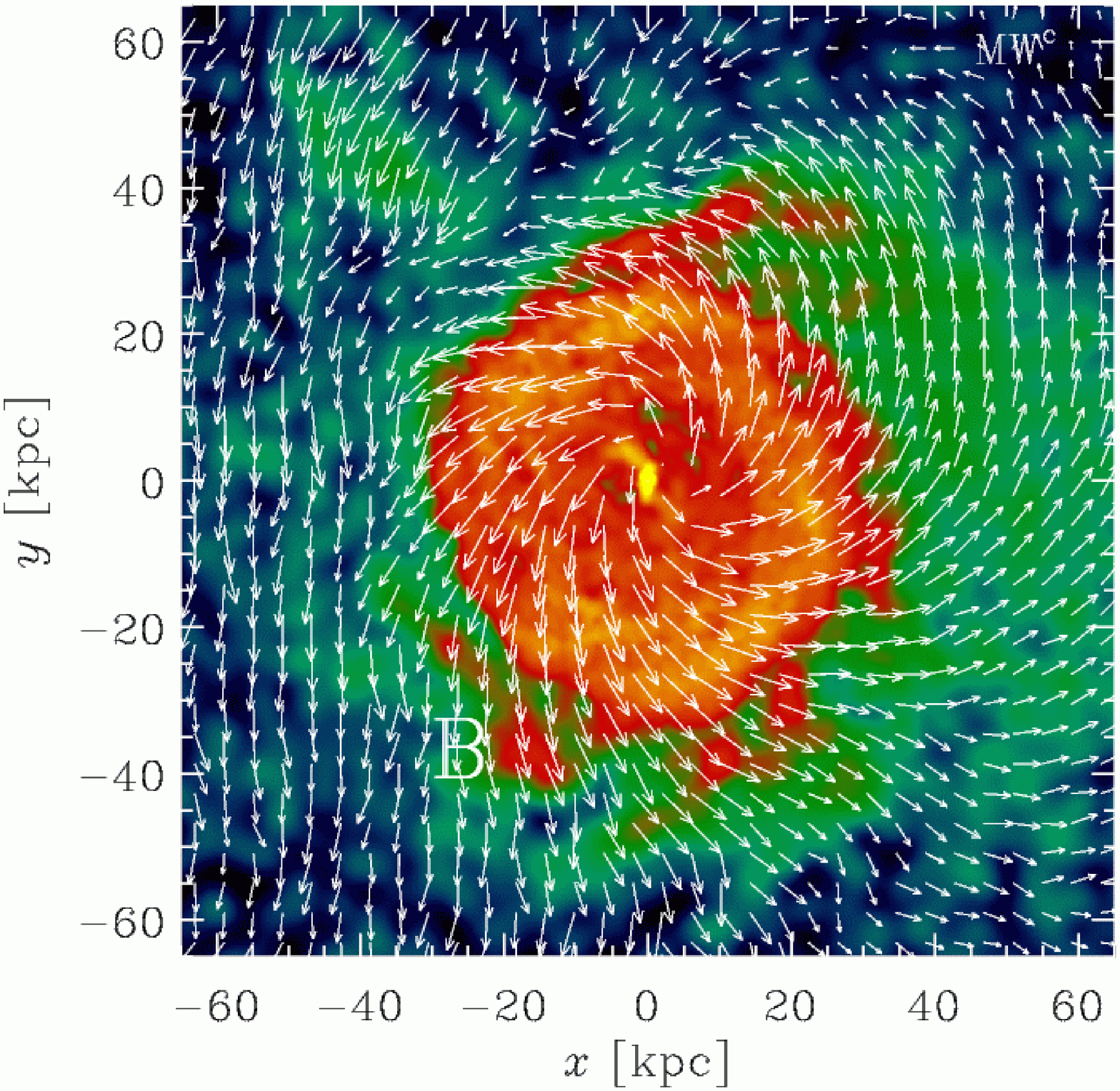}
\includegraphics[width=56mm]{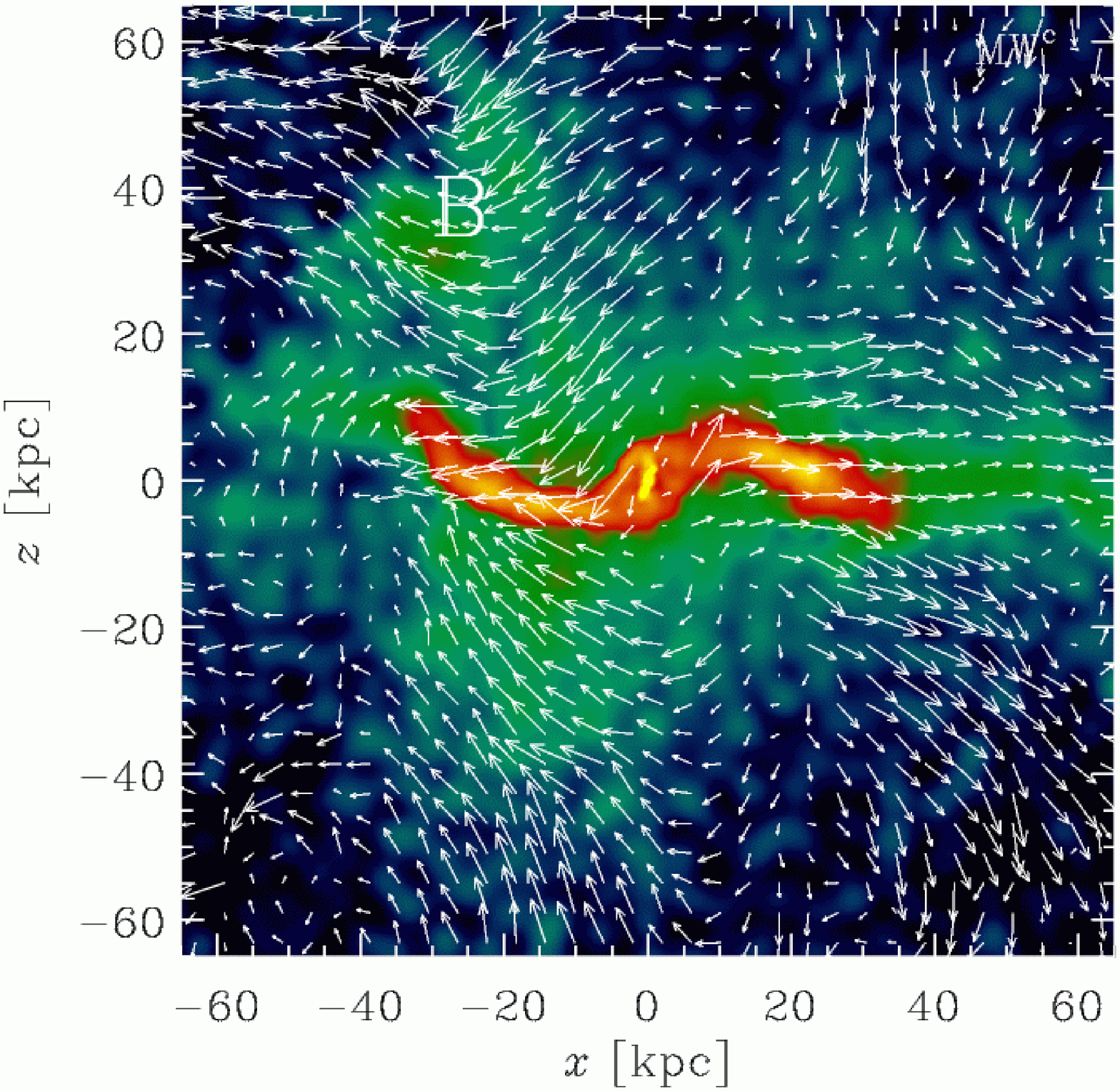}
\includegraphics[width=56mm]{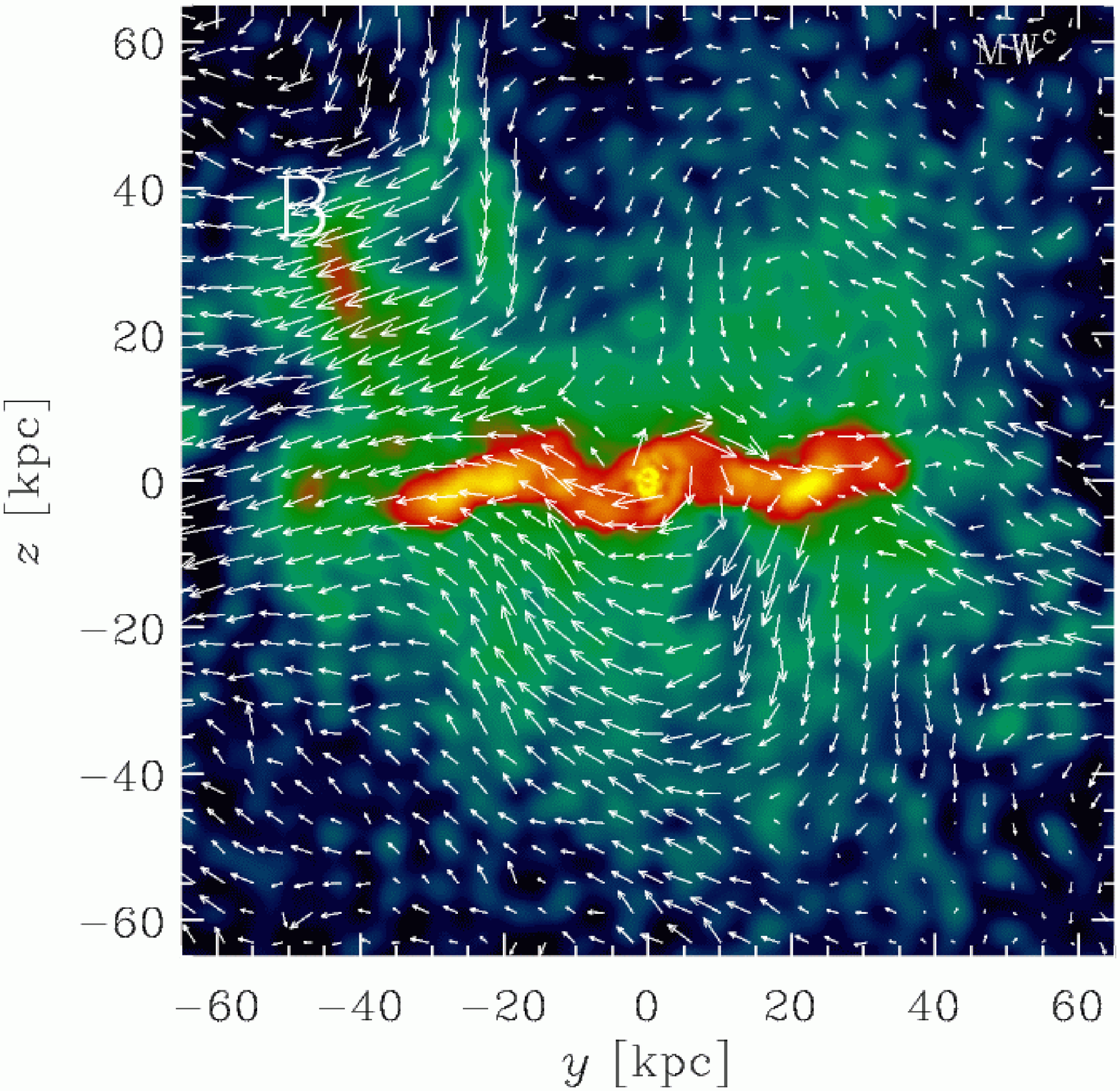}}

\caption{Zoom in of the projected gas density for  MW$^{\rm c}$, within the inner
$0.3 \times R_{\rm vir}$, and corresponding velocity field (arrows). 
The colour scale is the same as that of Fig.~\ref{MW}. 
See Section~\ref{all-sky-map_CF} for reference of label ``B''.}
\label{MW_central}
\end{center}
\end{figure*}

\begin{figure*}
\begin{center}
{\includegraphics[width=58mm]{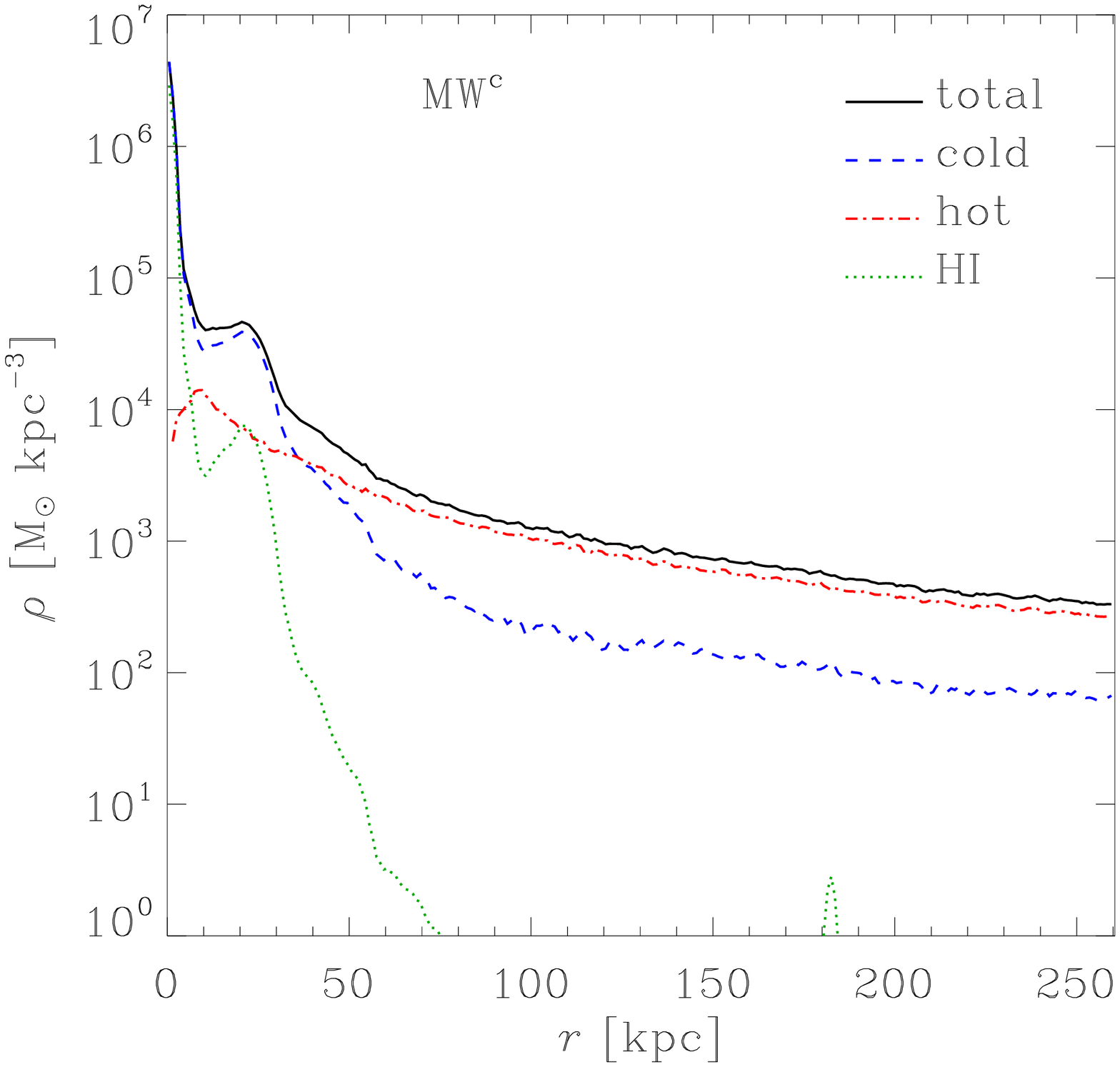}
\includegraphics[width=58mm]{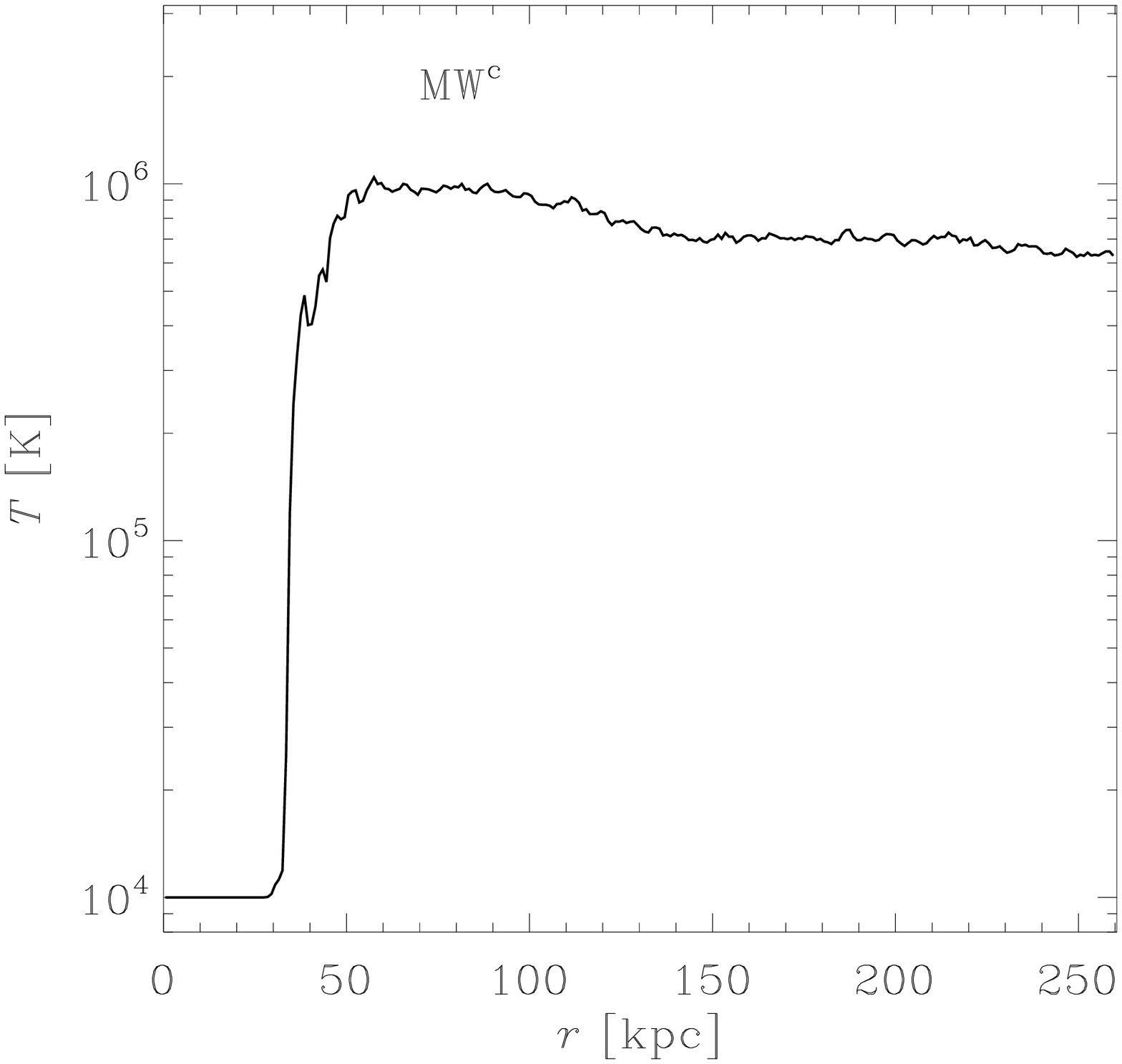}
\includegraphics[width=58mm]{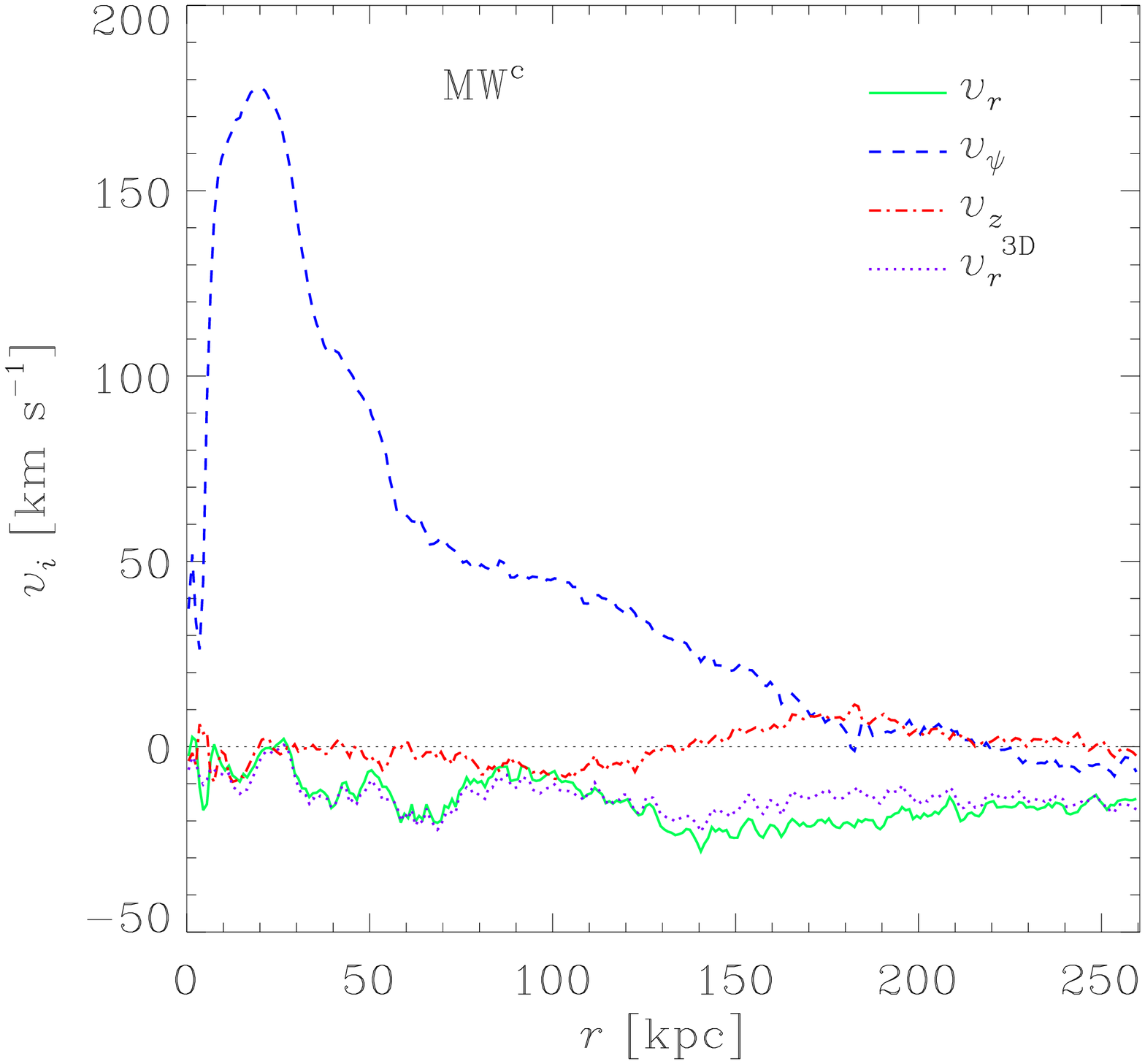}}

\caption{Density, temperature and velocity profiles for the  MW$^{\rm c}$, up to 
$1.2\times R_{\rm vir}$ (idem as Fig.~\ref{profiles_M31} but for the  MW$^{\rm c}$  case).}
\label{profiles_MW}
\end{center}
\end{figure*}

\begin{table}
\centering
\caption{Main properties of the  M31$^{\rm c}$ and MW$^{\rm c}$ simulated haloes: virial mass, virial radius, virial velocity, and total 
dark matter and gas masses. The latter correspond to values within the virial radius as defined in 
Section~\ref{sim_LG}. Masses, radii and velocities are given in units of 
$10^{10}$ M$_\odot$, kpc and km s$^{-1}$, respectively.}\label{Table_global}
\begin{tabular}{lccccc}
\hline\hline 

 &\hspace{0.1cm} $M_{\rm vir}$ \hspace{0.1cm}&\hspace{0.1cm}$R_{\rm vir}$\hspace{0.1cm} &\hspace{0.1cm}$V_{\rm vir}$\hspace{0.1cm} &\hspace{0.1cm}$M_{\rm DM}$ \hspace{0.1cm}&\hspace{0.1cm} $M_{\rm gas}$\hspace{0.1cm} \\

\hline

M31$^{\rm c}$  &  167.9 &  244.9 &   171.9 & 153.3 & 6.7 \\\\
MW$^{\rm c}$   &   125.1  & 222.2   &   155.8 & 113.1 & 5.7 \\

\hline\hline
\end{tabular}

\end{table}

\section{General properties of the gas within the LG}
\label{general}

In this section, we discuss some basic properties of the gas within
the LG, as well as in the  MW$^{\rm c}$ and M31$^{\rm c}$ haloes at $z=0$. 
We first give  a general description of the large-scale environment where  MW$^{\rm c}$  and M31$^{\rm c}$ 
reside and later discuss these systems in more detail.

\subsection{Large-scale environment}

We first focus our attention on the large-scale environment where our MW$^{\rm c}$ and M31$^{\rm c}$ 
candidates reside. Throughout this work we adopt a simple criterion 
to separate gas among cold and hot phases: a temperature threshold of $10^5\,$K. We also study 
the neutral gas component in terms of its H\,{\sc i} content. At every time, we solve for the amount 
of H\,{\sc i} assuming ionization equilibrium for all hydrogen and helium species 
including photoionization by a radiation background that accounts for the 
integrated UV field of external sources. In this work, we do not consider radiation 
generated by stars belonging to our simulated galaxies.

Fig.~\ref{large_scale_env} shows one projection of 
the gas density distribution for the hot, cold and H\,{\sc i} gas 
components in a box of $6\,$Mpc on a side centred in the location of the simulated LG. 
The virial radii of our  MW$^{\rm c}$ and M31$^{\rm c}$ galaxy candidates are indicated as small circles 
(dashed and solid lines respectively). As expected, the cold gas phase (middle panel) displays 
a more filamentary structure than the more diffuse, hot component (left-hand panel). 
The gas density 
distribution of H\,{\sc i} is shown in the right-hand panel. 
The distribution of neutral gas is much more concentrated, being  mainly located 
within the very centres of the  MW$^{\rm c}$ and M31$^{\rm c}$ haloes, and also 
in individual complexes in the intergalactic medium not related to satellite galaxies. 
We analyse in more detail the H\,{\sc i} distribution 
around the  MW$^{\rm c}$ and M31$^{\rm c}$ galaxies in Section~\ref{comp_model_obs}. 

From our simulation we can easily evaluate the fraction of gas in the different components 
within the simulated LG. This can be seen in Fig.~\ref{LG_gas_fractions} where we plot 
the fraction of hot, cold and H\,{\sc i} gas as a function of distance to the geometrical centre of 
the  MW$^{\rm c}$/M31$^{\rm c}$ system. 
For the sake of completeness, we also show the baryon fraction normalized to the 
universal value. As expected, fluctuations are higher for the smallest scales
as a result of the clumpiness of matter aggregations. The features located at $\sim$$300\,$kpc 
are owing to the presence of the MW$^{\rm c}$ and M31$^{\rm c}$ galaxies which increase (decrease) the 
fraction of cold (hot) gas. 

At dynamically relevant distances for the  MW$^{\rm c}$/M31$^{\rm c}$ interacting 
system the hot component dominates the gas mass as expected for merging haloes. 
At a scale of $1\,$Mpc the hot gas fraction reaches a value of $\sim$$80$\%. 
For larger distances, the hot gas fraction decreases slightly as the 
fraction of cold gas increases reaching a value of about $30\,$\% at a radius of $2\,$Mpc. 
These results indicate that a considerable 
fraction of the mass in the LG is most likely in the form of a large, difficult to detect, hot gas reservoir. 
In Section~\ref{comp_model_obs} we will focus on the mass content of hot gas within the haloes of our  MW$^{\rm c}$  and M31$^{\rm c}$ 
candidates and their comparison with observations. 

As mentioned above, we also evaluated the baryon fraction of our simulated LG as a function of distance. 
Interestingly, for a typical LG radius of $2\,$Mpc, we found that the baryon fraction 
has almost converged to the universal value, while at smaller distances it is slightly lower. 
For instance, if we evaluate the baryon fraction at a distance of $1\,$Mpc, which already comprises the two main 
members of the LG, the  MW$^{\rm c}$ and M31$^{\rm c}$, the baryon fraction is roughly $\sim$$20\,$\% smaller than 
the universal value showing that cosmic variance plays a role at Mpc scales. 

In the next two sections we turn our attention to the two main simulated galaxies within the LG. 
For each galaxy we rotate the system of reference in order to align 
the total angular momentum of the gas (within the inner $30\,$kpc) 
with the $z$-direction. In this way, projecting the systems in the $xy$ 
plane will result in a face-on view, while doing so in the $xz$ or $yz$ 
planes will correspond to the edge-on views.

\subsection{Andromeda galaxy candidate}

In Fig.~\ref{M31} we show the face-on (upper panels) and edge-on (middle and lower panels)
maps of density and temperature for the simulated
M31$^{\rm c}$. The right-hand panels show the corresponding projected
 velocity field, in absolute value (colours) and direction (arrows).
We plot peculiar velocities with respect to the corresponding
centre of mass velocity.
The plots cover a cubic region of $1.2$ times the virial radius of M31$^{\rm c}$ which is 
indicated by the circles. 
For every projection, arrows point in the direction towards the 
 MW$^{\rm c}$  galaxy.

From the density and temperature plots, we can see that 
the simulated M31$^{\rm c}$ system has an extended 
disc\footnote{This simulated galaxy does 
show a prominent stellar disc up to $z\approx0.3$, which is 
significantly reduced at $z=0$ as a result of the interaction with 
2 intermediate-mass satellites. For more details see Scannapieco 
et al. (in preparation).} of cold gas, and a more diffuse hot gas halo. 
The disc is thin and it shows the presence of an 
extended spiral arm-like feature which has a slightly higher
temperature than the typical $10^4\,$K of the disc. 
We also see that some of the gas around the disc, in
particular at positive values of the $z$ component
in the edge-on views, has slightly higher temperature
than the halo gas in the immediate vicinity. This
is owing to supernova feedback, which heats up
gas and drives winds, mainly in directions perpendicular
to the disc (see \citealt{Scann06}).

As a result of the combined and complex effects of cooling and supernova heating,
the gas distributions in the most central regions of the simulated M31$^{\rm c}$ 
show significant asymmetries. This can be better seen from Fig.~\ref{M31_central},
where we show zoom-in versions of the projected density of M31$^{\rm c}$, together with
the velocity field (arrows). The rotation of the gas is clearly seen
in the face-on view, while important supernova-driven outflows are present,
particularly in the positive $z$-axis region.

We found that some of the gas in the halo of M31$^{\rm c}$ is 
flowing in, as can be seen from the right-hand panels of Fig.~\ref{M31}. In particular, 
inflowing gas with a velocity of $\sim$$100\,$km s$^{-1}$ 
is seen coming from outside the virial radius all the way down to the galactic 
disc. In the central regions, however, there are important outflows, as
already evident from Fig.~\ref{M31_central}.
In general, the gas velocity structure is complex and extremely asymmetric,
as expected in a cosmological context where both internal (i.e., star formation and supernova
feedback) and external (e.g., gas accretion, interactions and mergers) effects shape the 
gas distribution in a non-trivial manner.

Fig.~\ref{profiles_M31} shows the spherically-averaged gas density, temperature and velocity 
profiles for the simulated M31$^{\rm c}$. 
The density profile is split into the hot ($T\geq10^5\,$K), cold (otherwise), and neutral gas components. 
As can be seen from the left-hand panel, the density increases significantly 
within the disc region, has a slight increase owing to the spiral arm-like feature located at a 
distance of about 60 kpc, and declines smoothly for larger radii.

As expected, in the central region, the material is mainly star-forming, cold gas, while outside 
the disc it is in a hot, diffuse phase. 
This can be seen in the middle panel of Fig.~\ref{profiles_M31} where we show the temperature profile 
of the gas. The gas disc has a temperature 
of $10^4\,$K, with a typical extent of about 40 kpc. For larger scales the temperature 
increases very quickly up to about $10^6\,$K staying pretty much constant through the hot gas 
halo (the drop in temperature between $50$ and $70\,$kpc is 
indicative of the arm-like feature discussed earlier). 
As for the neutral gas, it is the most concentrated component, and declines very
steeply after $r\sim 70\,$kpc.

The gas velocity profile can be seen in the right-hand panel of Fig.~\ref{profiles_M31} 
where we show the profiles of radial, tangential and vertical velocity as measured 
in a cylindrical coordinate system. For the vertical velocity, we adopt $V_z^*\equiv v_z$sign$(z)$ 
to easily differentiate between inflow ($V_z^*\leq0$) and outflow ($V_z^*>0$) gas velocities.
As expected, the gas velocity in the disc region is dominated by the tangential component, where 
the bump observed at $\sim$$60\,$kpc is related to the spiral arm-like feature 
described before. 
For distances approaching the virial radius, tangential velocity decreases indicating that 
the rotational support is not significant at these larger radii. 
Furthermore, we can see that, in the disc plane, there is a net
radial infall of gas which, however, has a small velocity, of the order of 10 
${\rm km\,s}^{-1}$ in the central region and reaching about 30 ${\rm km\,s}^{-1}$ near 
the virial radius.

\begin{figure*}
\begin{center}
{\includegraphics[width=75mm]{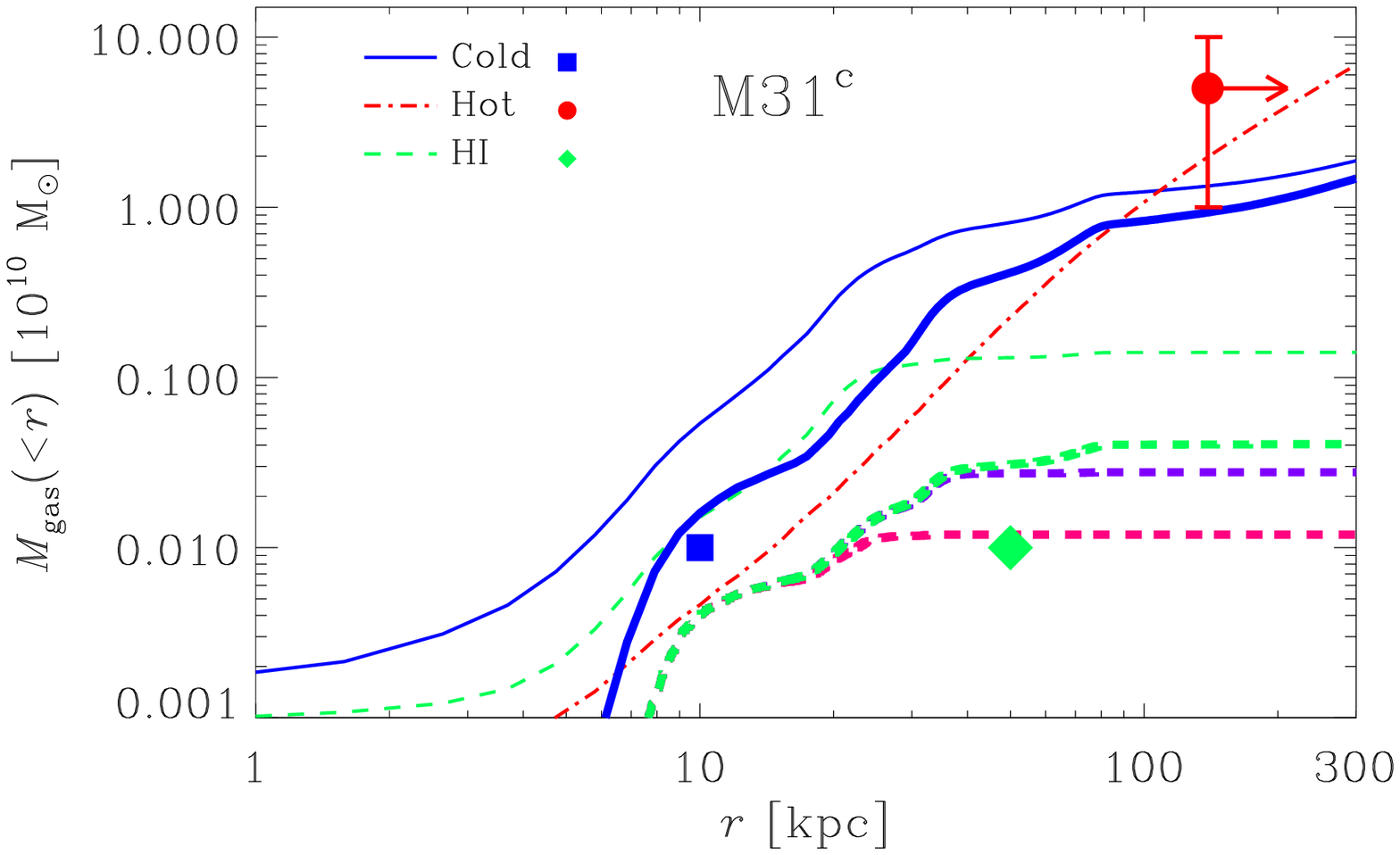}
\hspace{0.7cm}
\includegraphics[width=75mm]{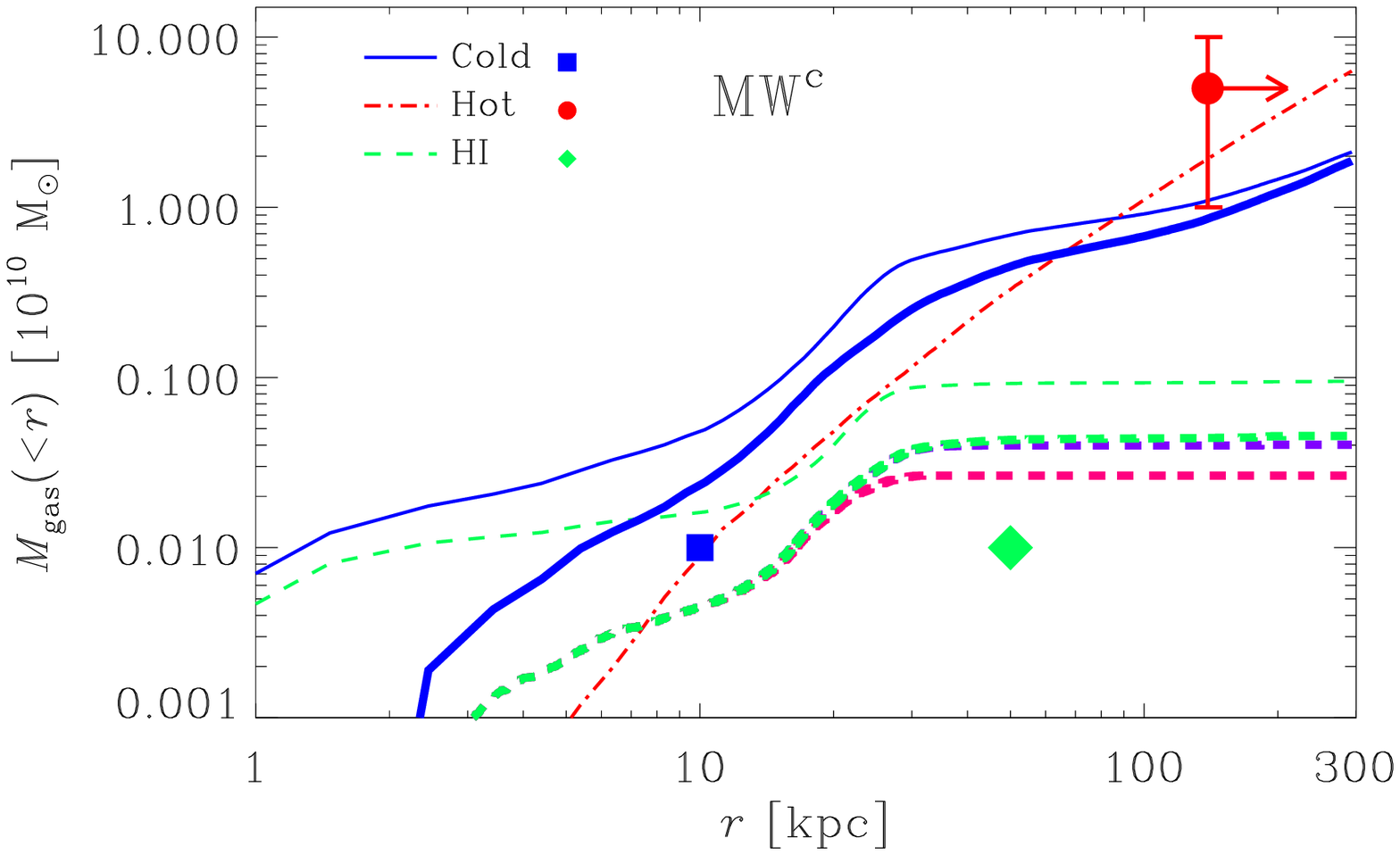}}

\caption{Cumulative mass profiles of different gas components (i.e., cold, hot and H\,{\sc i}) 
  for the M31$^{\rm c}$ and  MW$^{\rm c}$  simulated galaxies (left- and right-hand panels respectively). 
  The cold and H\,{\sc i} contributions excluding the gaseous disc are shown as thicker lines. 
  In the case of H\,{\sc i} we also plot the cumulative mass profiles for column density limits 
  of $N_{\rm HI}\geq 10^{19}\,$cm$^{-2}$ (blue dashed lines) and 
  $N_{\rm HI}\geq 10^{20}\,$cm$^{-2}$ (magenta dashed lines) excluding the gaseous disc.
  Symbols show Milky Way gas-mass estimates from observations for material 
  within $10\,$kpc \citep[cold; see e.g.][]{Shull09,Lehner11}, $50\,$kpc \citep[H\,{\sc i}; see][]{Richter12} 
  and $139\,$kpc \citep[hot; see][]{Gupta12}. In the latter, the 
  arrow indicates a lower-limit distance.}
\label{mass_profiles}
\end{center}
\end{figure*}

\subsection{Milky Way galaxy candidate}

Now we turn our attention to the simulated  MW$^{\rm c}$  system. As explained above,
 MW$^{\rm c}$  is slightly less massive than M31$^{\rm c}$, but we expect a similar behaviour
of the gas, specially in the halo region. 
Fig.~\ref{MW} shows the face-on (upper panels) and edge-on (middle and lower panels) 
projections of density, temperature and projected velocity for the simulated  MW$^{\rm c}$.
In these plots, the colour scale for each quantity is the same as that used for M31$^{\rm c}$, 
in order to highlight the possible differences and similarities of the two galaxies.

As we found for M31$^{\rm c}$, the simulated  MW$^{\rm c}$ system also has 
a disc\footnote{The associated stellar disc 
has a radius of about $12\,$kpc and a mass of $2.83\times10^{10}\,\Msun$ 
according to the kinematical decomposition method. For more details see 
Scannapieco et al. (in preparation).} of cold, dense gas.
In this case, no large spiral arms are seen, but we do observe a significant
asymmetry in the gas density, evident in both the face-on and edge-on maps,
specially near the disc region. The temperature distribution is 
extremely complex; although most of the gas in the halo is at temperatures 
close to $T \sim 10^6\,$K, there are regions with somewhat colder temperatures.
The most salient features in our simulated  MW$^{\rm c}$  are the presence
of a relatively large region of significant accretion of $\sim$$10^5\,$K  gas (to the
upper-left direction in the face-on view), and a more localized accretion
region to the upper-right corner in the $yz$ view near the virial radius.
The typical velocities of these inflows are $\sim$$100$--$150$ km s$^{-1}$.

From the right-hand panels of Fig.~\ref{MW}, we
can observe a high degree of complexity, as we have seen
for M31$^{\rm c}$. Again, most of the gas is inflowing, except for a few regions
dominated by supernova outflows. Fig.~\ref{MW_central} shows a zoom-in
in the disc region of the  MW$^{\rm c}$, where the left-hand panel shows a face-on 
view, while the middle- and right-hand panels show edge-on views. 
From the edge-on projections, it is clear that the
gas disc, instead of showing the usual rectangular-like shape, presents 
some asymmetries. The disc is also not fully symmetric in the 
outer disc regions, as particularly evident in the face-on
view (left-hand panel), at $x>40\,$kpc.
As in M31$^{\rm c}$, such complexity is the result of the non trivial
interplay between the different physical processes (cooling,
heating, exchange of gas between the halo and the intergalactic
medium) involved in the assembly of $\Lambda$CDM galaxies.

Fig.~\ref{profiles_MW} shows the spherically-averaged profiles of gas density (left-hand panel), 
temperature (middle panel) and velocity (right-hand panel) for the  MW$^{\rm c}$. 
In this case, the gas disc is smaller in extent, reaching a radius of $\sim$$30\,$kpc.
The neutral gas is, as we found for M31$^{\rm c}$, the most concentrated
component.  
The gas temperature at the disc is of about $10^4\,$K. At larger distances, it peaks 
at $\sim$$10^6\,$K to gradually decrease to about $6\times 10^5\,$K near the virial radius. 
As before, the velocity of gas belonging to the disc is dominated by their tangential components, 
reaching similar velocity values than those found for M31$^{\rm c}$. However, contrary to the case of M31$^{\rm c}$, we detect slightly higher 
vertical velocities in comparison. More importantly, we found that the radial velocities
are negative with absolute values ranging from $10-20$ km s$^{-1}$ near
the disc edges, and up to $20-30$ km s$^{-1}$ in the halo. This indicates
the presence of radial gas infall that occurs mainly in the galactic disc plane.

\section{Hot, Cold and Neutral gas phases: comparison with observations}
\label{comp_model_obs}

We have so far discussed the general distribution and properties of the gas within
the simulated  MW$^{\rm c}$  and M31$^{\rm c}$ haloes. In this section, we study the different gas phases 
present in the simulation (i.e., cold, hot or H\,{\sc i}) and compare our
results with available observations. 
As mentioned before, we use a simple threshold temperature 
of $10^5\,$K to separate gas among cold and hot phases. 
Additional information of our simulations is the amount of neutral hydrogen,
which is a useful quantity to compare with observations
given the large amount of available $21\,$cm data. 

The cumulative mass profiles of the different gaseous components of our two simulated galaxies, 
together with observational estimates for the hot, cold and H\,{\sc i} gaseous phases in the 
Milky Way, are shown in Fig.~\ref{mass_profiles}. 
Additionally, Table~\ref{table_Mgas} shows the amount of gas in the 
different phases for the MW$^{\rm c}$ and M31$^{\rm c}$ within several radii; 
results are presented for the virial radius, as well as for $10$, $50$ 
and $100\,$kpc from the centre, respectively. These scales enable us to compare 
our results with observations in a more consistent way.

\begin{table*}
\centering
\caption{Total mass of cold, hot and H\,{\sc i} gas of the simulated M31$^{\rm c}$ and MW$^{\rm c}$ 
galaxies within different radii $R$. We also show the gas 
mass in the different components within $10\,$ and $50\,$kpc, but excluding the
disc region. Additional column density cuts are considered in the 
H\,{\sc i} case. All masses are given in units of $10^{8}\,$M$_\odot$.}.
\label{table_Mgas}
\begin{tabular}{lcccccccccc}
\hline\hline
Radius (kpc) & \multicolumn{2}{c}{$M_{\rm cold}(\leq$$R)$} & \multicolumn{2}{c}{$M_{\rm hot}(\leq$$R)$} & \multicolumn{2}{c}{$M_{\rm HI}(\leq$$R)$} & \multicolumn{2}{c}{$M_{\rm HI}(\leq$$R)^\dagger$} & \multicolumn{2}{c}{$M_{\rm HI}(\leq$$R)^\ddagger$}\\
                  & M31$^{\rm c}$ &  MW$^{\rm c}$ 
                  & M31$^{\rm c}$ &  MW$^{\rm c}$
                  & M31$^{\rm c}$ &  MW$^{\rm c}$
                  & M31$^{\rm c}$ &  MW$^{\rm c}$  
                  & M31$^{\rm c}$ &  MW$^{\rm c}$  \\         

\hline
$10$            &  4.84 &  4.59   & 0.41   & 0.76 &  1.38  &    1.58  &    1.38  &    1.58  &    1.38   &   1.58 \\
$10$  (no disc) &  1.42 & 2.16    & 0.27   & 0.56 &  0.40  &   0.45  &   0.40  &   0.45   &  0.40  &   0.45\\
$50$            &  81.07  & 68.33 & 22.04  & 32.32  & 13.06   &   9.21   &   12.71   &   8.90   &   10.03   &   6.44\\
$50$  (no disc) & 40.97 & 44.48   & 21 .44 & 30.84  & 3.08  &   4.09  &  2.73  &  3.98  &  1.19   &  2.65 \\
$100$           & 123.26 & 91.22  & 106.09 & 109.60 & 14.01 &   9.30  &  12.75 &  8.90  &  10.03  &  6.44  \\
$R_{\rm vir}$   & 166.56 & 159.45 & 499.19 & 408.94 & 14.04 &   9.45  &  12.75 &  8.90  &  10.03  &  6.44  \\

\hline\hline\\
\end{tabular}\\
\vspace{-0.3cm}
\tiny \hspace{-8cm} $^{\dagger}N_{\rm HI}\geq 10^{19}\,$cm$^{-2}$, $^{\ddagger}N_{\rm HI}\geq 10^{20}\,$cm$^{-2}$\\
\normalsize
\end{table*}

\subsection{Warm-hot gaseous haloes}

Recent observations have allowed us to estimate a number of quantities
related to the amount and nature of the gas within the haloes
of the Milky Way and Andromeda galaxies \citep[][]{Sembach03,Gupta12}.
In particular, studying the X-ray absorption features 
of O\,{\sc vi} and O\,{\sc vii}, \citet{Gupta12} inferred the presence of a
large warm-hot gas reservoir surrounding the Milky Way, with a temperature 
$T\gtrsim 10^6\,{\rm K}$. 
These authors also give an estimate of the amount of mass in this warm-hot phase
traced by O\,{\sc vii} absorbers  in the Milky Way, which is $M_{\rm hot}\gtrsim6.1 
\times 10^{10}\,\Msun$. 
This estimate, however, strongly depends on several model assumptions,
such as the metallicity of the gas and the O\,{\sc vii} ionization fraction, and
thus should be interpreted with caution.
The value  given above corresponds to a path-length of $l>139\,$kpc, where the 
oxygen solar abundance of \citet{AndersGrevesse89} was used. 
Adopting newer estimates of the oxygen solar abundance
results in even larger path-lengths, thus decreasing the lower bound 
for the mass of the Milky Way hot gaseous halo. For instance, the use
of the \citet{Asplund09} oxygen solar abundance results in an estimate of 
$M_{\rm hot}\gtrsim1.2 \times 10^{10}\,\Msun$ 
\citep[see][for a discussion of the different assumptions involved in these 
kind of estimations]{Mathur12}. 
Results of other researchers also support values for
the total  mass of the Milky Way hot coronal gas in the range 
$M_{\rm hot}=10^{10}-10^{11}\,\Msun$ 
\citep[e.g.,][]{Fang13}. However, we note that not all authors 
agree with these scalings for the size and mass of the Milky Way's hot gaseous corona. 
For a critical discussion concerning this point see 
e.g. \cite{Collins05}, \cite{Bregman07}, \cite{Yao08} and \cite{Anderson10}.

\begin{figure*}
\begin{center}
\includegraphics[width=65mm]{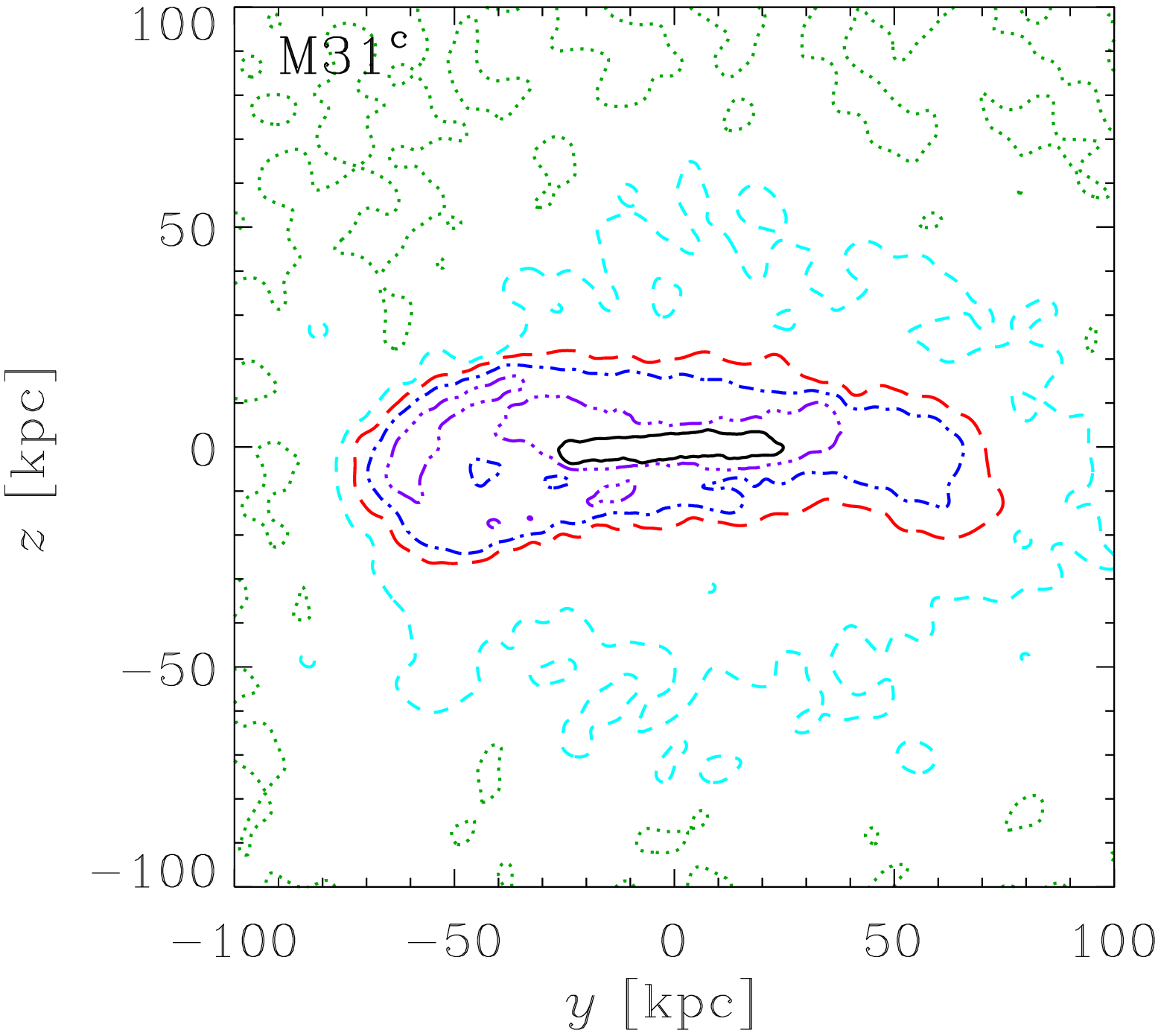}
\hspace{0.7cm}
\includegraphics[width=65mm]{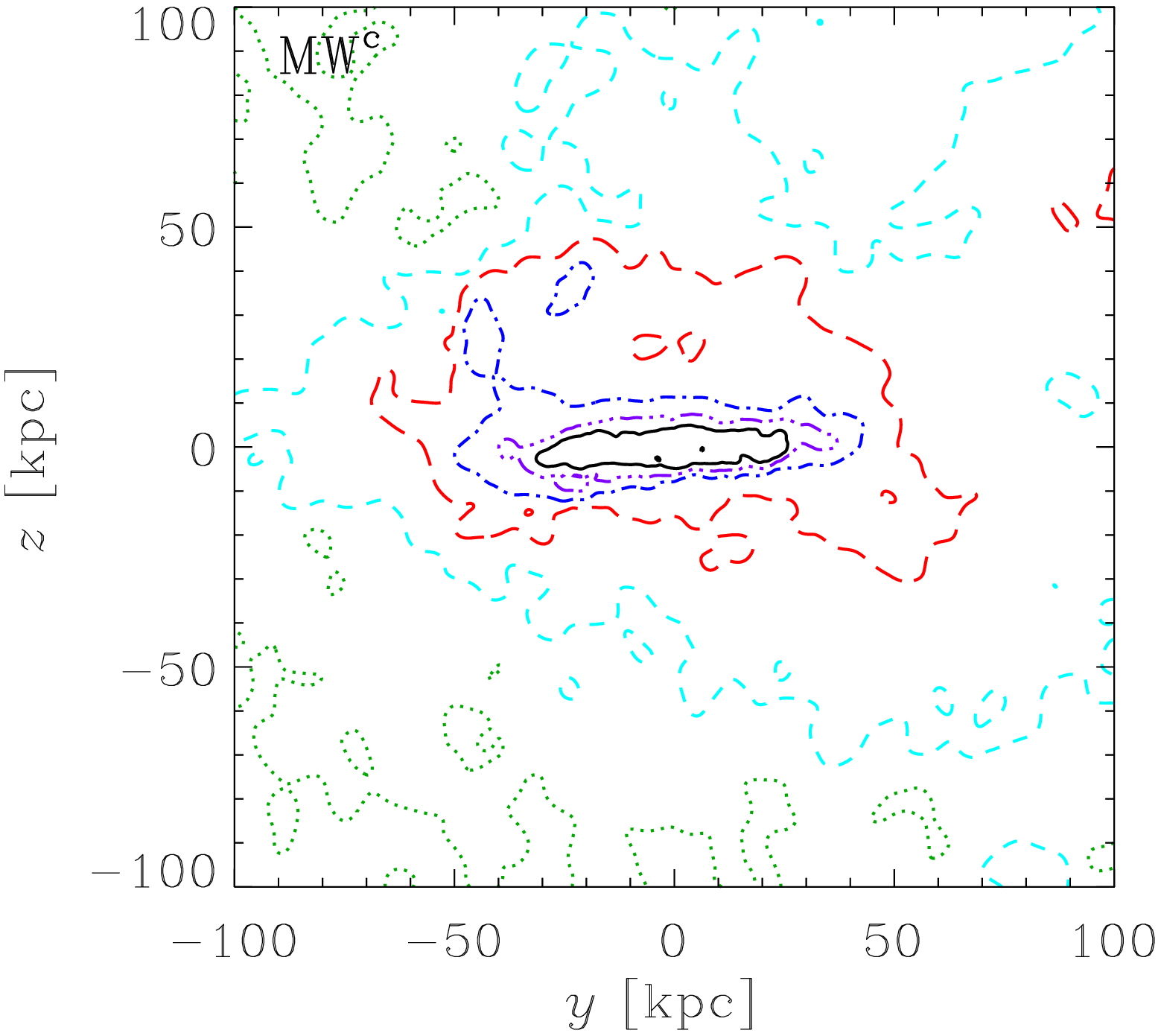}
\caption{Contours of the H\,{\sc i} gas distribution around the simulated M31$^{\rm c}$ (left-hand panel) and  MW$^{\rm c}$  (right-hand panel) for two arbitrary edge-on views. 
The contours indicate column densities of $N_{\rm HI}\geq10^{15}\,$cm$^{-2}$ (dotted lines); $N_{\rm HI}\geq10^{16}\,$cm$^{-2}$ (short-dashed lines); 
$N_{\rm HI}\geq10^{17}\,$cm$^{-2}$ (long-dashed lines); $N_{\rm HI}\geq10^{18}\,$cm$^{-2}$ (dotted-dashed lines); $N_{\rm HI}\geq10^{19}\,$cm$^{-2}$ (triple-dot-dashed lines), 
and $N_{\rm HI}\geq10^{20}\,$cm$^{-2}$ (solid lines).}
\label{HVCs_simulated_contours}
\end{center}
\end{figure*}

As shown in the previous section, the haloes of our simulated
 MW$^{\rm c}$  and M31$^{\rm c}$ systems, in particular at
distances greater than $\sim$$50\,$kpc, have temperatures close to $10^6\,$K 
(middle panels of Figs.~\ref{profiles_M31} and~\ref{profiles_MW}). 
Fig.~\ref{mass_profiles} shows the cumulative mass profile of 
the hot ($T\geq10^5\,$K) component
of our simulated galaxies (dot-dashed lines).
When we consider all the gas up to the virial radius,
we find that the cumulative mass is
$M_{\rm hot} \approx 5 \times 10^{10}$ 
and $4.1 \times 10^{10}\,\Msun$ for M31$^{\rm c}$ and  MW$^{\rm c}$, respectively 
(see Table~\ref{table_Mgas}). 
These values are similar to the observational estimates discussed above. 
If we consider only gas enclosed within $100\,$kpc the resulting gas masses 
lower to $\sim$$1.1 \times 10^{10}\,\Msun$ for both simulated
galaxies. These quantities are again of the same 
order of magnitude than the observational
estimates corresponding to the Milky Way.

\subsection{Cold-warm gas component}
\label{cold_data}

Observations of circumgalactic gas with temperatures below $\sim$$10^5\,$K may occur either as 
a diffuse, ionized medium, or as a neutral one. The neutral gas fraction in these structures 
is too small for the H\,{\sc i} line to be detected, however, their neutral hydrogen content 
can be estimated from the analysis of the H\,{\sc i} Lyman series absorption 
\citep[e.g.,][]{Richter09}. In our simulation, we found that most of the material 
corresponding to the cold gas component is mainly composed of ionized gas, in agreement with 
observational studies.

Diffuse, ionized gas in the extended Milky Way halo has been studied in
UV absorption of intermediate and high ions of carbon, silicon, and other
elements (e.g., C\,{\sc iii}, C\,{\sc iv}, Si\,{\sc iii}, Si\,{\sc iv}). 
UV observations with {\it FUSE} and {\it Hubble Space Telescope}/STIS indicate that 
the warm ionized gas in the Milky Way halo comes in individual gas complexes and 
coherent structures, thus similar to those found for H\,{\sc i} structures 
\citep[see, e.g.,][]{Collins09,Shull09,Winkel11,Lehner12,Herenz13}.
Such diffuse, ionized halo structures have large sky covering fraction 
of more than 70 percent \citep[][]{Shull09,Herenz13}, suggesting
that this gas phase is widespread and it may harbour a significant fraction of
the gas mass with $T<10^5\,$K in the Milky Way halo.

\citet{Shull09} and \citet{Lehner11} estimated a total
mass of this warm, ionized gas component in the Milky Way halo, of $\sim$$10^8\,\Msun$ 
at a distance of $\sim$$10\;$kpc above the galactic plane.
As for the coronal gas, it has to be noted, however, that this estimate 
depends critically on the (uncertain) ionization conditions and the chemical 
composition of the gas and thus must be regarded as a rough estimate. 
Nevertheless, we can compare this result with the simulation 
outcomes. Fig.~\ref{mass_profiles} shows the
cumulative mass profile of  warm-cold  gas ($T<10^5\,$K)
 in the simulated  MW$^{\rm c}$  and M31$^{\rm c}$, which we use as a proxy
for the amount of warm, ionized mass. Note that 
the fraction of neutral gas mass is always a very small fraction of the
total amount of cold gas, and therefore we can use the cold gas safely
to compare with observational results.
Because observations only count gas outside the discs, we show 
in Fig.~\ref{mass_profiles} the profiles  including and excluding the 
galactic discs  (thin and thick solid lines).
In each case, to identify the galactic disc, we first rotate the galaxies 
to get the disc plane, and make a cylindrical cut of $30$ kpc radius and 4 kpc 
height (more details will be presented in Section~\ref{HI_data}). 
After excluding the gaseous disc contributions, we found, for distances within $10\,$kpc, 
similar results as in observations: the MW$^{\rm c}$ and M31$^{\rm c}$ systems show a 
cold gas amount of $M_{\rm cold}(\leq10\,{\rm kpc})\approx2.1\times10^8\,\Msun$ and 
$2.6\times10^8\,\Msun$ respectively. Interestingly, our 
cold gas mass estimates at larger scales go also in line with the results 
of the ``COS-Halos'' sample \citep{Werk14}: a set of 38 quasar sightlines 
passing $L \sim L^{*}$ galaxies at $z\sim0.2$. On average, these authors found 
a lower limit of cold material of $2\times10^{10}\,\Msun$ within a distance 
of $160\,$kpc, whereas we get a value of $\sim$$1.6\times10^{10}\,\Msun$ within 
the virial radii of our simulated galaxies as it is shown in Table~\ref{table_Mgas}.

\subsection{H\,{\sc i} gas component}
\label{HI_data}

We turn our attention to the distribution of H\,{\sc i} gas in the
simulated galaxies, and compare our results with observations.
In and around the Milky Way cold gas that is predominantly 
neutral can be observed through radio observations of the 
$21\,$cm line if the total H\,{\sc i} column density 
lies above a threshold of $\sim$$7\times 10^{17}\,$cm$^{-2}$ \citep{Wakker04}. 

Gas-rich spiral galaxies such as the Milky Way and Andromeda are known to 
have extended H\,{\sc i} discs with radii up to a few dozen kpc and 
typical H\,{\sc i} column densities $N_{\rm HI} \gtrsim 10^{20}\,$cm$^{-2}$ 
\citep[e.g.,][]{Rosenberg03,Zwaan05,Levine06,Levine08}. 
While the radial extent of the Milky Way H\,{\sc i} disc 
is not possible to be measured directly owing to the interior 
vantage point as an observer, \citet{Braun09} show that Andromeda has an H\,{\sc i} 
disc with a radial extension of $\sim$$30\,{\rm kpc}$ for column densities
above $N_{\rm HI} \gtrsim 2 \times 10^{20}\,$cm$^{-2}$ 
\citep[however, see e.g.][for a detailed analysis of 
the structure of the H\,{\sc i} disc in our Galaxy]{Levine06,Levine08}.

\begin{figure*}
\begin{center}
\includegraphics[width=120mm]{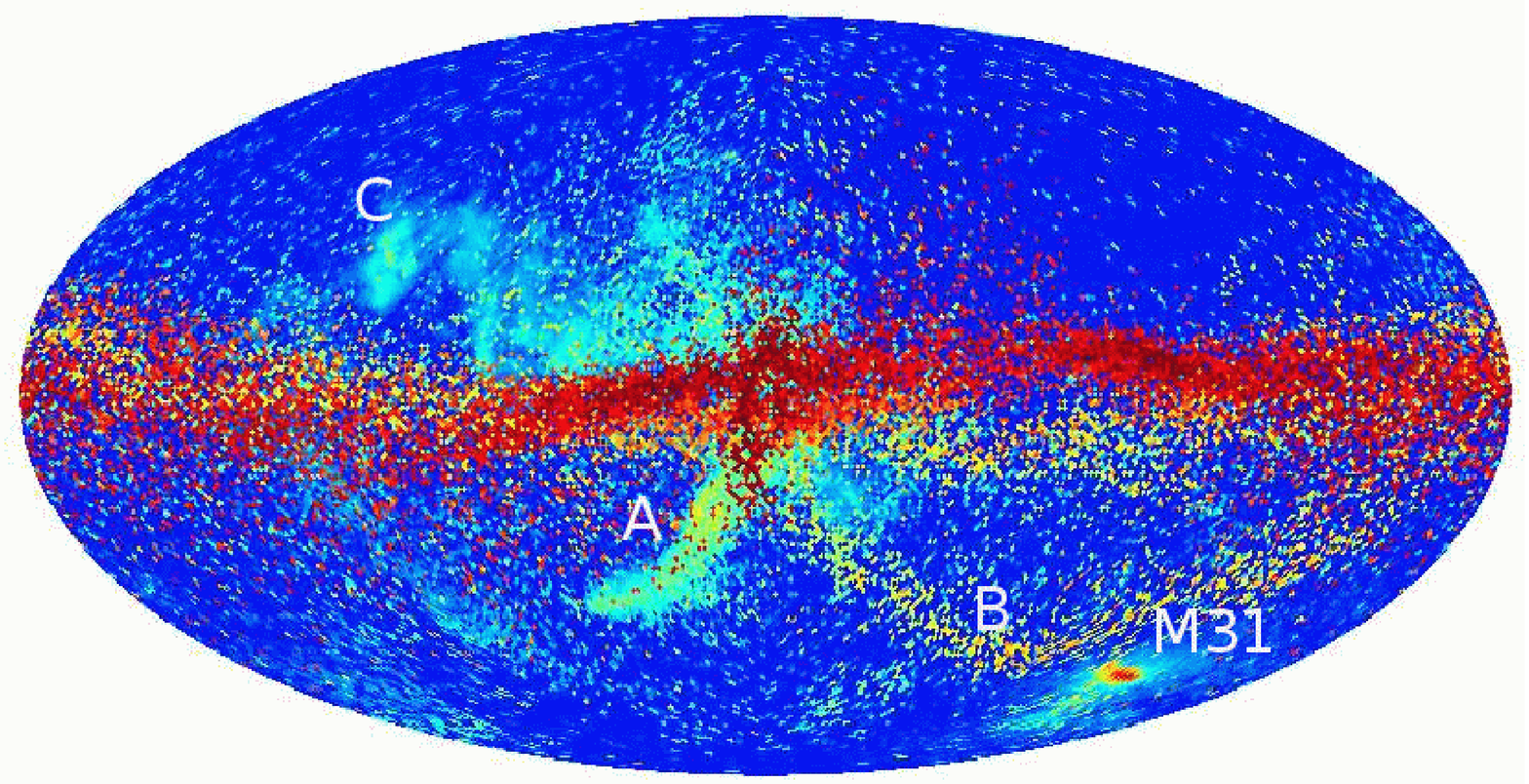}\hspace{1cm}\includegraphics[width=18mm]{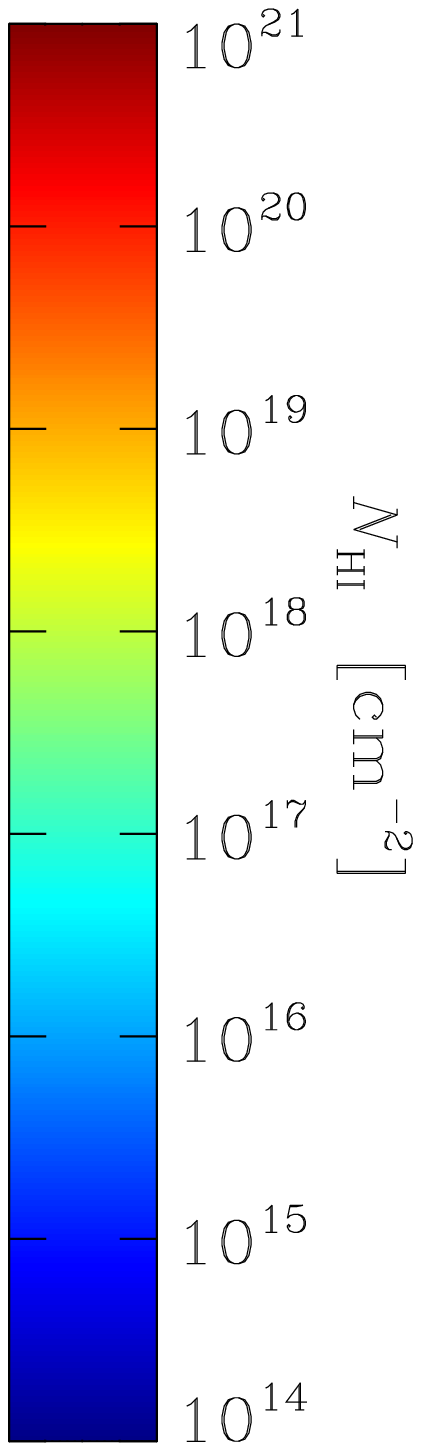}

\caption{Column density of H\,{\sc i} for the gas in our simulation with respect to the local standard of rest 
 of the MW$^{\rm c}$ galaxy in a Mollweide projection \citep[for a similar map from the H\,{\sc i} Galactic LAB survey see Fig. 3 of][]{Kalberla05}. 
 The observer is located at $(r_{\rm LSR},\phi_{\rm LSR})=(8\,{\rm kpc},0)$ in the cylindrical coordinate system defined by the MW$^{\rm c}$ disc. 
 The M31$^{\rm c}$ galaxy can be see in the lower-right region. In addition, gaseous features 
 labeled A, B and C are also shown (see the text for details).
}
\label{column_density_mollweide}
\end{center}
\end{figure*}

\begin{figure*}
\begin{center}
{\includegraphics[width=50mm]{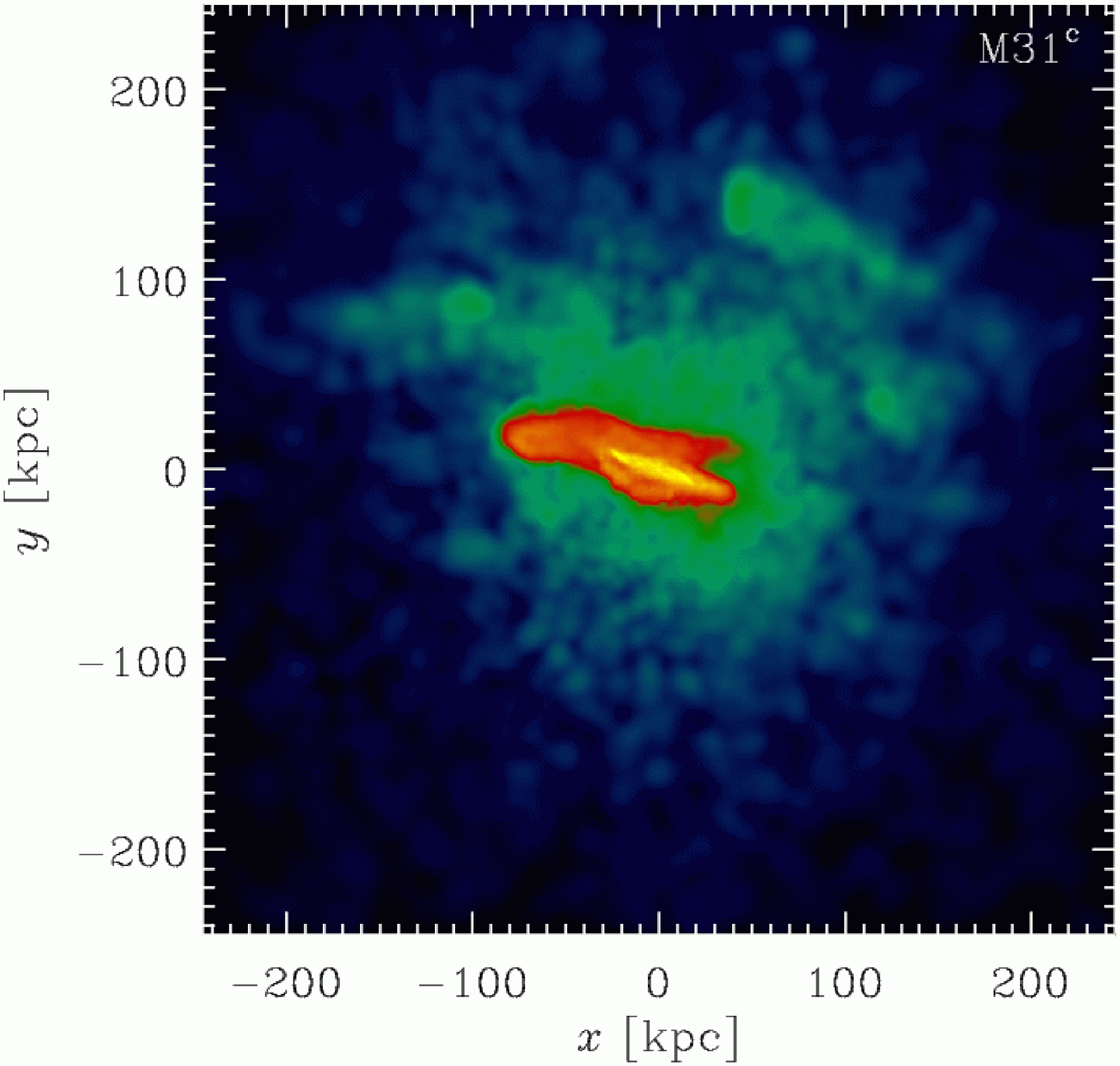}\hspace{5mm}\includegraphics[width=50mm]{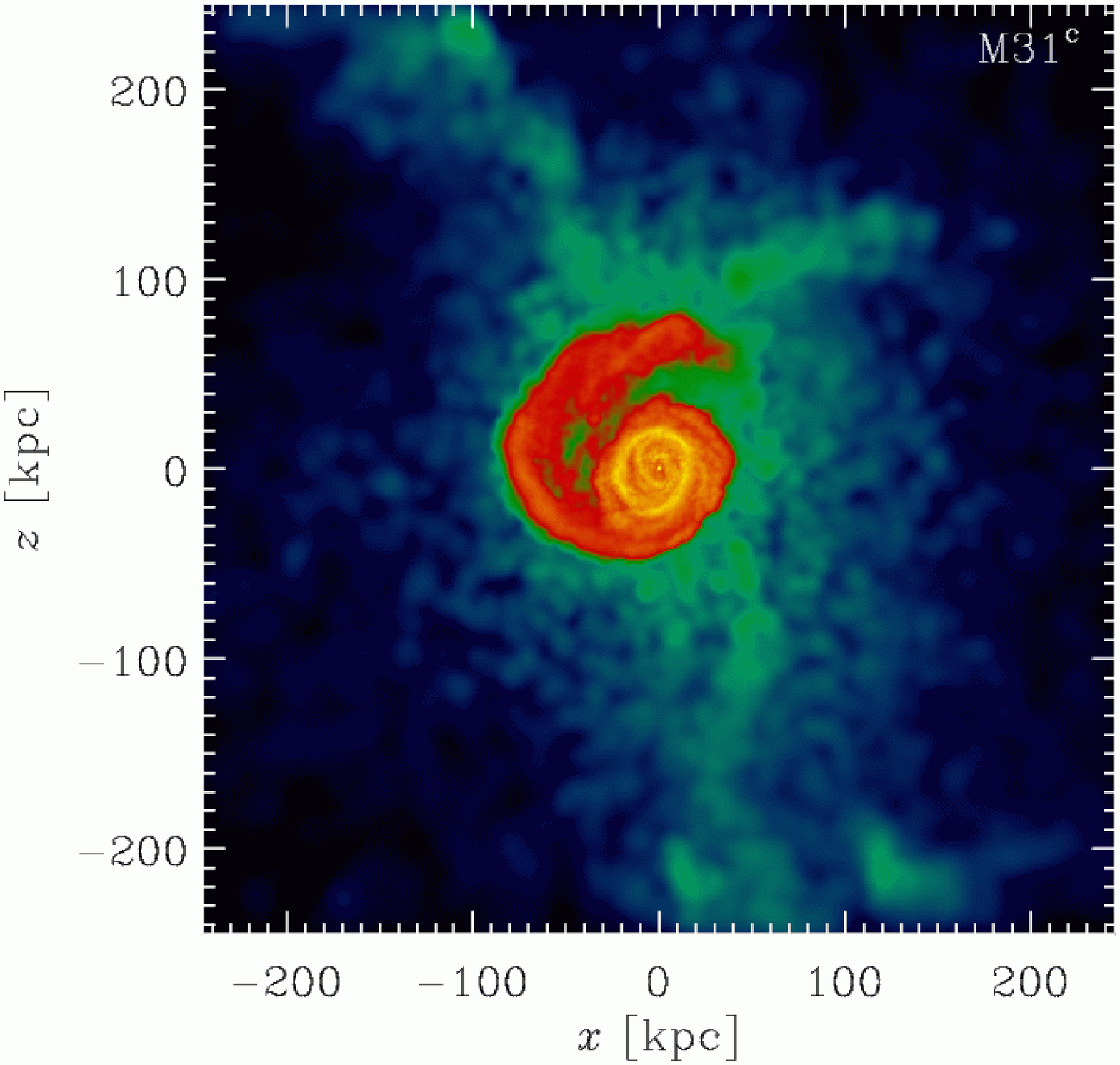}\hspace{5mm}\includegraphics[width=50mm]{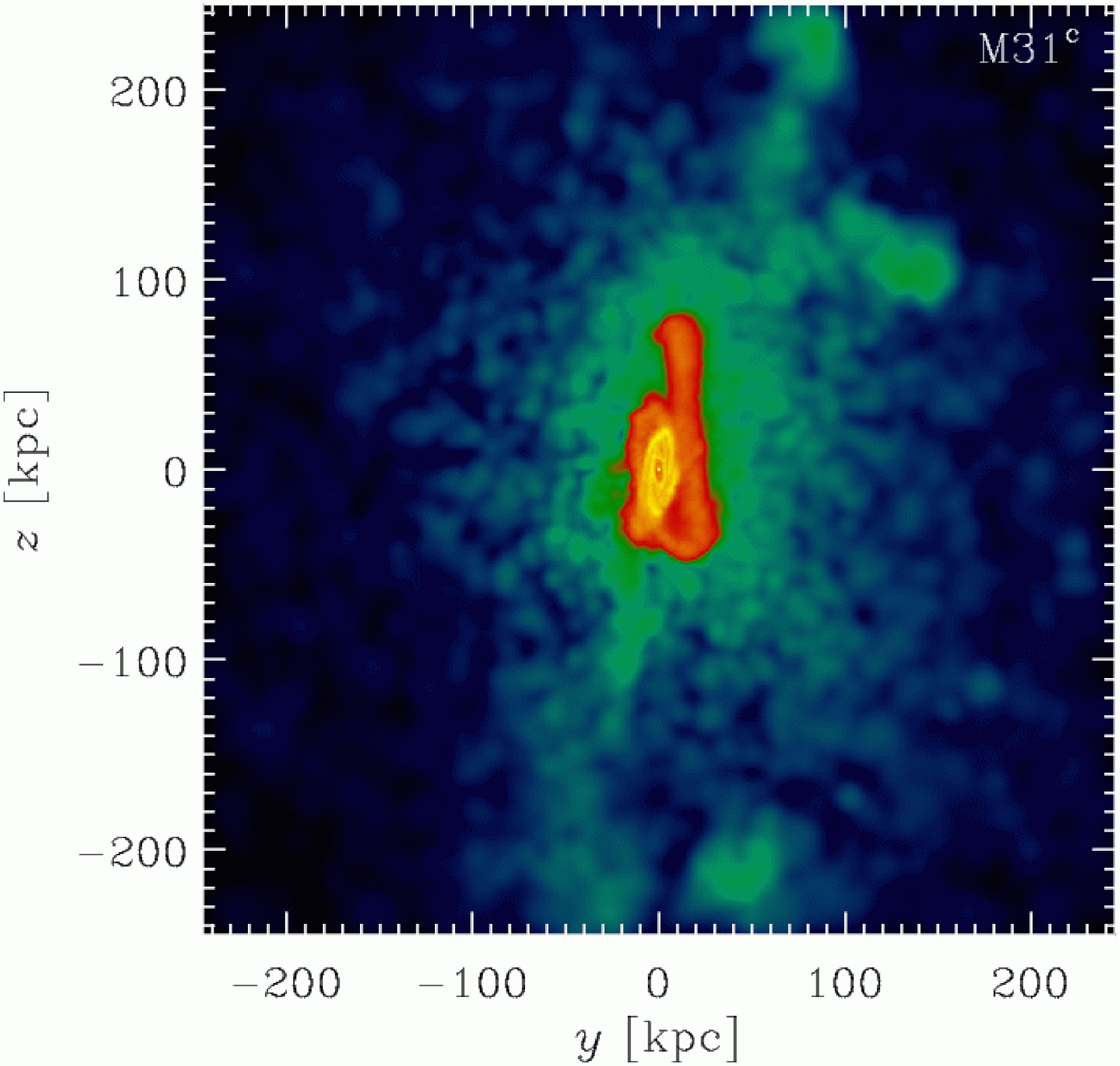}}

{\includegraphics[width=50mm]{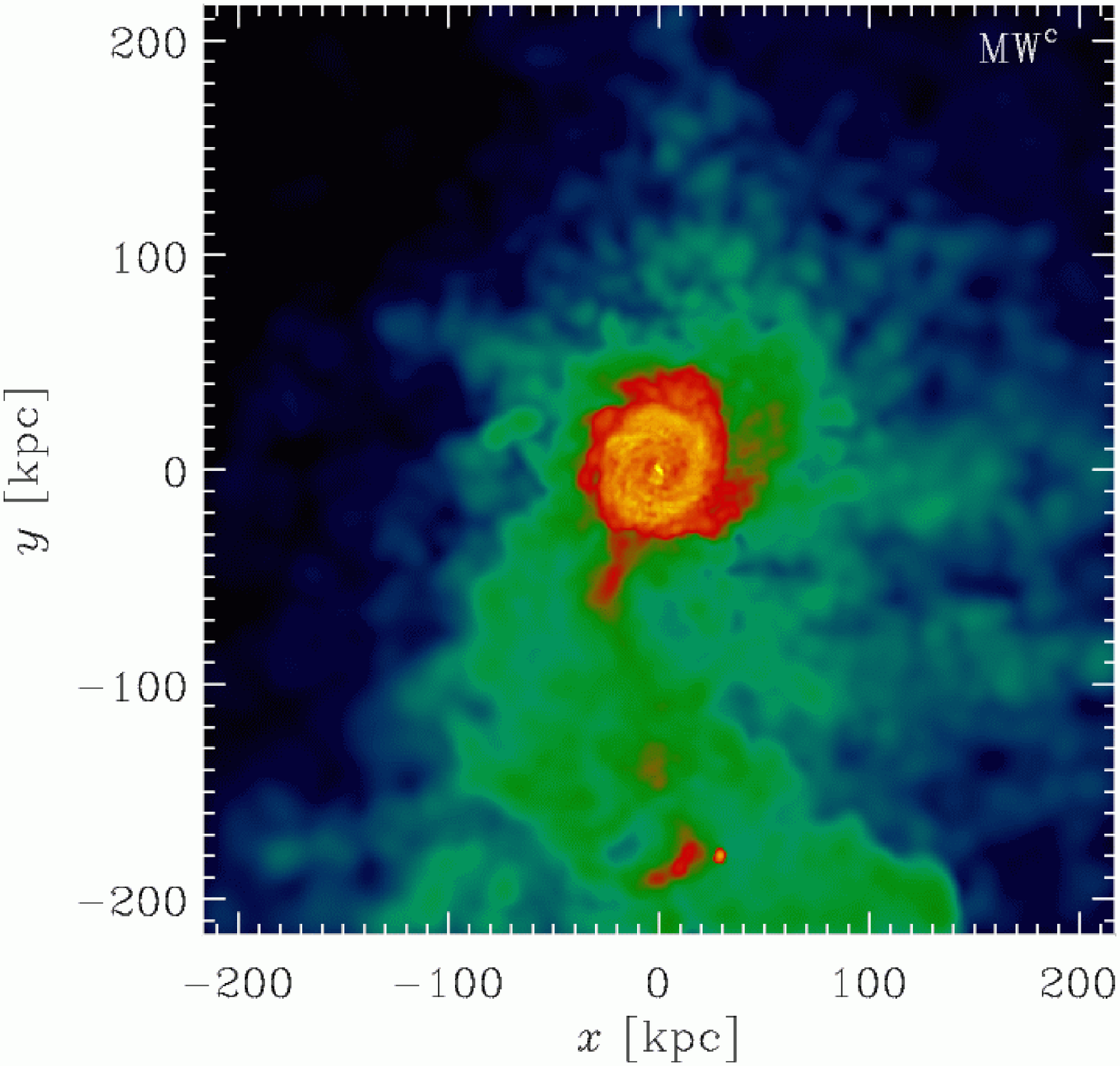}\hspace{5mm}\includegraphics[width=50mm]{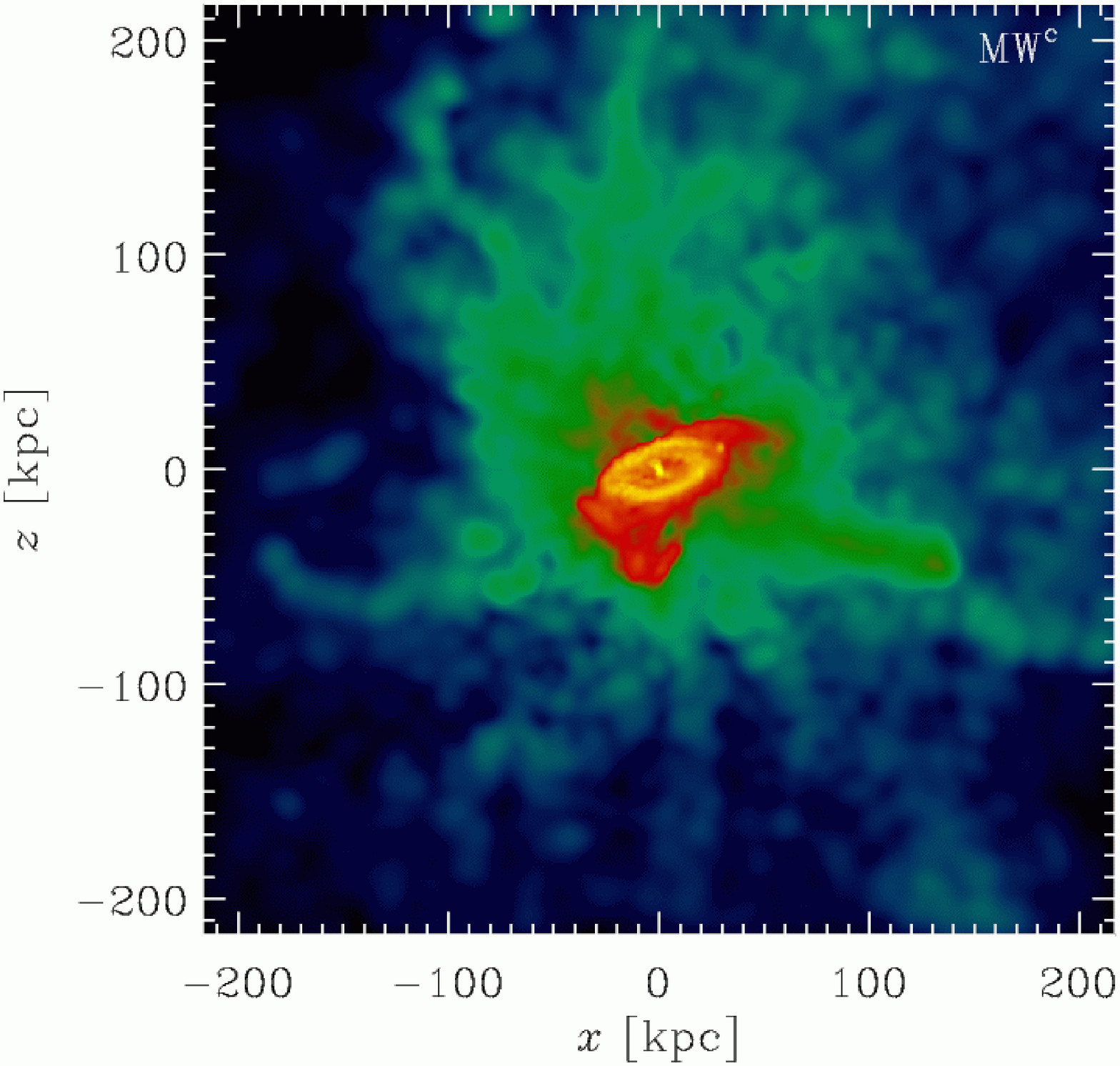}\hspace{5mm}\includegraphics[width=50mm]{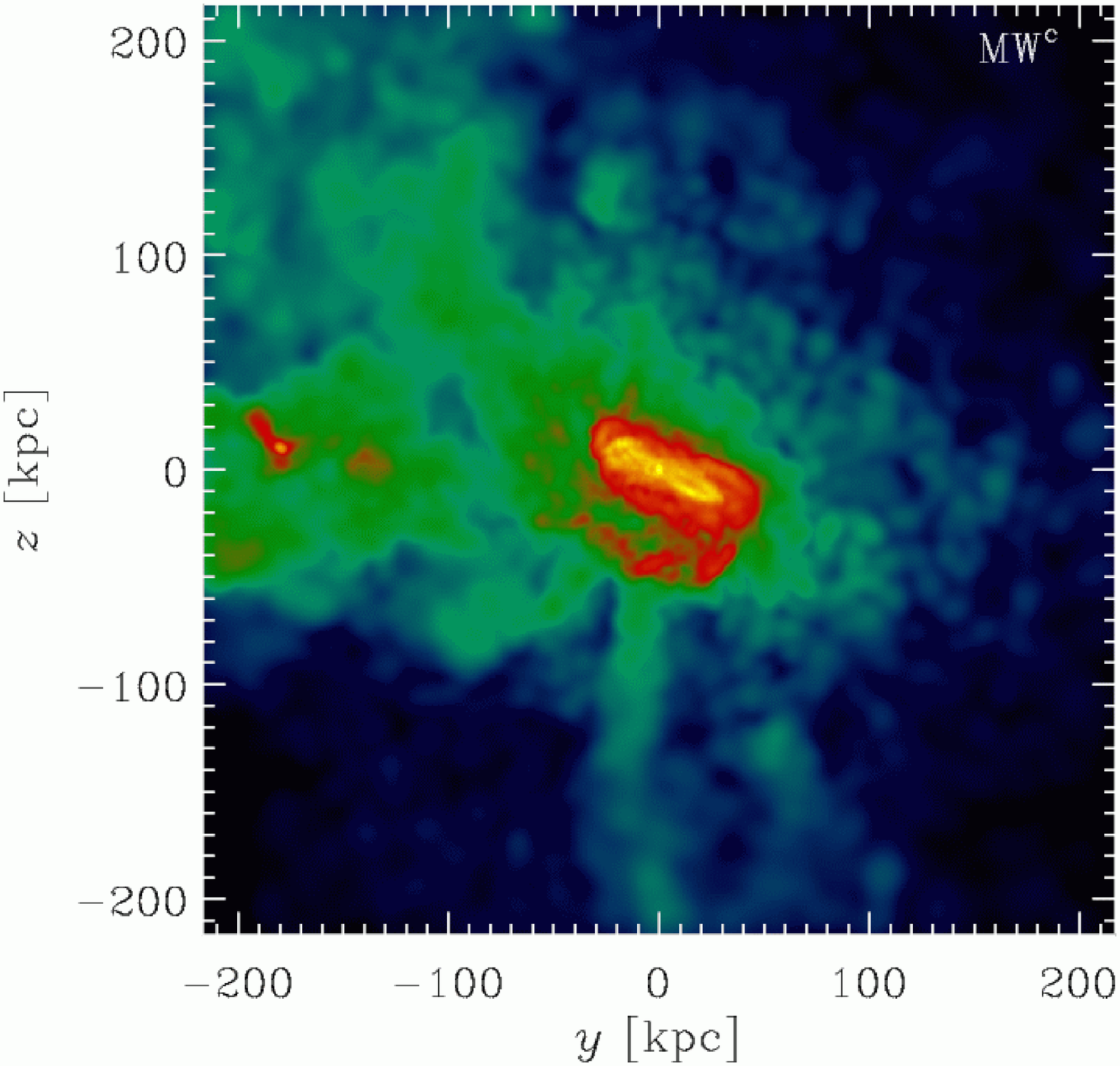}}

{\hspace{5mm}\includegraphics[width=50mm]{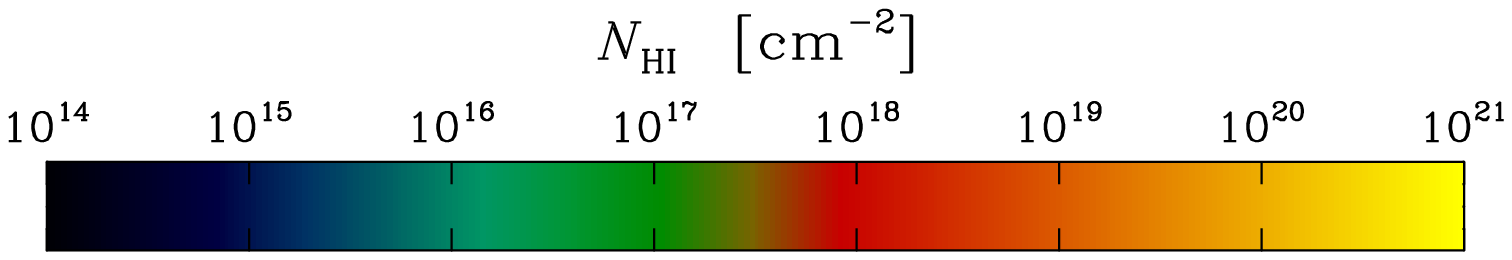}}

\caption{Column density of neutral hydrogen for the simulated
galaxies. Upper panels: results for M31$^{\rm c}$; the left-hand
panel shows the H\,{\sc i} distribution as seen from the MW, while the middle and 
right ones display the remaining two perpendicular views.
Lower panels: idem as the upper panels but in the case of the MW$^{\rm c}$. 
The plots show column densities up to the corresponding virial radius.}
\label{column_density_views}
\end{center}
\end{figure*}

The H\,{\sc i} discs of our simulated M31$^{\rm c}$ and MW$^{\rm c}$ systems have a 
similar extent to those given by observations, as can be
inferred from Figs.~\ref{column_density_views} and~\ref{cov_frac_profile}, where 
H\,{\sc i} column density maps and covering fractions in different planes are shown 
(see the next section for a detailed description of these figures).

Observations of the Milky Way and other nearby galaxies carried out 
in the $21\,$cm line 
have shown that spiral galaxies commonly also exhibit 
{\it extraplanar} H\,{\sc i} structures that indicate inflows, outflows,
and merger processes in these systems 
\citep[see reviews by][]{Wakker97,Richter06,Putman12}. 
In the Milky Way, such extraplanar neutral gas structures manifest themselves 
as the so-called high-velocity clouds (HVCs), which have column densities 
within the range $N_{\rm HI}\sim 10^{18}-10^{20}\,$cm$^{-2}$. 
These are $21\,$cm structures observed at high galactic 
latitudes that show high radial velocities that are inconsistent
with a simple galactic rotation model. In Andromeda, \citet{Thilker04} 
have mapped a similar population of extraplanar 
H\,{\sc i} features that can be regarded as HVC analogues.

Based on $21\,$cm observations of the Milky Way and Andromeda galaxies, R12
developed a three-dimensional model aimed at describing the infall of 
extraplanar H\,{\sc i} structures on to these galaxies,
which allows them to estimate the amount of
neutral gas in  HVCs as well as their distribution. Their modelling suggests a characteristic 
radial extension of $\sim$$50\,{\rm kpc}$ for the HVCs location within 
galactic haloes, as well as a total H\,{\sc i} mass in HVCs of the order of $\sim$$10^8\,\Msun$.

In order to compare our simulation with the R12 model, it is necessary to exclude 
the H\,{\sc i} gas in the disc region in our simulated galaxies, as we did in
the previous section (we exclude 
particles belonging to a flat cylinder in the disc plane 
with a radius of $30\,{\rm kpc}$ and a height of $\pm2\,{\rm kpc}$).
We note that, in our simulations, we lack the necessary
resolution to resolve features like individual HVCs; however,
we can use the total amount of extraplanar neutral gas as a proxy of the
neutral gas mass content. 
The H\,{\sc i} cumulative mass profile of our simulated galaxies including/excluding the 
galactic disc can be seen in Fig.~\ref{mass_profiles} (thin/thick dashed
lines) where we have additionally imposed two column density limits to better compare our 
results with observations. If no limit is applied, 
the resulting H\,{\sc i} masses within $50\,$kpc (excluding the gaseous disc) 
give $M_{\rm HI}(r\leq50\,{\rm kpc})\approx3.1\times10^{8}$ 
and $4.3\times10^{8}\,\Msun$ for the MW$^{\rm c}$ and M31$^{\rm c}$ galaxies
respectively. In general, for higher column density limits, the mass of H\,{\sc i} decreases 
by a factor of a few displaying a better match with observational mass estimates from HVCs. 
Additionally, the maximum spatial extent of the material with higher column
density cuts tends to decrease. 
We checked that the precise radius assumed to exclude the discs 
has a negligible impact on the final neutral gas masses.
For instance, for an H\,{\sc i} disc of $40\,$kpc radius, the
resulting mass differences are only of the order of 5\%. 
It is worth mentioning that most of the neutral material surrounding the galaxies does not 
belong to satellites orbiting the MW$^{\rm c}$ and M31$^{\rm c}$ systems. 
Rather, it is mainly part of the diffuse phase, and is probably associated 
with ambient gas. In Table~\ref{table_Mgas} we present these results 
for all H\,{\sc i} gas (where, essentially, most of the mass can be obtained 
for $N_{\rm HI}\geq10^{18}\,$cm$^{-2}$), as well as considering column density 
cuts of $N_{\rm HI}\geq10^{19}$, $10^{20}\,$cm$^{-2}$, at several relevant distances.

\begin{table}
\centering
\caption{All-sky covering fractions of H\,{\sc i} for different column density limits, as calculated
using Mollweide projections from the point of view of both MW$^{\rm c}$ and M31$^{\rm c}$ galaxies 
(Fig.~\ref{column_density_mollweide} shows the all-sky MW$^{\rm c}$ view).}
\label{table_cov_fraction}
\begin{tabular}{lcccccc}
\hline\hline
$N_{\rm HI}$ (cm$^{-2}$) &   \multicolumn{2}{c}{$> 7\times 10^{17}$} & \multicolumn{2}{c}{$> 10^{16}$} & \multicolumn{2}{c}{$> 10^{15}$}\\
                  & M31$^{\rm c}$ & MW$^{\rm c}$ 
                  & M31$^{\rm c}$ & MW$^{\rm c}$    
                  & M31$^{\rm c}$ & MW$^{\rm c}$ \\         \hline

all gas           & 0.29 &  0.33 &   0.42   & 0.51  &  0.59   &  0.69 \\
no disc           & 0.19 &  0.21 &   0.37   & 0.46  &  0.56   &  0.66 \\

\hline\hline
\end{tabular}
\end{table}

To characterize the properties of the neutral gas we studied its spatial location, 
kinematics and column density distribution. 
An impression of the H\,{\sc i} distribution around the galaxies for different 
column densities above $N_{\rm HI}>10^{15}\,$cm$^{-2}$ is given in Fig.~\ref{HVCs_simulated_contours}. 
This figure shows contour plots of the edge-on neutral gas distribution 
 of the MW$^{\rm c}$ and M31$^{\rm c}$ galaxies. The solid contours represent 
column densities of  $N_{\rm HI}\geq10^{20}\,$cm$^{-2}$, 
indicating the  presence of the galactic disc, while 
the remaining line styles stand for column densities in the 
range $N_{\rm HI}=10^{15}-10^{19}\,$cm$^{-2}$ (see caption for the line style reference). 
From these plots, it can be clearly seen that most of the material 
with $N_{\rm HI}\sim10^{18}\,$cm$^{-2}$ is located at 
projected distances $\lesssim50\,$kpc from the galaxy centres in accordance with 
observations. In particular, we found that the mean (three-dimensional) distance of the 
accreting neutral gas with column densities of $N_{\rm HI}\geq7\times10^{17}\,$cm$^{-2}$ 
is about $60\,$ and $70\,$kpc for the MW$^{\rm c}$ and M31$^{\rm c}$, respectively. 
In general, we found that neutral material with the lowest column densities tend 
to be distributed in a more spherical manner than larger column density gas. This 
goes in line with recent H\,{\sc i} observations around nearby edge-on galaxies. 
In particular, the neutral gas contours shown in Fig.~\ref{HVCs_simulated_contours} 
are similar to those of NGC 891 as given by \citet{Oosterloo07} (see their Fig. 1).

We have also studied the kinematics of the high column density gas complexes, 
finding that the mean (three-dimensional) radial velocity of the neutral 
gas with $N_{\rm HI}\geq7\times10^{17}\,$cm$^{-2}$ is of 
the order of $45\,$km s$^{-1}$ with respect to the centre of the galaxies.

Our results are reasonably close to observational estimates; however, 
if no H\,{\sc i} column density limit is applied, they show somewhat 
larger mass values in comparison to the R12 model. 
This can be explained by the fact that H\,{\sc i} mass 
estimates around the Milky Way and Andromeda are mostly driven by HVC
observations which represent the high-density peaks of the halo gas distribution. 
In fact, as can be seen in Fig.~\ref{mass_profiles}, after applying column density limits 
above those of typical HVCs, our H\,{\sc i} mass estimates get smaller,
improving the match with the R12 result. 
In particular, this is most noticeable in the case of our M31$^{\rm c}$ simulated galaxy.

\subsection{Covering fractions}
\label{HI_CovFracs}

Among the large amount of available observational data of $21\,$cm line in the Milky Way,
the spatial distribution of neutral gas provides an observational
ground against which we can further test the results of our simulations. 
Such maps of the H\,{\sc i} distribution are available for the 
whole sky and also in the direction of Andromeda \citep[e.g.,][]{Braun04,Braun09}. 
In what follows we will study the all-sky covering fractions from the 
point of view of our simulated MW$^{\rm c}$, as well as for the 
projected MW$^{\rm c}$ and M31$^{\rm c}$ galaxies, and compare with 
observations.

\subsubsection{All-sky map}
\label{all-sky-map_CF}

In Fig.~\ref{column_density_mollweide}, we show an all-sky Mollweide projection of the
H\,{\sc i} column density as seen from the local standard of rest (LSR) reference
system of our simulated MW$^{\rm c}$. We placed the observer at coordinates 
$(r_{\rm LSR},\phi_{\rm LSR})$ in the cylindrical coordinate system defined by the MW$^{\rm c}$ 
disc, where $r_{\rm LSR}=8$ kpc and $\phi_{\rm LSR}=0$. We have checked that
different choices for $\phi_{\rm LSR}$ have no impact on the resulting all-sky map 
beyond the galactic disc. 

In this view, the MW$^{\rm c}$ disc appears as an approximately horizontal distribution
of very high column density gas, with typical values of $N_{\rm HI}\gtrsim
10^{20}\,$cm$^{-2}$. The M31$^{\rm c}$ galaxy is also evident as a high H\,{\sc i} column density
region ($N_{\rm HI}\gtrsim 10^{20}\,$cm$^{-2}$)
in the lower-right end of the map.
A number of other interesting features are evident from this plot.  
A stream of cold gas, which starts close to (and within) the MW$^{\rm c}$ virial radius and extends to
$\sim$$400\,$kpc from its centre, is responsible for the feature labelled ``A''.
The stream appears  mainly below the disc, but also partly above it, as a result
of the projection. The feature labelled ``B'' comes from a gas complex right outside the MW$^{\rm c}$ disc,
located at a distance of $r\sim 40-50\,$kpc. 
Finally, we see a $N_{\rm HI}\sim10^{17}\,$cm$^{-2}$ structure, labelled ``C'', 
which originates well outside the MW$^{\rm c}$ and M31$^{\rm c}$ galaxies
(at about $550\,$kpc from the MW$^{\rm c}$ centre, in the direction opposite to M31$^{\rm c}$).
For reference, we included labels ``A'' and ``C'' in our  Fig.~\ref{both_galaxies} and a 
label ``B'' in Fig.~\ref{MW_central} (note that in this figure, however, the
maps consider all gas and not only the neutral component). 
While feature ``B'' corresponds to nearby 
gas, probably associated with accreting gas, both ``A'' and ``C'' structures are located at much larger 
distances. Therefore, our results suggest that it is possible to find neutral gas cloud complexes 
even outside galaxy haloes. Note, however, that some of these 
structures (e.g., our simulated gas cloud ``C'') could be difficult to observe in emission 
as a result of their low column densities. Interestingly, after observing the Andromeda/M33 sky region 
at high-resolution, \citet{Wolfe13} report the existence of H\,{\sc i} material located far beyond these 
galaxies with column densities $N_{\rm HI}\gtrsim10^{17}\,$cm$^{-2}$ which are reminiscent of 
our gaseous features.

\begin{figure*}
\begin{center}
{\includegraphics[width=75mm]{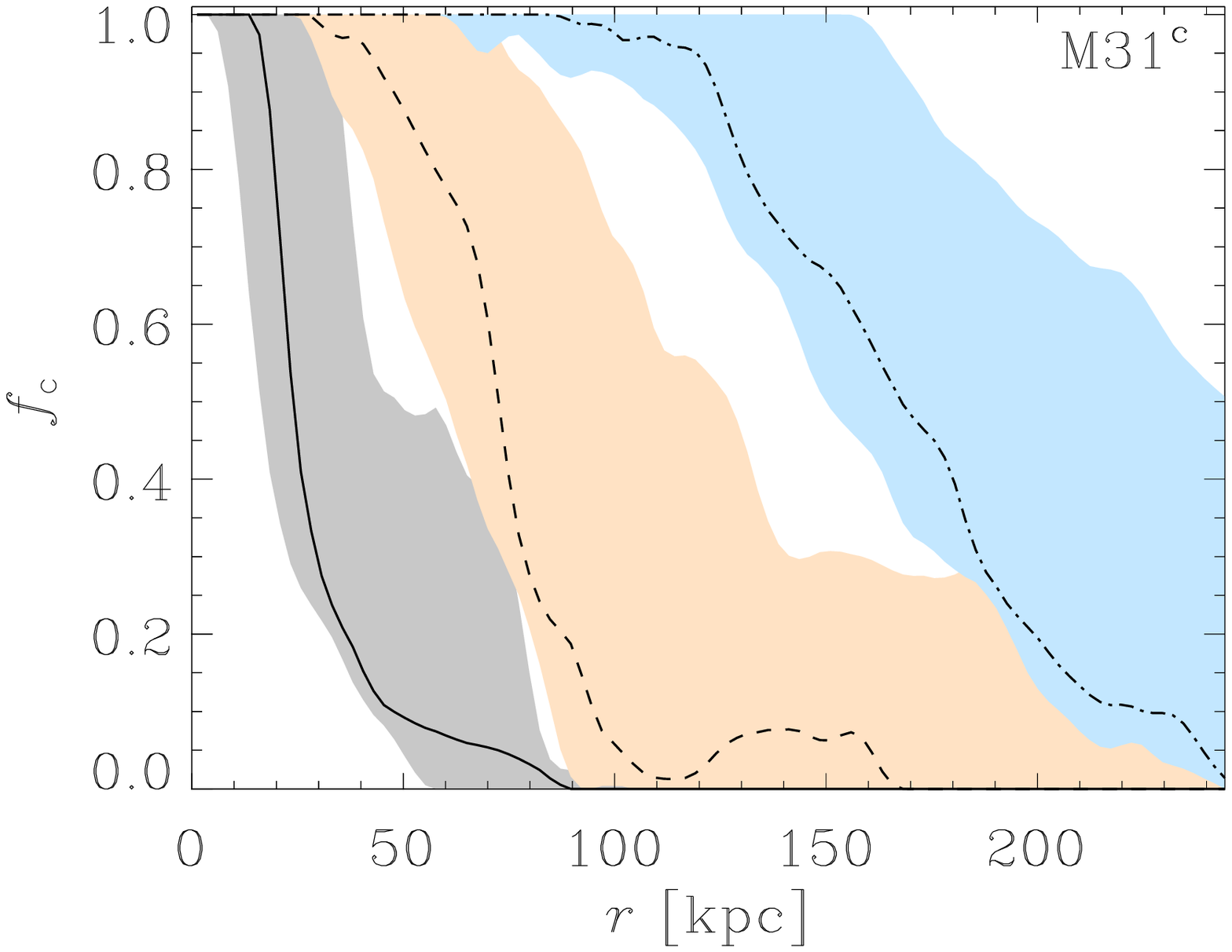}\hspace{1cm}\includegraphics[width=75mm]{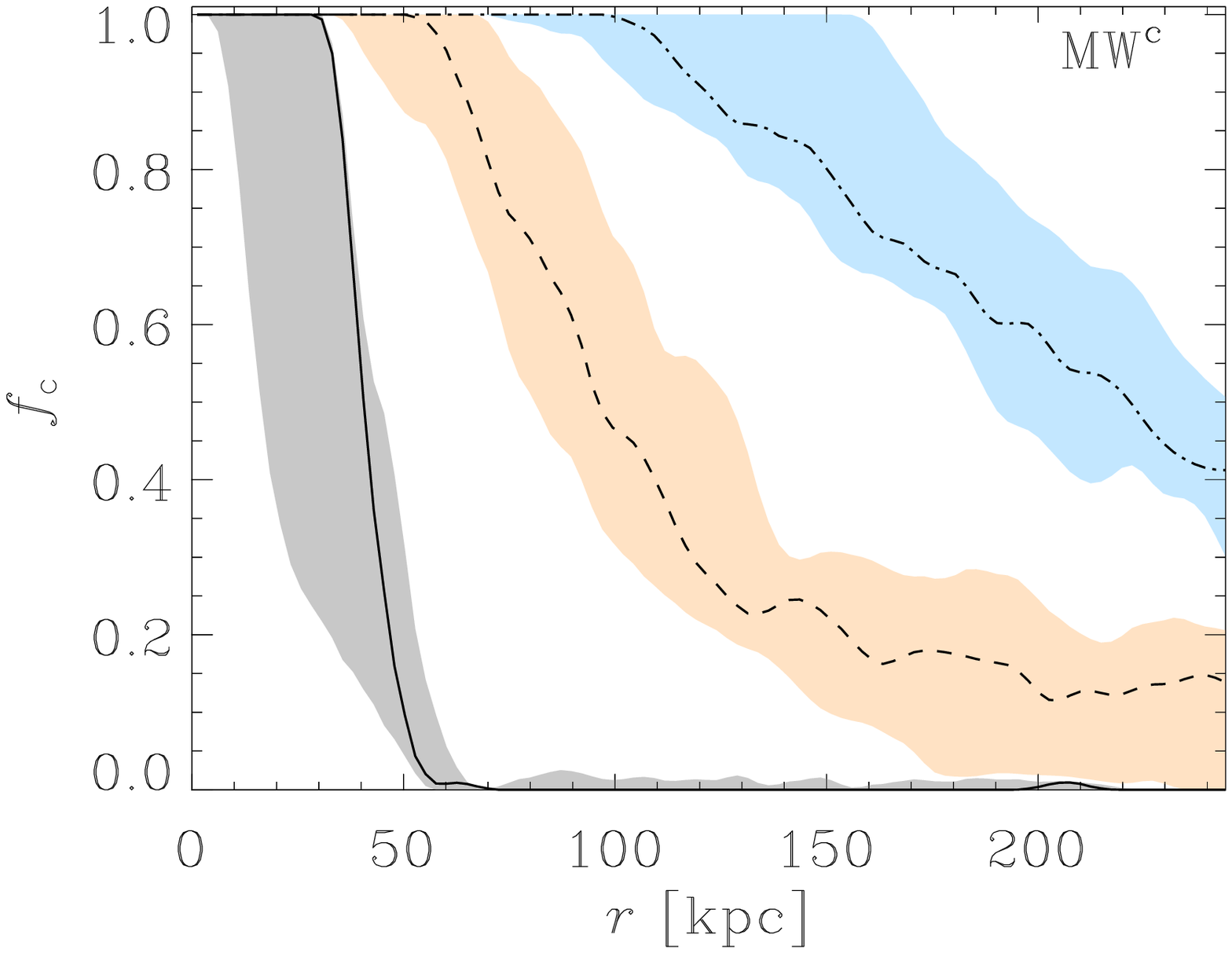}}

\caption{Covering fraction profile as a function of projected distance for the simulated M31$^{\rm c}$ (left-hand
panel) and MW$^{\rm c}$ (right-hand panel) galaxies. Shown are different limits of H\,{\sc i} 
column densities: $N_{\rm HI}>10^{15}\,$cm$^{-2}$ (dot-dashed lines), $N_{\rm HI}>10^{16}\,$cm$^{-2}$ (dashed lines)
and $N_{\rm HI}>7\times 10^{17}\,$cm$^{-2}$ (solid lines). The shaded areas indicate the 
range of possible covering fraction values after choosing different viewing angles.}
\label{cov_frac_profile}
\end{center}
\end{figure*}

\begin{figure*}
\begin{center}

\includegraphics[width=160mm]{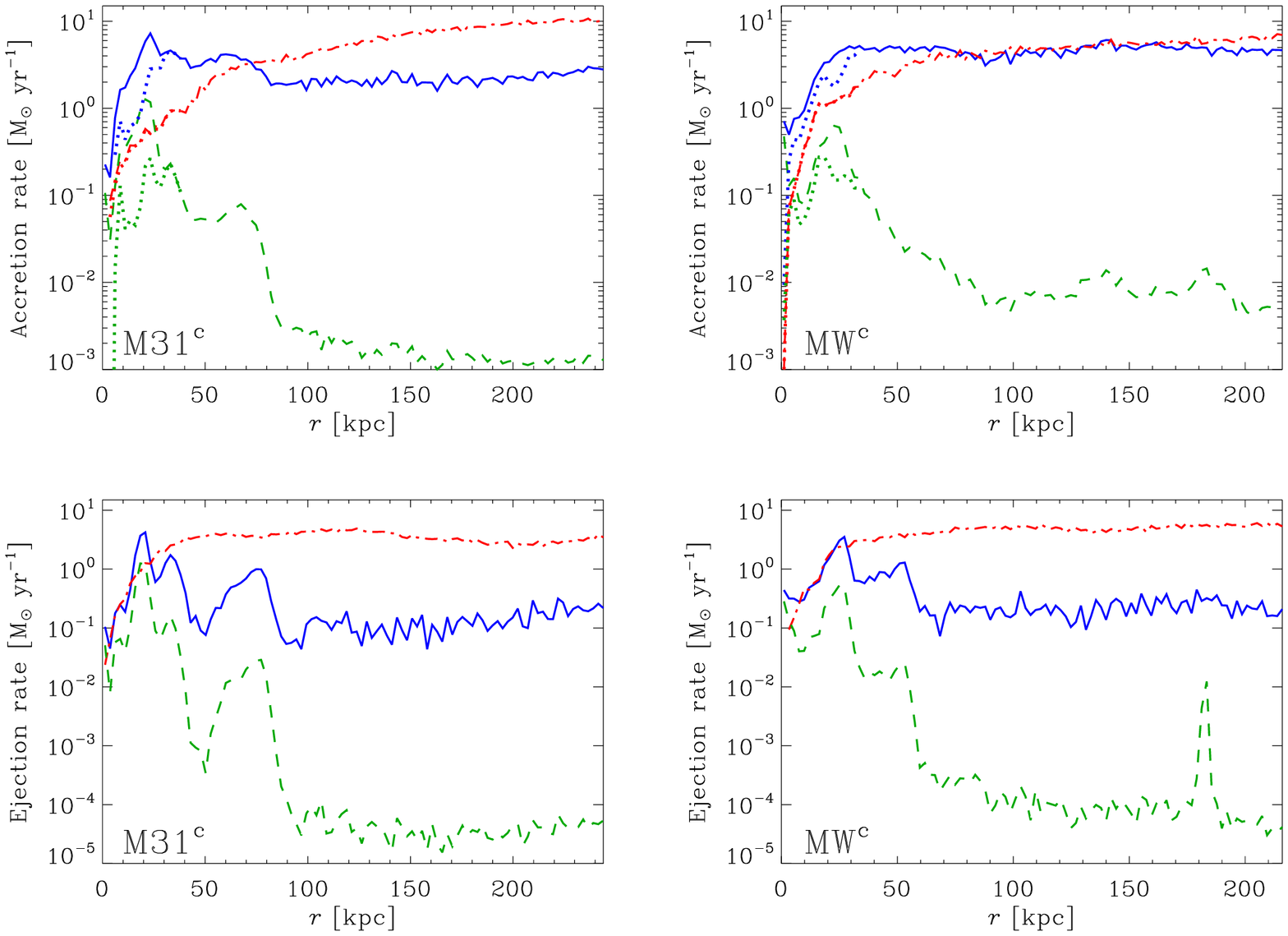}

\includegraphics[width=160mm]{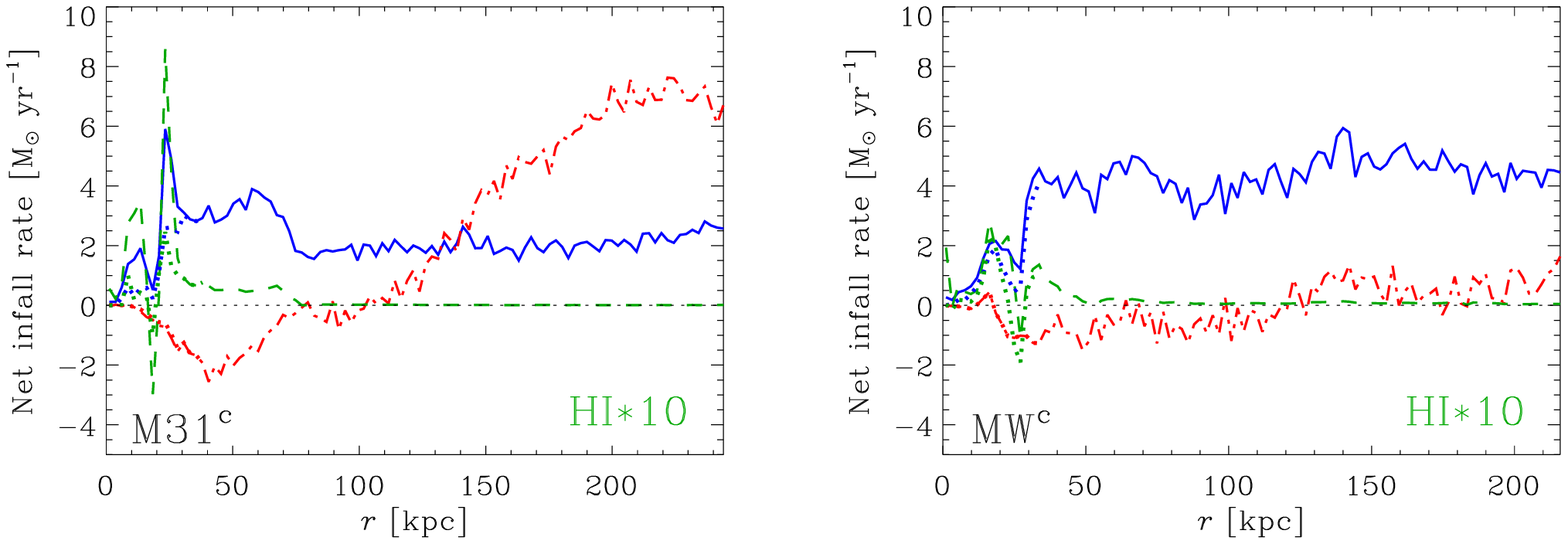}

\caption{Accretion, ejection and net infall rate profiles (spherically-averaged; upper, middle and lower panels, respectively) 
 for gas belonging to the M31$^{\rm c}$ (left-hand panels) and MW$^{\rm c}$ (right-hand panels) 
 simulated galaxies. Different gas components are shown: hot (dot-dashed lines), cold (solid lines) 
  and H\,{\sc i} (dashed lines). For the accretion case, we also include rates
  excluding the gaseous discs (dotted lines, as explained in Section~\ref{HI_data}). 
  Note that the latter is only relevant for $r\lesssim30$ kpc. Also note that
the net infall rate of HI has been multiplied by 10 to allow a proper visualization.}
\label{acc_eje_both_galaxies}
\end{center}
\end{figure*}

Table~\ref{table_cov_fraction} presents the all-sky covering fraction of 
H\,{\sc i} from the Mollweide projection for several 
column density limits: $N_{\rm HI}>10^{15}\,$cm$^{-2}$, 
$N_{\rm HI}>10^{16}\,$cm$^{-2}$, and $N_{\rm HI}>7\times 10^{17}\,$cm$^{-2}$. 
For completeness, we also computed the 
all-sky covering fractions seen from M31$^{\rm c}$ for an observer 
located at coordinates $(r,\phi)=(8\,{\rm kpc},0)$ as MW$^{\rm c}$ 
and M31$^{\rm c}$ show similar properties in terms 
of their neutral gas content. 
The resulting values range from about 0.3 up to 0.7 as the column density limit 
decreases. In particular, for $N_{\rm HI}>7\times 10^{17}\,$cm$^{-2}$, our results 
are similar to those inferred from the observations of 
\citet{Lockman02} and \citet{Wakker04} who found $f_{\rm c}\approx0.37$ and 
$f_{\rm c}\approx0.3$ respectively. Interestingly, if we compute the all-sky covering 
fraction of H\,{\sc ii}, we get larger values for a given column density limit, 
as suggested by observations. For instance, for a threshold 
of $N_{\rm HII}\gtrsim3\times 10^{19}\,$cm$^{-2}$ 
we get an all-sky covering fraction of $0.86$. In comparison, \citet{Shull09} 
estimated a Milky Way value of $f_{\rm c}=0.81$ 
for $N_{\rm HII}\gtrsim6\times 10^{18}\,$cm$^{-2}$. 
In this respect, to improve the agreement between our model and observations, 
we would need to increase the spatial resolution of our simulation, as well as 
to include relevant physical processes at smaller scales 
(e.g., radiative transfer effects).

\subsubsection{Projected galaxy fields}
\label{zoomed-in_CF}

In order to further investigate the H\,{\sc i} column densities 
in a way that is comparable to actual zoomed-in field views, we show in
the left-hand panels of Fig.~\ref{column_density_views} the
distribution of H\,{\sc i} column density for M31$^{\rm c}$ as seen from the MW$^{\rm c}$ (upper panel)
and for the MW$^{\rm c}$ as seen from M31$^{\rm c}$ (lower panel). 
These maps allow us to estimate H\,{\sc i} projected covering fractions,
which is a useful quantity to compare to observational results.
From our plots, it is clear that, when we see M31$^{\rm c}$ from the MW, 
it  appears more or less
edge-on;  in contrast, the MW, as seen from M31$^{\rm c}$, appears
almost face-on. This difference will strongly affect
estimations of covering fractions, in particular
for high column densities, e.g. $N_{\rm HI}> 7\times 10^{17}\,$cm$^{-2}$, since
the disc regions (which are the dominant H\,{\sc i} component)
always fulfill this condition and the area covered by the disc strongly 
depends on the viewing angle.

In Fig.~\ref{cov_frac_profile} we show, for M31$^{\rm c}$ (left-hand panel) and MW
(right-hand panel), the radial profiles of the covering fraction up to the 
virial radius of each galaxy for three  H\,{\sc i} column density limits; i.e., 
$N_{\rm HI}>10^{15}\,$cm$^{-2}$ (dot-dashed lines); $N_{\rm HI}>10^{16}\,$cm$^{-2}$ 
(dashed lines), and $N_{\rm HI}>7\times 10^{17}\,$cm$^{-2}$ (solid lines). 
We found that, when the limit $N_{\rm HI}>7\times10^{17}\,$cm$^{-2}$ is
considered, the covering fractions $f_{\rm c}$ decrease quickly with radius as we go outside
the disc region: $f_{\rm c}$ goes from
$1$ to values close to zero at  $r\approx 90\,$kpc for M31$^{\rm c}$ and $r\approx 60\,$kpc for the MW$^{\rm c}$. 
If we take reference values at $r=30$ and $50\,$kpc,
we found that $f_{\rm c}(r=30\,$kpc$)\approx 0.3$ and $f_{\rm c}(r=50\,$kpc$)\approx 0.1$ for M31$^{\rm c}$,
and that $f_{\rm c}(r=30\,$kpc$)\approx 1$ and $f_{\rm c}(r=50\,$kpc$)\approx 0.1$ 
for MW$^{\rm c}$. 
The fact that the covering fraction at $30\,$kpc is close to $1$ and $0.3$ for the MW$^{\rm c}$ and
M31$^{\rm c}$, respectively, can be understood in terms of the projections shown in Fig.~\ref{column_density_views}:
the MW$^{\rm c}$ appears almost face-on and, therefore, displays very high covering fractions
up to larger radii, only decreasing near the disc edge; whereas M31$^{\rm c}$ appears almost edge-on, thus decreasing the covering fraction for smaller radii.
Our results, in particular those for M31$^{\rm c}$, show a similar 
trend to the estimates of R12 who determined covering fractions of 
$f_{\rm c}(r=30\,$kpc$)\approx 0.15$ and $f_{\rm c}(r=50\,$kpc$)\approx 0.05$ 
for Andromeda.

In our simulation, the relative position between the galactic discs turned 
out to be almost perpendicular. However, the simulation gives us the
possibility to construct mock H\,{\sc i} observations for each galaxy to `observe'
them in any possible viewing angle. For example, in the middle- and right-hand 
panels of Fig.~\ref{column_density_views} we show the results for two
additional projections, i.e., perpendicular to those shown in the corresponding left-hand panels.
From this plot it is clear that the covering fractions vary significantly 
when seeing the galaxies in different projections showing that 
this effect will be certainly important in any observational
estimation of H\,{\sc i} covering fractions.

\begin{table*}
\centering
\caption{Integrated accretion and ejection rates of the simulated MW$^{\rm c}$ and M31$^{\rm c}$ galaxies 
for the cold, hot and H\,{\sc i} gas phases within distances of $r=10$ and $50\,$kpc from the corresponding 
galactic centres. The inflow/outflow characteristic time of the different components has been obtained by 
dividing the gas distance with its radial velocity (see Section~\ref{acc_eje_obs}).
Quantities including the gaseous galactic discs are shown within 
parentheses; as expected, the hot component is the one least affected by the disc exclusion. For 
$r\leq 50\,$kpc, we also show results for the accretion rates, excluding the gaseous discs, for 
column density limits of $N_{\rm HI}>10^{18}$, $10^{19}$ and $10^{20}$ cm$^{-2}$.}
\label{acc_eje_rates}
\begin{tabular}{lcccc}
\hline\hline
Rate ($\Msun\,{\rm yr}^{-1}$) & \multicolumn{2}{c}{Accretion} & \multicolumn{2}{c}{Ejection}\\
                              & M31$^{\rm c}$ & MW$^{\rm c}$ 
                              & M31$^{\rm c}$ & MW$^{\rm c}$ \\       
\hline

$\dot{M}(r\leq10\,$kpc): & & & &\vspace{0.05cm}\\
Hot          & 0.16 (0.21) & 0.09 (0.19) & 0.11 (0.23) & 0.17 (0.25)\\
Cold         & 0.33 (1.68) & 0.57 (2.38) & 0.05 (0.59) & 0.35 (1.26)\\
H\,{\sc i}   & 0.04 (0.62) & 0.11 (1.12) & 0.01 (0.28) & 0.10 (0.66)\\

\hline

$\dot{M}(r\leq50\,$kpc): & & & &\vspace{0.05cm}\\
Hot          & 1.18 (1.25) & 2.14 (2.40) & 2.59 (2.82) & 3.03 (3.31)\\
Cold         & 3.75 (7.80) & 5.10 (8.11) & 0.75 (2.64) & 1.51 (3.08)\\
H\,{\sc i}   & 0.22 (1.40) & 0.34 (1.54) & 0.09 (0.77) & 0.23 (0.90)\\
H\,{\sc i} {\tiny ($N_{\rm HI}>10^{18}$ cm$^{-2}$)}   &  0.21  &  0.33  & -- & --\\
H\,{\sc i} {\tiny ($N_{\rm HI}>10^{19}$ cm$^{-2}$)}  &  0.16  &  0.28  & -- & --\\
H\,{\sc i} {\tiny ($N_{\rm HI}>10^{20}$ cm$^{-2}$)} &  0.06  &  0.15  & -- & --\\

\hline\hline
\end{tabular}
\end{table*}

To quantitatively test the effect of different viewing angles on the covering
fractions,  we ``observe'' each simulated galaxy from different locations. 
The shaded areas in Fig.~\ref{cov_frac_profile} show the range
of all possible covering fraction values as a result of the varying viewing angle. 
As expected, we found that the M31$^{\rm c}$ covering fractions, as seen from the MW, 
are close to the minimum possible values since M31$^{\rm c}$ is almost edge-on in this view. 
In contrast, as seen from M31$^{\rm c}$, the simulated MW$^{\rm c}$ is close to face-on and, consequently, 
displays a covering fraction that is similar to the maximum possible values.

Fig.~\ref{cov_frac_profile} also shows the covering fractions 
(as well as the corresponding shaded areas; see above) for two other 
H\,{\sc i} column density limits, namely, $N_{\rm HI}>10^{16}$ and
$N_{\rm HI}>10^{15}\,$cm$^{-2}$. As expected, in these cases, the covering fractions 
remain at $f_{\rm c}\approx1$ up to much larger radii. In the case of M31$^{\rm c}$, the 
covering fraction for $N_{\rm HI}>10^{15}\,$cm$^{-2}$ decreases significantly, 
being close to zero at the virial radius. In contrast, for 
the MW$^{\rm c}$, the covering fraction is $f_{\rm c} \approx 0.4$ even at the virial radius, indicating 
that the gas distribution extends beyond it. 
A similar behaviour is seen for the resulting covering fractions when 
a column density limit of $N_{\rm HI}>10^{16}\,$cm$^{-2}$ is adopted.
A notable bump at $r\sim150\,$kpc demonstrates that
the covering fraction can be non-zero for any radius, as
long as there are clumps of neutral gas and/or infalling neutral material 
and their associated tidal gas features within the halo.

\section{Accreted and ejected gas}\label{accretion}

In this section, we  study the accretion and ejection rates for the simulated
MW$^{\rm c}$ and M31$^{\rm c}$ galaxies, and compare our results with recent estimates 
from observational data. 

We calculate the gas accretion/ejection rates at $z=0$ by considering the velocity of 
gas particles lying at a spherical surface located at a given distance from the centre of the 
simulated galaxies and up to the corresponding virial radius. 
We estimate the amount of accretion (ejection) of gas using particles 
that are infalling (outflowing) near this boundary taking into account the time it would take them 
to cross it inwards (outwards). We do this for all gas, but also separately for the hot, cold 
and H\,{\sc i} gas components. 

Fig.~\ref{acc_eje_both_galaxies} shows the accretion, ejection and net infall rates for
M31$^{\rm c}$ (left-hand panels) and MW$^{\rm c}$ (right-hand panels) in the case of 
cold, hot, and H\,{\sc i} gas. For accretion and net infall rates we also show results 
for the cold and H\,{\sc i} phases excluding (dotted lines) and 
including (otherwise) the gaseous disc region in each 
case (as explained in Section~\ref{HI_data}). In this way, we can discriminate infalling 
material associated with the disc from that belonging to the gaseous galactic halo which 
is the one we are mainly interested in.
However, as expected, differences between these two measures 
are only appreciable at distances smaller than $\sim$$30\,$kpc, i.e. close to 
the radius of the removed disc region. 
In what follows, for such small scales, we will always refer 
to estimates {\it excluding} the gaseous discs.

\begin{figure*}
\begin{center}
{\includegraphics[height=50mm]{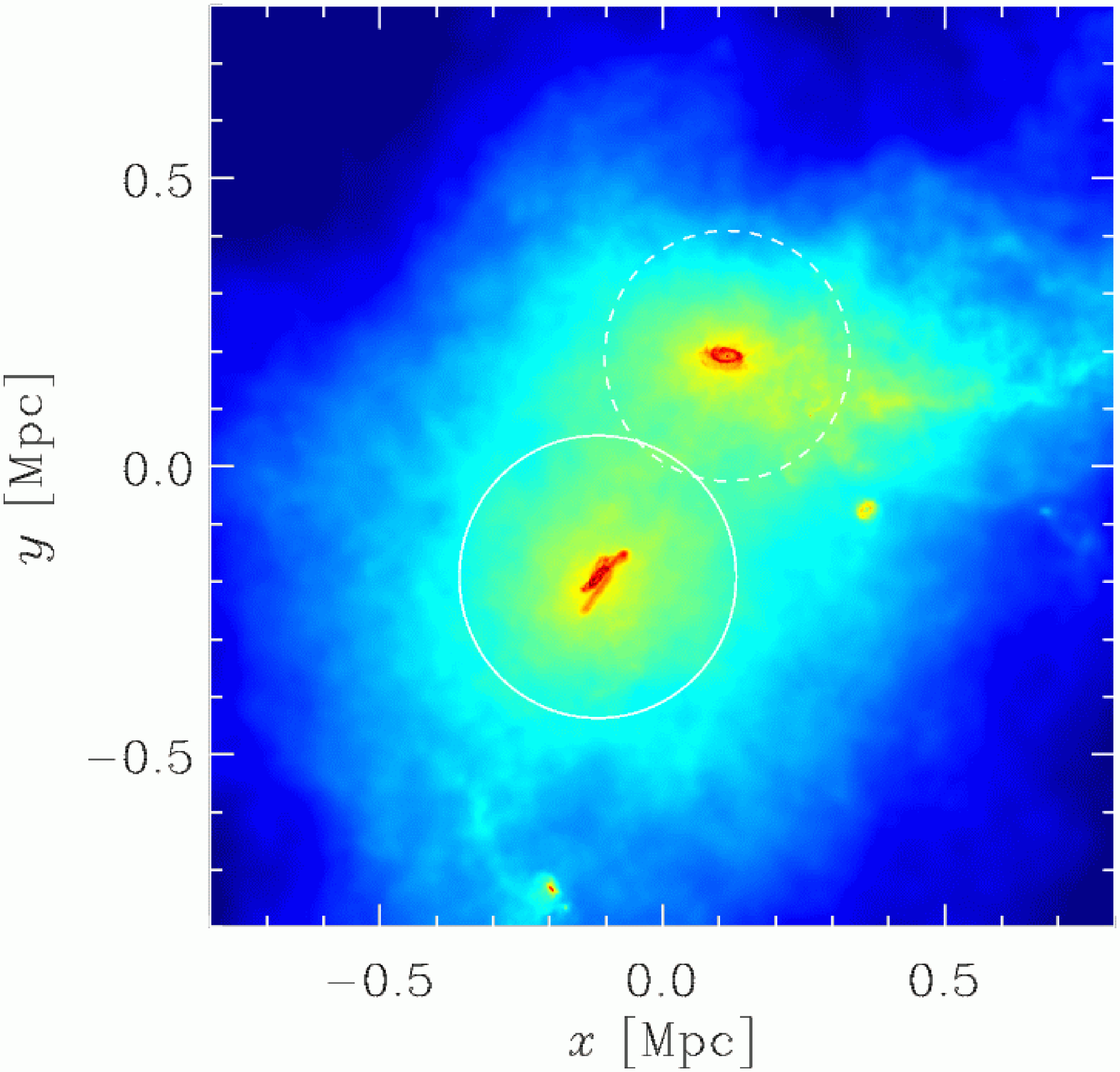}\includegraphics[height=50mm]{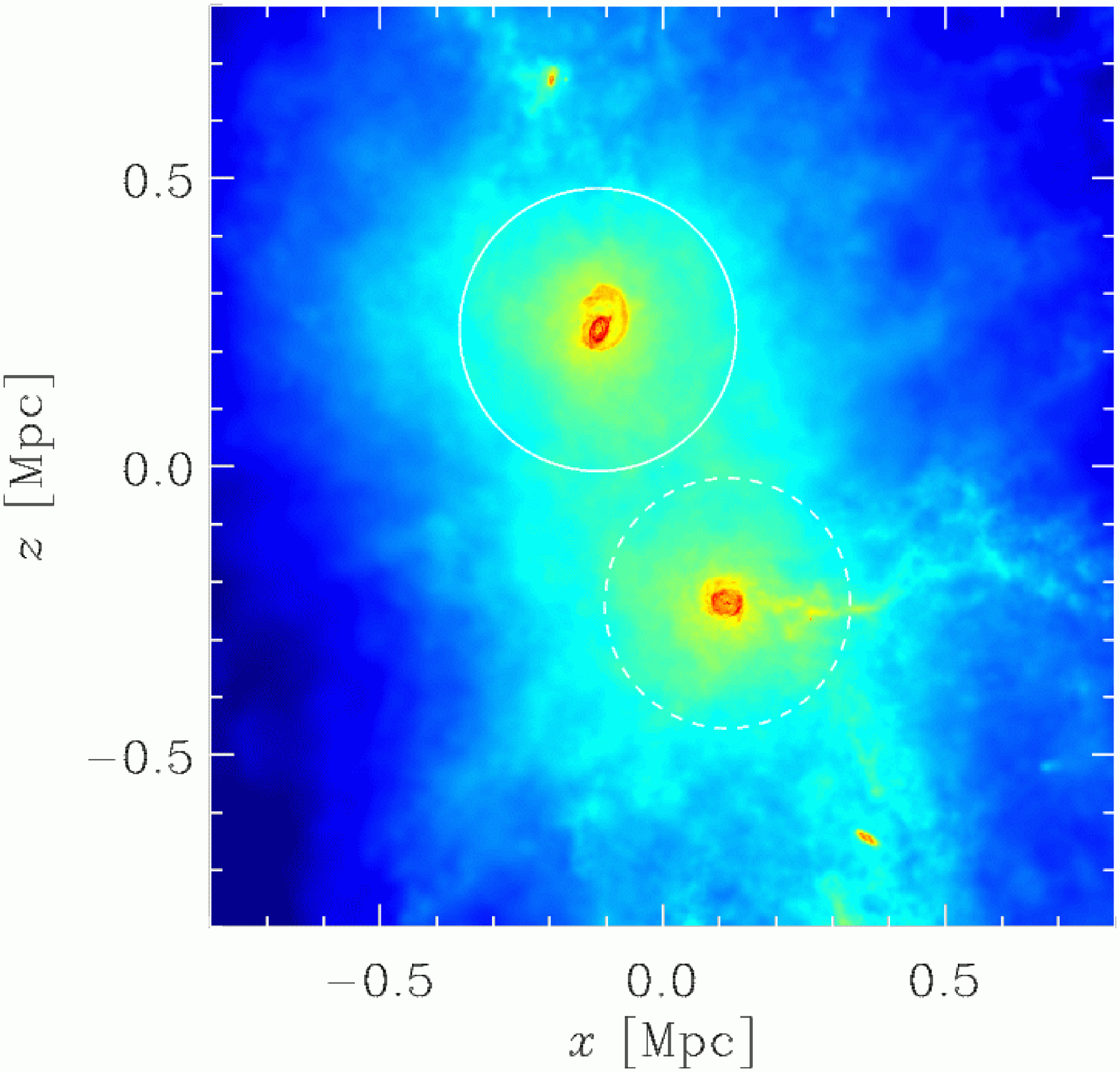}\includegraphics[height=50mm]{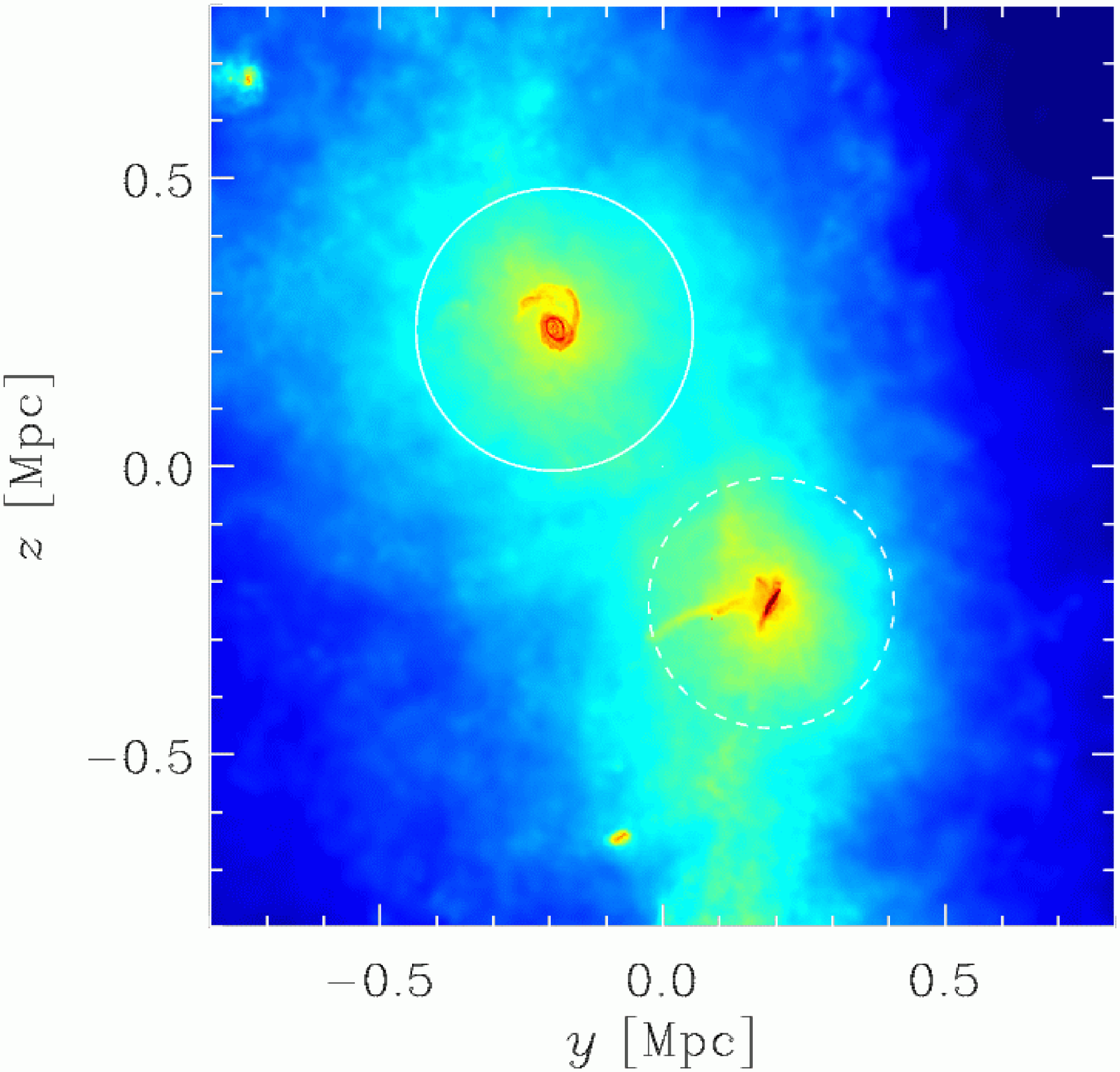}\includegraphics[height=50mm]{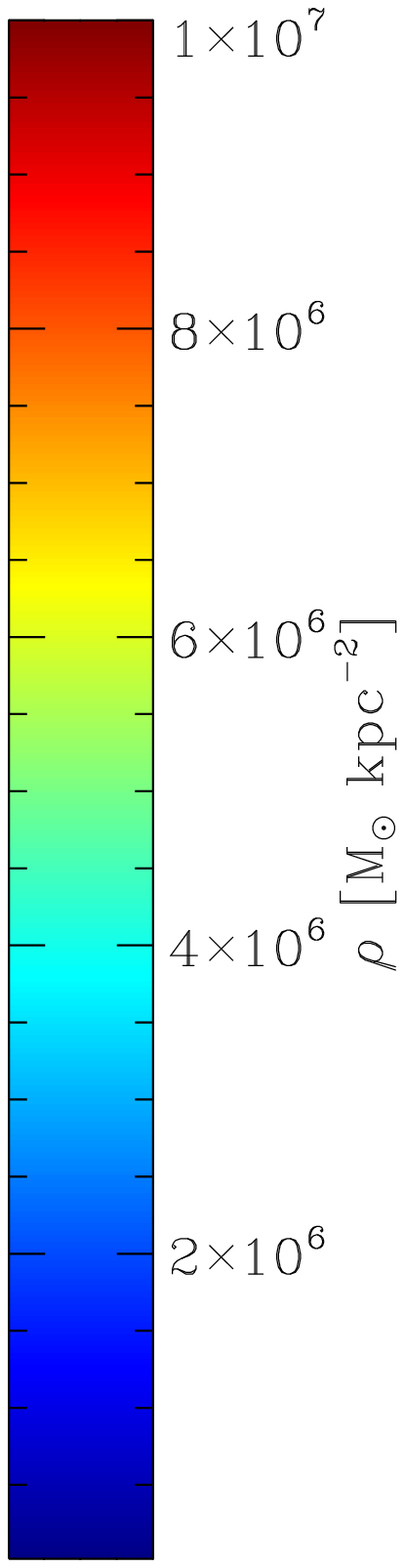}}
{\includegraphics[height=50mm]{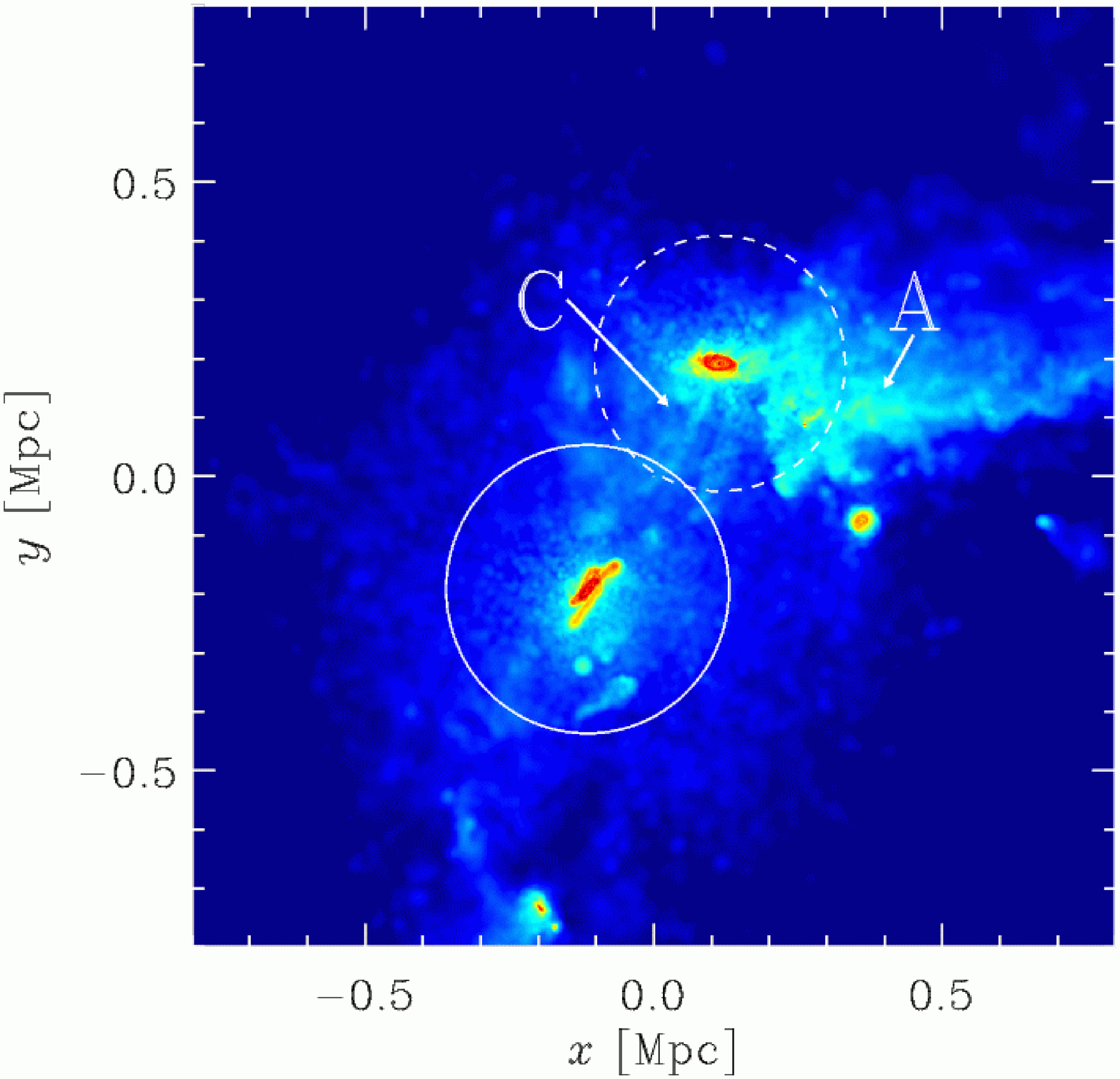}\includegraphics[height=50mm]{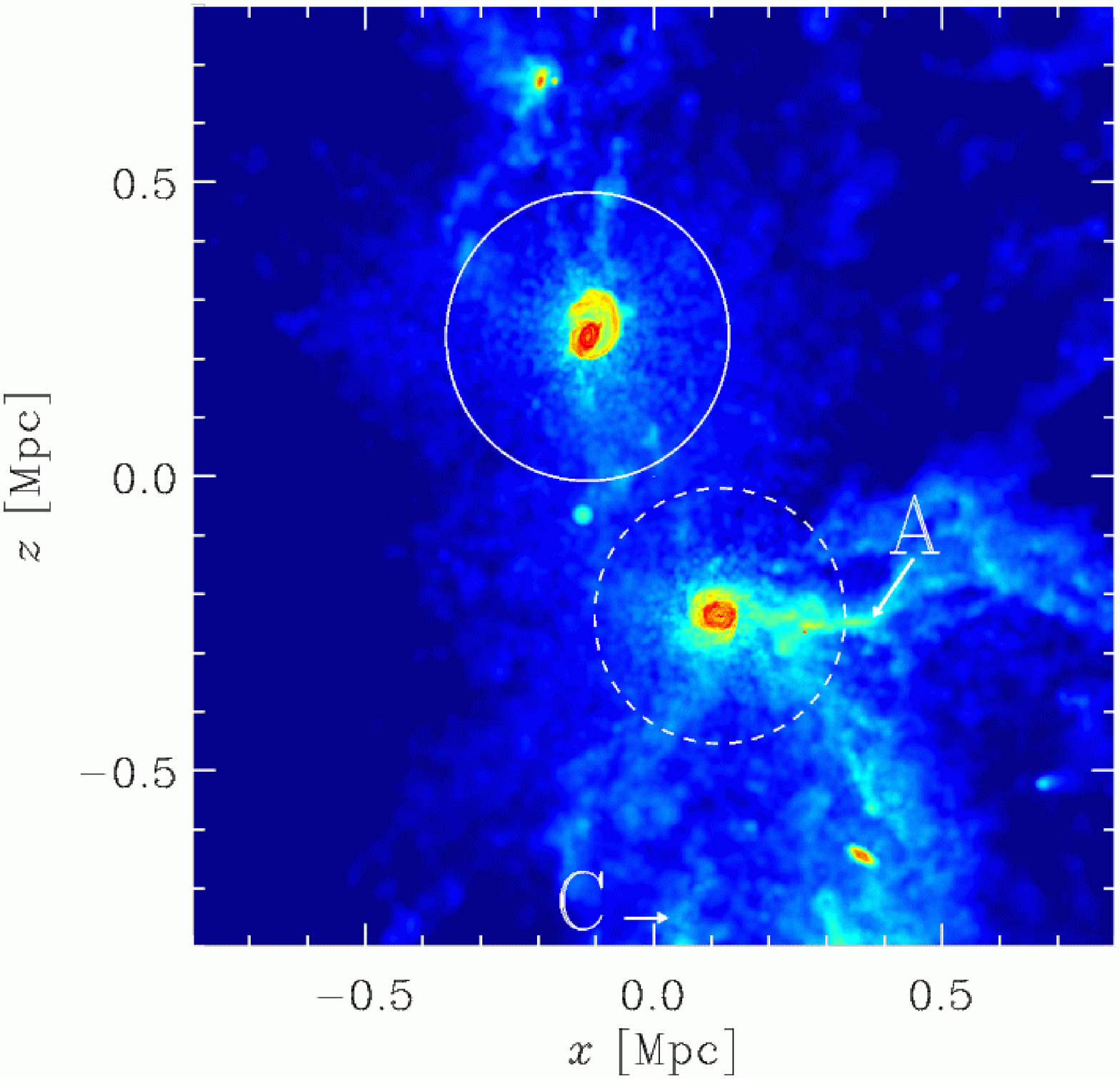}\includegraphics[height=50mm]{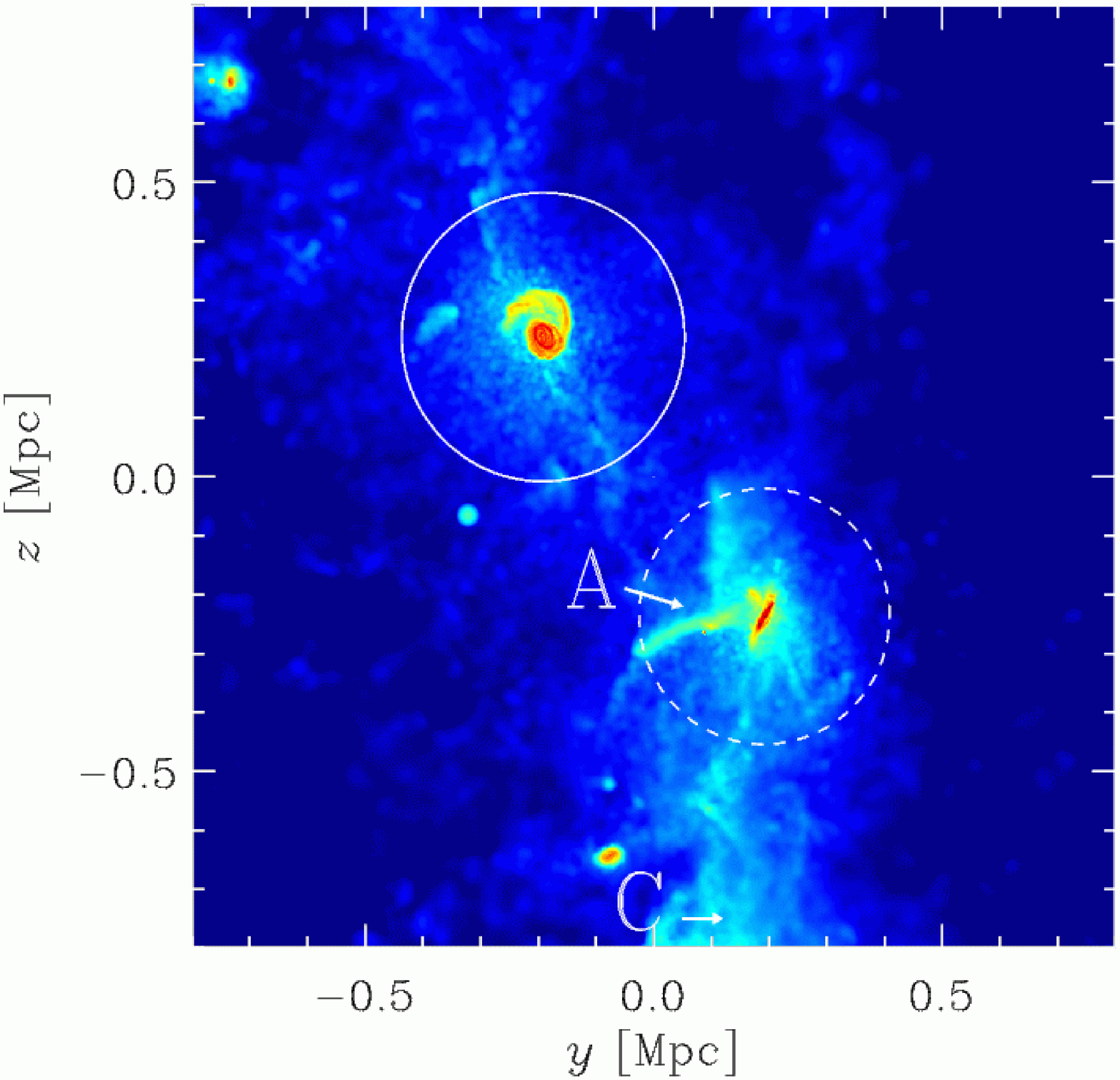}\includegraphics[height=50mm]{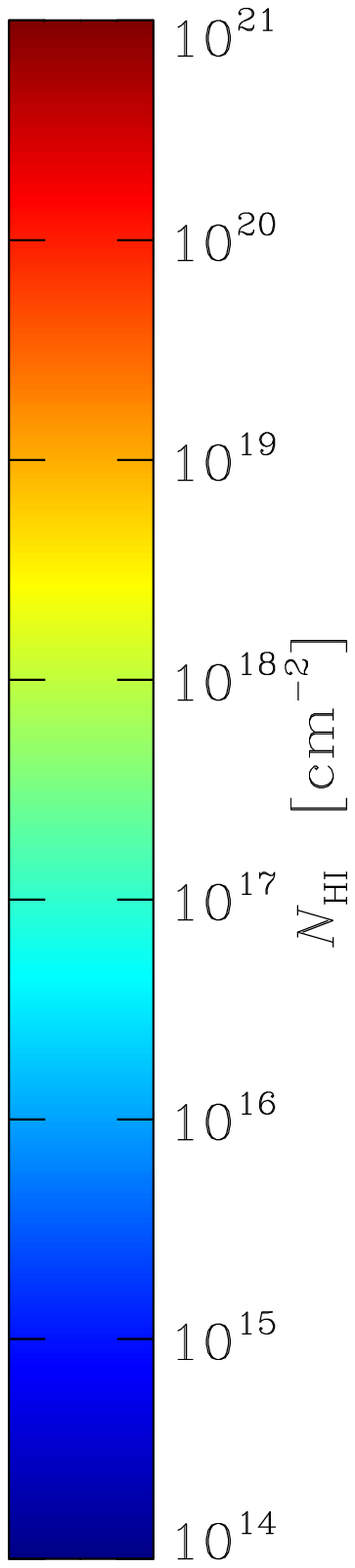}}
{\includegraphics[height=50mm]{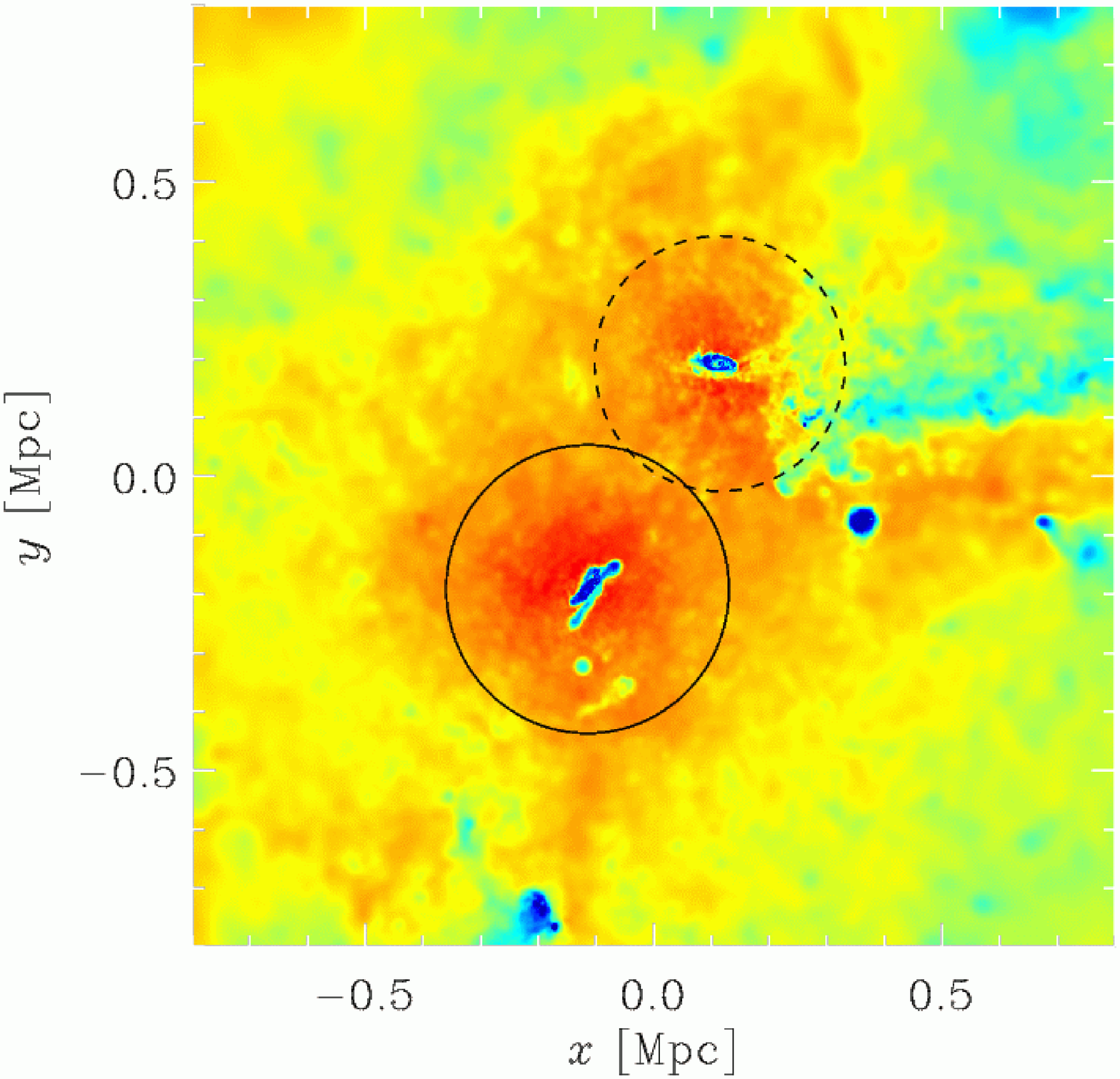}\includegraphics[height=50mm]{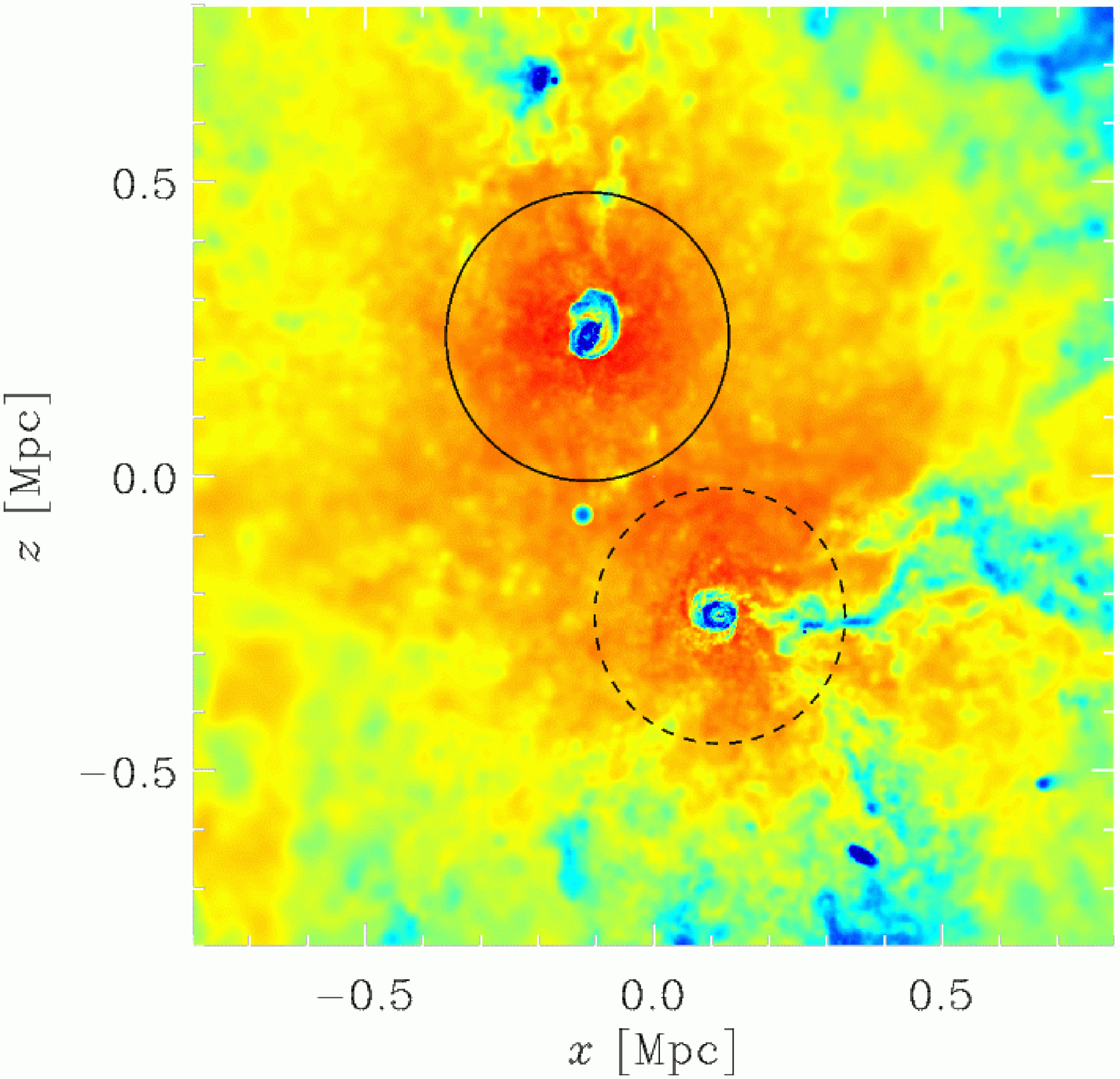}\includegraphics[height=50mm]{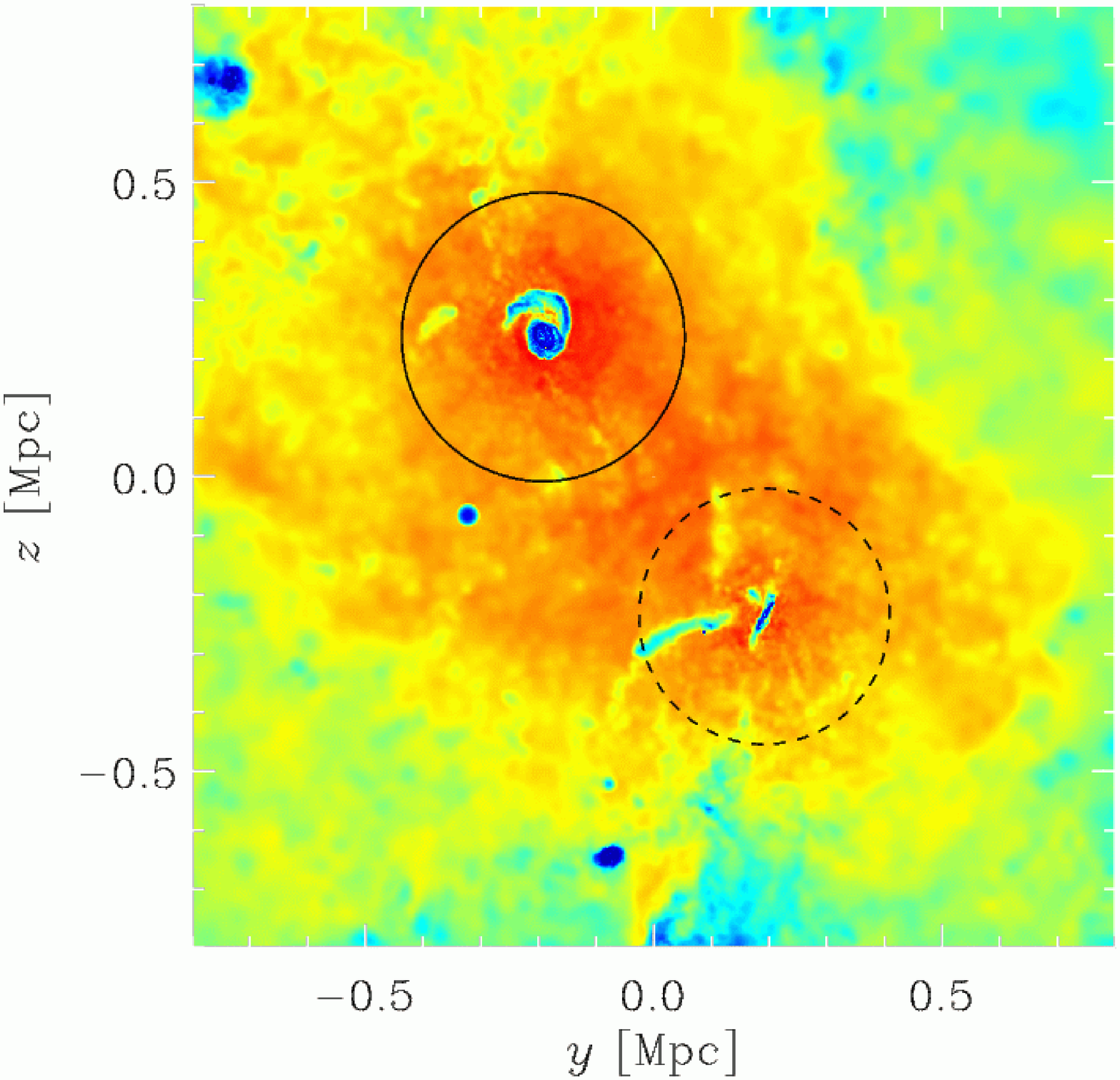}\includegraphics[height=50mm]{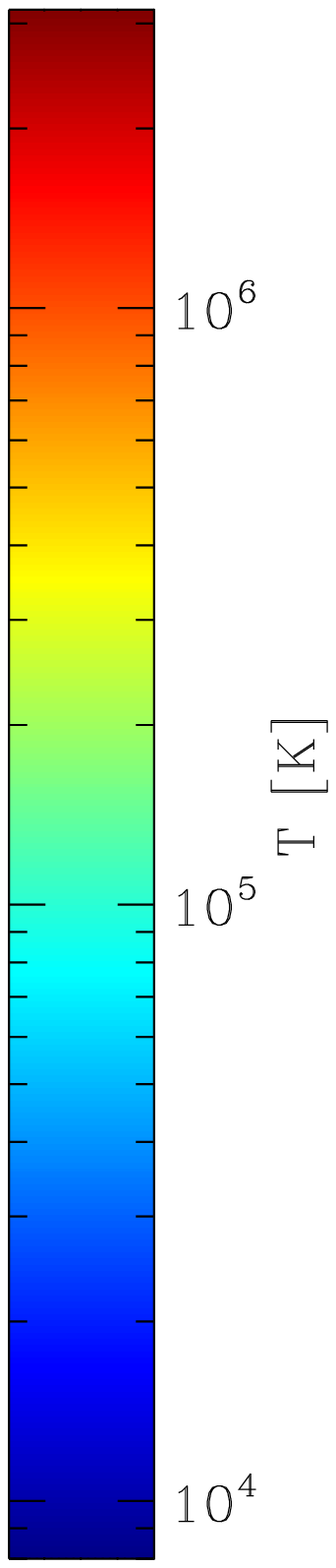}}

\caption{Gas density (upper panels), H\,{\sc i} column density (middle panels)
  and temperature (lower panels) maps for the M31$^{\rm c}$/MW$^{\rm c}$
  system. Each row shows three perpendicular projections of the corresponding
  quantity, in order to highlight the three-dimensional distributions. The
  virial radii of M31$^{\rm c}$ and the MW$^{\rm c}$ are indicated by the
  solid and dashed circles, respectively. Prominent neutral gas features are
  indicated (see text).
} 
\label{both_galaxies}
\end{center}
\end{figure*}

\subsection{Accretion rate profiles}

The accretion rate profiles of our simulated galaxies 
are shown in Fig.~\ref{acc_eje_both_galaxies} (upper panels). 
We found that the accretion rates at distances $r\gtrsim 70\,$kpc are
dominated by hot gas, reaching values of $7-10\,\Msun\,{\rm yr}^{-1}$ near the
virial radius, where the cold mode accretion has typical values of 
$3-5\,\Msun\,{\rm yr}^{-1}$. At smaller distances, the main contribution 
is, conversely, owing to the cold gas component, with 
typical rates of a few $\Msun\,{\rm yr}^{-1}$, followed 
by the hot and H\,{\sc i} accretion modes respectively. 
The transition between the dominance of the hot and cold modes 
is directly connected to the cooling efficiency
as this quantity -- and therefore the accretion rate -- 
is higher for gas in the central regions
than for material located in the haloes, which are at temperatures close to
$10^6\,$K (see Figs.~\ref{profiles_M31} and ~\ref{profiles_MW}).

The H\,{\sc i} accretion rates reach a peak of
$\sim$$0.25\,$M$_\odot\,$yr$^{-1}$ at $20\,$kpc, and decrease significantly at
larger radii, reaching very small values 
of $\sim$$10^{-3}-10^{-2}\,$M$_\odot\,$yr$^{-1}$ as one approaches the virial radius.   
Interestingly, in the case of M31$^{\rm c}$, most of the H\,{\sc i} 
accretion comes from gas located within a distance of $\sim$$50-70\,$kpc, 
in agreement with HVC data, as well as with the modelling of R12.

\subsection{Ejection rate profiles}

The ejection rate profiles can be seen in the middle panels of Fig.~\ref{acc_eje_both_galaxies} 
(note the different $y$-scale). As expected, most of
the ejected gas corresponds to the hot phase which is intimately related to
the process of supernova feedback. The presence of outflowing material is a
direct consequence of the mass-loaded galactic winds. In our model, such winds 
are naturally generated after supernova explosions, in some cases being able
to reach the virial radius and penetrate in the intergalactic medium (see
e.g., \citealt{Scann06}). 
The winds are particularly evident at distances 
$r\gtrsim50\,$kpc (i.e., greater than the size of the galactic discs) with 
typical hot gas ejection rates of the order 
of $\sim$$5\,\Msun\,{\rm yr}^{-1}$.

The ejection rates of cold and neutral material at distances larger than $\sim$$50\,$kpc 
are smaller compared to those of the hot phase: $\sim$$0.1$ and 
$\sim$$5\times10^{-5}\,\Msun\,{\rm yr}^{-1}$, respectively. 
At $r\lesssim50\,$kpc, we find higher ejection rates, although these are still
relatively low, particularly in the case of H\,{\sc i} material. 
As shown in Figs.~\ref{M31_central} and ~\ref{MW_central}, the gas 
distributions in the inner regions of galaxies are
extremely complex owing to the interplay between cooling, heating and accretion.
Outflowing cold/H\,{\sc i} gas is the result of a pressure support gain, either 
directly via energy feedback (even if they cool out rapidly afterwards) 
or via the local hot phase.

\subsection{Net infall rate profiles}

The net infall of gas for our two simulated galaxies can be calculated, as a function
of radius,  as the difference 
between the corresponding accretion and ejection rate profiles. These can be seen 
in the lower panels of Fig.~\ref{acc_eje_both_galaxies}, where, for clarity, 
we have multiplied the net 
infall of neutral gas by 10. The results indicate that there is a net 
infall of cold gas of $\sim$$2-5\,\Msun\,{\rm yr}^{-1}$ at distances greater 
than $\sim$$30\,$kpc, whereas the neutral (H\,{\sc i}) material accumulates
on to the galaxies at a rate of $\sim$$0.05-0.1\,\Msun\,{\rm yr}^{-1}$ for
distances smaller than $r\sim70\,$kpc, as indicated by the corresponding dotted lines 
(see the caption of Fig.~\ref{acc_eje_both_galaxies}).
 
In the case of hot gas, our simulated galaxies show a clear transition scale, 
located at about $100-120\,$kpc from the galactic centres, 
below (above) which hot gas is being ejected (accreted). This is an indication
of the strength of the supernova-driven winds, which weaken with increasing radius. 
The precise values of the net infall rates  depend on the particular galaxy considered. For instance, 
in the case of M31$^{\rm c}$, we found a net accretion (ejection) of hot gas 
$\lesssim7\,\Msun\,{\rm yr}^{-1}$ ($\lesssim2\,\Msun\,{\rm yr}^{-1}$) above 
(below) the transition scale, whereas, for the MW$^{\rm c}$, the resulting net 
rates (both of accretion and ejection) are 
more modest, namely, $\lesssim1\,\Msun\,{\rm yr}^{-1}$ 
(see dot-dashed lines in the bottom panels of
Fig.~\ref{acc_eje_both_galaxies}). The net accretion rates of {\it all} 
material (neutral$+$cold$+$hot) at the virial radius of the galaxies 
are about $6-8\,\Msun\,{\rm yr}^{-1}$.

In the next section we compare our predictions for the accretion rates of 
cold and neutral material with observations. Unfortunately, there are no available 
observational estimates on the hot gas accretion, as well as on the ejection rates, 
but, in relation to the latter, our results above indicate that outflowing material 
might be present near the virial radius of galaxy-sized haloes, although at low densities.

\subsection{Cold and H\,{\sc i} gas accretion rates}
\label{acc_eje_obs}

\subsubsection{Comparison with observations}

Observational estimates of the gas accretion rates on to the Milky Way galaxy 
are done using the cold and neutral gas components. In this section, we
compare the accretion rates of our simulated galaxies with such observations.
Table~\ref{acc_eje_rates} shows the {\it integrated} 
accretion/ejection rates of the simulated M31$^{\rm c}$ and MW$^{\rm c}$
for the cold and neutral gas phases (we also show results for the
hot phase,  for completeness). 
To resemble observational methodologies we compute, in each case, 
the characteristic inflow (outflow) time by dividing the distance of the gas clouds 
to the centre of the galaxy with their corresponding radial velocities. 
Results are shown for two scales of interest, including all 
material within a distance of $r=10$ and $50\,$kpc, 
and excluding the gaseous discs (as explained in Section~\ref{HI_data}), 
to meaningfully compare with observational results. 
This means that material within the boundaries of our  
gaseous discs will not contribute to our mass infall estimates. In particular, 
when considering scales $r\leq10\,$kpc, only material above and below the discs 
will be considered as a result of this approach.

As mentioned in Section~\ref{cold_data}, the mass of cold gas in the Milky Way halo 
at a distance of $\sim$$10\,$kpc above the galactic plane has been estimated in $\sim$$10^8\,\Msun$.  
If this gas would represent infalling material, its contribution to the gas-accretion rate of the Milky 
Way would be, as a first approximation, of the order of 
$\sim$$1\,\Msun$\,yr$^{-1}$ \citep[][]{Shull09,Lehner11}. 
In our simulation, the {\it integrated} accretion rates of cold gas within
$10\,$kpc (excluding the gaseous disc) are $0.33$ 
and $0.57\,\Msun\,{\rm yr}^{-1}$ for M31$^{\rm c}$ and MW$^{\rm c}$, respectively. 
Interestingly, the resulting MW$^{\rm c}$ value is of the same order of
magnitude as in observations. If the discs were not excluded, the rates for both galaxies would result 
a factor of $\sim4-5$ larger, as it is shown in Table~\ref{acc_eje_rates}.

Observational estimates for the infall of neutral gas in the form of HVCs in the 
Milky Way are within the range $\sim$$0.1-0.5\,\Msun\,{\rm yr}^{-1}$ 
when integrating the contributions of all known H\,{\sc i} gas complexes 
(e.g., \citealt{Wakker04,Wakker07,Wakker08,Putman12}). The three-dimensional 
modelling of R12 gives, instead, an accretion rate of 
$\sim$$0.7\,\Msun\,{\rm yr}^{-1}$ inside a spherical region of $\sim$$50\,$kpc
radius. In the case of neutral material in our simulations, we found that the 
accretion rates of {\it all} H\,{\sc i} gas within this distance 
are $\dot{M}_{\rm HI}(r\leq50\,{\rm kpc})\approx 0.22\,$$\Msun\,$yr$^{-1}$ and 
$0.34\,\Msun\,{\rm yr}^{-1}$ for M31$^{\rm c}$ and MW$^{\rm c}$, respectively, 
i.e. consistent with the range of observed accretion rate values for the Milky Way 
but a factor $2-3$ smaller than the R12 modelling. 
When repeating the same calculation but imposing column density cuts to the accreting 
material we obtain similar results: H\,{\sc i} gas with projected densities of
$N_{\rm HI}\geq 10^{18}, 10^{19}, 10^{20}\,$cm$^{-2}$ displays accretion rates of 
$\dot{M}_{\rm HI}(r\leq50\,{\rm kpc})\approx 0.21, 0.16, 0.06\,$$\Msun\,$yr$^{-1}$ 
and $0.33, 0.28, 0.15\,$$\Msun\,$yr$^{-1}$ for the M31$^{\rm c}$ and MW$^{\rm c}$ simulated 
galaxies, respectively. For column density cuts corresponding 
to HVCs with $N_{\rm HI}\geq 10^{19}\,$cm$^{-2}$ the {\it integrated} H\,{\sc i} 
accretion rates decrease by less than $30\%$, still remaining consistent with observations. 
For the largest column density values the rates show a significant decrease. 
This could indicate the need for a higher spatial resolution in our simulation 
to better describe the most compact gas complexes.

\subsubsection{Comparison with star formation rate}

At this point, it is interesting to compare the accretion 
rate of gas on to the simulated galaxies with their star formation rate 
(SFR) as accreting material is believed to be responsible of sustaining 
the star formation activity. 
As mentioned before the integrated accretion rate of neutral gas within $50\,$kpc 
is about $0.2-0.3\,$M$_\odot\,$yr$^{-1}$ (see Table~\ref{acc_eje_rates}), which is 
consistent with the local rate estimates obtained at the termination of the discs.  
These values are smaller than the typical SFRs of our simulated galaxies at $z=0$, 
which are about $1\,$M$_\odot\,$yr$^{-1}$ (see Scannapieco et al., in preparation). However, 
as stated before, most of the accreting cold gas is not in the neutral phase, thus 
providing more fuel that can eventually be transformed into stars. In fact, as 
can be seen in Table~\ref{acc_eje_rates}, typical cold gas accretion rates within 
a $50\,$kpc region are about $4-5\,$M$_\odot\,$yr$^{-1}$, although only a
fraction of it may undergo star formation. 
This suggests that, in this way, it could be possible to maintain the 
continuous SFR required in observed present-day galaxies.

\section{Gas distribution in the simulated Milky Way/Andromeda system}
\label{sec:between}

We turn our focus to the study of the gas distribution in the MW$^{\rm c}$/M31$^{\rm c}$ 
system and, in particular, in the region between the two galaxies.
The main question we  want to address is whether there is a gas excess between
M31$^{\rm c}$ and MW$^{\rm c}$, in relation to the amount of gas in any other random 
direction.

Fig.~\ref{both_galaxies} shows maps of total gas density, H\,{\sc i} column
density and temperature for the MW$^{\rm c}$/M31$^{\rm c}$ system, at $z=0$, in a cubic box of 1.6 Mpc on a side. 
The origin is located in the geometric centre of the two galaxies, 
and the circles show the virial radii of M31$^{\rm c}$ (solid lines) and MW$^{\rm c}$ (dashed lines).
Clearly, the gas distribution of the two haloes 
overlap suggesting that the gaseous haloes are interconnected, and the presence
of a gas bridge in that particular direction. Interestingly, some of the neutral 
gas clouds between the galaxies have column densities 
of $N_{\rm HI}\sim10^{17}\,$cm$^{-2}$ (middle panels), 
thus resembling those observed by \cite{Wolfe13} in the direction linking 
Andromeda and M33 galaxies.
The temperature maps also show that there is no clear separation of the gas in the two galaxies,
neither there is significant temperature difference between the
two haloes (see also Section~\ref{general}).

\begin{figure}
\begin{center}
{\includegraphics[width=90mm]{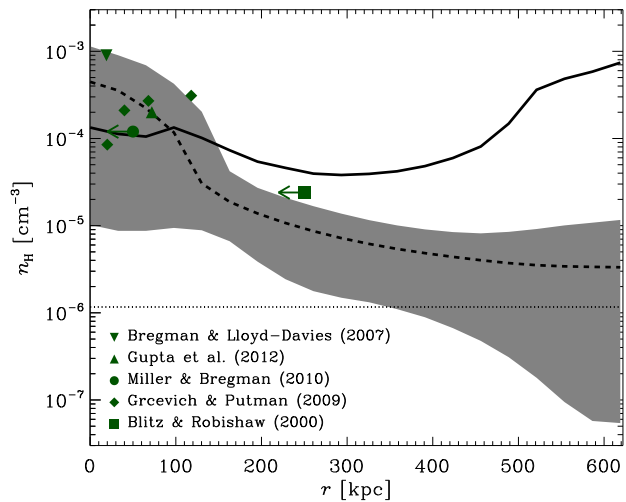}}
\caption{Hydrogen number density profile as function of distance to the MW$^{\rm c}$ mass centre for
  different random directions. The mean value is indicated as a dashed line,
  while the shaded region corresponds to the standard deviation. 
  The thick black line corresponds to the direction towards the M31$^{\rm c}$ galaxy. 
  For comparison, we show the cosmological Hydrogen number density as a dotted line. 
  Filled symbols indicate estimates of the observed {\it electron} number density 
  at different distances from the Milky Way which have been included as reference values. 
  Arrows indicate an upper limit for the corresponding path-length.
  } 
\label{density_profiles}
\end{center}
\end{figure}

To further investigate this, we plot
in Fig.~\ref{density_profiles} the Hydrogen number density profile as a function
of radius, after centring our reference system in the MW$^{\rm c}$ galaxy. 
The plot is done up to a distance of $652$ kpc, where the M31$^{\rm c}$ 
galaxy is located. We first constructed the density profiles, from the MW$^{\rm c}$ centre, 
along random line of sights, and we plotted the mean over these profiles
(dashed line), as well as the corresponding $\pm 1\sigma$ regions (shaded region). For comparison, 
the mean cosmological number density of Hydrogen is shown as a 
horizontal dotted line. Interestingly, for distances smaller than $\sim$$100\,$kpc, 
our simulated density values are consistent, within a 1$\sigma$ deviation, 
to the average {\it electron} number density limits derived by several authors 
\citep[][see Fig.~\ref{density_profiles}]{BregmanLloyd07,Grcevich09,Gupta12,Miller13}. 
For distances close to the virial radius of the MW$^{\rm c}$, our 
simulation gives a similar result to that inferred by \citet{Blitz00}, 
which estimate an electron number density of 
$n_{\rm e}\approx 2.4\times10^{-5}\,$cm$^{-3}$ within a distance of 
$\sim$$250\,$kpc from the Sun by studying the depletion of cold gas in dwarf 
spheroidal galaxies as a result of ram pressure striping. 
We note, however, that this comparison must be taken with 
caution as the \citet{Blitz00} estimation is not free of uncertainties.

The thick black line in Fig.~\ref{density_profiles} shows the 
Hydrogen number density profile corresponding 
to the line-of-sight towards M31$^{\rm c}$. 
We found a clear gas excess in this direction: 
the Hydrogen number density between the two galaxies is about 
$n_{\rm H}\sim 4\times10^{-5}\,$cm$^{-3}$, which is well 
above the gas density at this distance for a random line-of-sight 
(i.e., about $10^{-5}\,$cm$^{-3}$; see the dashed line). If we consider the spread 
of the random sightline values the significance of the gas excess 
detection at a distance of $\sim$$300\,$kpc from the MW$^{\rm c}$ is 
close to $20\sigma$. 
As we further approach to the simulated M31$^{\rm c}$, the gas density 
profile rapidly increases owing to the presence of 
the M31$^{\rm c}$ gaseous halo, while for any 
other random direction the mean number density profile slightly decreases approaching 
to the mean cosmological value.

A gas bridge between the Milky Way and Andromeda 
may be  supported by observations of high-velocity O\,{\sc vi} absorption
in the Far-UV. \citet{Sembach03} and \citet{Wakker03} 
(as shown by their Figs.\,$10-12$) found an excess of relatively strong O\,{\sc vi} 
absorption with high negative radial velocities ($v_{\rm LSR}\approx-400$ to
$-150$ km\,s$^{-1}$) at $l\approx20^{\circ}-140^{\circ}$ 
and $b\lesssim30^{\circ}$, i.e. including the position of Andromeda. 
A possible scenario explaining these observations 
could be a general excess of hot gas towards this region, as implied by our simulation. 
In such a case, the observed O\,{\sc vi} absorption features may indicate 
somewhat cooler gas patches at $T\sim3\times 10^5$ K that condense out 
of the ambient hot ($T\sim10^6$ K) medium which comprises 
the bulk of the gas that interconnects the Milky Way and Andromeda galaxies 
(see Fig.~\ref{both_galaxies}).

However, it is worth noting that the interpretation of the 
\citet{Sembach03} and \citet{Wakker03} observations is not straightforward, 
as O\,{\sc vi} absorption at high negative radial velocities is ubiquitous 
at $l<180^{\circ}$, a sky region also occupied by 
many prominent $21\,$cm HVCs with $v_{\rm LSR}< 100\,$km s$^{-1}$ 
(e.g., the Magellanic Stream and Complexes A, C, H; 
see Fig. 1 in \citealt{Richter09}). Therefore, the observed 
O\,{\sc vi} absorption could also be produced by the existence of 
transition-temperature gas in an extended boundary between the 
neutral body of the mentioned halo clouds and the surrounding 
hot halo gas.

After repeating the exercise presented in Fig.~\ref{density_profiles} 
for different redshifts we discovered that the gas 
excess is first seen in the simulation at $z\sim 1$ as a result of the kinematical 
evolution of the LG, when the two haloes overlap and are part of the same filament. 
However, a detailed analysis of the gas distribution as a function of 
redshift will be presented in a forthcoming work.

\section{Comparison with recent work}
\label{comp_with_amr}

In this section, we  compare our results with the work of 
\cite{Fernandez12} and \citet{Joung12} (hereafter F12 and J12, respectively). 
These authors studied the neutral and ionized gas components of a 
galaxy with a halo mass of $1.4\times10^{12}\,$M$_{\sun}$ 
that formed in a cosmological simulation, run with 
the adaptive mesh refinement code {\sc enzo}. 
In general, our results are in reasonable agreement with those of
F12 and J12, as we explain below. Note, however, that variance in
the galaxy properties is expected, even for a given halo  
mass. Therefore, disagreement can only be established when differences are 
larger than the expected scatter that naturally arises in hierarchical 
galaxy formation models\footnote{It is worth noting also that, ultimately, 
an unbiased comparison between the two sets of simulations 
is not entirely possible owing to the different hydrodynamical and 
star formation/feedback implementations adopted in each case.}.

We first compare the total H\,{\sc i} masses, excluding the gaseous disc
and within the inner 50 kpc, as to be consistent with the outcomes of F12.
These authors found a lower H\,{\sc i} mass of about $10^8$M$_\odot$, when 
compared to our results, 
for which we found H\,{\sc i} masses of $3-4\times10^8\,$$\Msun$. 
Overall, the resulting mass profiles are similar, although in F12 
there is a larger number of peaks, presumably owing to their better 
spatial resolution. The gaseous halo of F12 agrees well
with ours in terms of their typical temperature, which is always of the order
of $10^6\,$K. The most significant difference between F12 and our work is on the
size of the gaseous neutral discs. In F12, they found a typical
radial extension of about $15-20\,$kpc, significantly smaller than
the $30-40\,$kpc we found for our galaxies. In this respect, our results
are in better agreement with observations, and allow
the formation of more extended stellar discs, as shown in 
Scannapieco et al. (in preparation).

Concerning the accretion rates given in F12, they quoted 
an H\,{\sc i} accretion rate on to the disc of $\sim$$0.2\,$$\Msun\,$yr$^{-1}$, 
at a radius of $20\,$kpc and excluding the disc. 
If we do the same calculation, we find
a similar value of $\sim$$0.25\,$$\Msun\,$yr$^{-1}$ 
for our simulated galaxies (see the dotted lines in 
the top panels of Fig.~\ref{acc_eje_both_galaxies}). 
If we take, instead, a radius of $30\,$kpc, which is closer to the size 
of our simulated discs, we obtain smaller accretion rates 
of $\sim$$0.1-0.2\,$$\Msun\,$yr$^{-1}$.

J12 focused on the net infall rates of the different
gas phases, excluding always the gas disc in their analysis and adopting
a similar criterion to ours (nevertheless, note that the disc
exclusion region is different, as a result of its smaller size).
Our simulation shows that the gas inflow at distances
larger than the disc sizes is dominated by ionized material belonging both 
to the cold ($T<10^5\,$K) and hot ($T>10^5\,$K) gas phases, 
in line with the findings of J12. We find, however, a significant 
discrepancy in the net accretion rates for the hot phase (which 
corresponds to that of ``warm-hot'' and ``hot'' phases of J12)
in the inner regions (i.e., $r\lesssim50-100\,$kpc):
whereas we get a net {\it ejection} rate of about $1-2\,$M$_\odot$yr$^{-1}$, J12
find a net {\it accretion} rate of about $1-3\,$M$_\odot$yr$^{-1}$. 
This indicates that the implementation of the feedback mechanism in the two simulations
works differently. Our model is much more efficient in
driving supernova outflows, which is also the reason for getting
more extended gaseous/stellar discs. At larger distances, 
i.e. $r\gtrsim100\,$kpc, the net accretion rates between the simulations are more similar: 
we find rates within the range $\sim$$1-6\,$M$_\odot$yr$^{-1}$ 
compared to the $\sim$$1-5\,$M$_\odot$yr$^{-1}$ obtained by J12.
For the cold phase, the net accretion rates of our simulated galaxies 
are within the range $\sim$$2-5\,$M$_\odot$yr$^{-1}$ in comparison with 
$\sim$$0.5-2\,$M$_\odot$yr$^{-1}$ in the J12 case.
Note that our simulation, having two galaxies of a similar mass,
gives an estimate of the (minimum) variations expected from  
halo-to-halo, and, therefore, agreement must be gauged 
taking this into account.

\section{Summary and Conclusions}
\label{concl}

We have studied the gas distribution within and between the two largest members 
of a simulated LG of galaxies using a constrained cosmological simulation.
The two simulated galaxies, dubbed MW$^{\rm c}$ and M31$^{\rm c}$, approximately 
resemble the real Milky Way and Andromeda galaxies 
in terms of their masses, separation, inclination and relative velocity and are surrounded by a large-scale
environment similar to the one observed in the real LG. The numerical experiment was carried out 
using the simulation code {\sc gadget3} with the extensions of \citet{Scann05,Scann06} that include 
a multiphase gas model, a treatment of supernova (thermal) feedback, chemical enrichment and 
metal-dependent cooling. 

We studied the main properties (e.g., masses, distribution, accretion/ejection rates, covering fractions) 
of the gaseous material belonging to our simulated galaxies for different gas phases 
(i.e., hot: $T\geq10^5\,$K; cold: $T<10^5\,$K, and H\,{\sc i} neutral gas) and compared our results 
with available observations. 

The main results of our work can be summarized as follows: 

\begin{itemize}

\item The gas distribution within the simulated MW$^{\rm c}$ and M31$^{\rm c}$ haloes
is characterized by the presence of a 
dense ($\rho\gtrsim 5\times10^4\,\Msun\,$kpc$^{-3}$) 
region with $T\sim10^4\,$K located in the central parts ($r\lesssim30\,$kpc) plus an additional 
gaseous halo component (which for our analysis we divide into two phases according to a 
temperature criterion). 
Generally speaking, within the halo, the cold gas phase dominates 
over the hot one for $r\lesssim 50\,\rm kpc$, while the opposite is true at larger scales. 
The mean temperature of the simulated gaseous coronae at the virial 
radius is about $7\times10^{5}\,$K. This value is close to the estimate given 
by \citet{Gupta12} for the warm-hot material surrounding the Milky Way 
inferred through O\,{\sc vii} absorption lines. 
The amount of hot gas within the virial radii of our simulated 
MW$^{\rm c}$ and M31$^{\rm c}$ galaxies is of the order 
of $\sim$$5\times10^{10}\,$M$_{\sun}$. Interestingly, this value is also 
consistent with the findings of \citet{Gupta12}. Under plausible assumptions, 
these authors claim the existence of a large hot gas reservoir of 
mass $\gtrsim1-6\times10^{10}\,$M$_{\sun}$ that corresponds to 
absorption path-lengths $l>139\,$kpc, i.e. far beyond the central regions of
the Galaxy. Our results show that 
the hot gas phase in galaxies like the Milky Way and Andromeda is an 
ideal reservoir of baryons that can comprise up to $\sim$$35\%$ of the total 
baryonic content of the galaxies, and up to $\sim$$75\%$ of the
gas content. 
We also computed the amount of gas in the cold phase (within the inner 10 kpc
and excluding the gaseous disc, as to be comparable with available observations),
and found values of $\sim$$10^8\,$M$_\odot$ for both simulated galaxies, in agreement with
observational results.
\\

\item We have studied the amount and distribution of the H\,{\sc i} 
neutral gas phase. The majority of neutral gas around our simulated galaxies
has typical column densities in the range $N_{\rm HI}\sim 10^{17}-10^{18}\,$cm$^{-2}$,
and is mainly located within a distance of about $50-70\,$kpc from their
centre resembling the observed distribution of 
HVCs in the Milky Way and Andromeda (see Fig.~\ref{HVCs_simulated_contours}). 
Higher column density clouds, i.e. those with 
$N_{\rm HI}\sim 10^{18}-10^{20}\,$cm$^{-2}$, are also present within the haloes 
although their distribution is presumably underestimated owing to the limited 
resolution of our simulation. After excluding the galactic discs, the
resulting mass content of {\it all} H\,{\sc i} surrounding the 
simulated MW$^{\rm c}$ and M31$^{\rm c}$ is 
$M(r\leq 50\,\rm kpc)\approx3-4\times10^8\,$M$_{\sun}$, in line with
observations. We found that this gas is typically not associated with satellite systems, but 
belongs to the ambient medium. Its spatial distribution tends to be 
more spherical than that of gas with higher column densities. 
If we, instead, consider H\,{\sc i} material 
with higher column density values 
these mass estimates can decrease, at most, by a factor of a few. 
Typical mean velocities of accreting neutral gas with $N_{\rm HI}\gtrsim
10^{18}\,$cm$^{-2}$ are of the order of $\sim$$50\,$km s$^{-1}$ with 
respect to the centre of the galaxies. 
\\

\item We calculated the covering fraction of neutral gas given by our simulation 
in two different ways, in order to compare with different available observations. 
First, we projected the H\,{\sc i} distribution of M31$^{\rm c}$ as 
seen from the MW$^{\rm c}$, as well as the other way around. 
We found that the projected covering fractions $f_c$  decline with radius; however, the
steepness of the profile depends on the viewing angle. 
For example, the 
projected covering fractions, assuming  $N_{\rm HI}>7\times10^{17}\,$cm$^{-2}$, 
and at a projected distance of $30$ kpc from the corresponding centre,
are $\sim$$0.3$ and $\sim$$1$ for M31$^{\rm c}$ and the MW$^{\rm c}$, respectively. 
Clearly, the projected covering fraction of M31$^{\rm c}$ is much smaller than that 
of MW$^{\rm c}$ at this distance. The reason 
is that, as seen  from the MW$^{\rm c}$, M31$^{\rm c}$ appears almost edge-on (therefore the
region covered by the high column density gas looks smaller), 
while the MW$^{\rm c}$, as seen from M31$^{\rm c}$, appears almost face-on (and covers a much larger
projected area). 
We can compare the covering fractions of M31$^{\rm c}$ at $30$ and $50\,$kpc with
the modelling of R12. We obtain $f_c(30\,$kpc$)=0.3$ and $f_c(50\,$kpc$)=0.1$, 
in line with the trend found by R12 for Andromeda. 
We also calculated the covering fraction of neutral gas from an all-sky H\,{\sc i} map, as ``observed'' 
from the mock LSR of the MW: we found, for $N_{\rm HI}>7\times10^{17}\,$cm$^{-2}$, 
a covering fraction of $f_{\rm c,all-sky}\approx 0.3$ ($0.2$) when including (excluding) 
the MW$^{\rm c}$ neutral gas disc. 
These values are similar to the covering fractions obtained by 
\citet{Lockman02} and \citet{Wakker04} by means of Milky Way H\,{\sc i} observations.  
\\

\item We calculated the accretion rates, as a function of radius, for the
hot, cold and H\,{\sc i} components of the simulated MW$^{\rm c}$ and M31$^{\rm c}$.
We found that, at distances close to the virial radii of the galaxies, the gas 
accretion rate is dominated by hot material, reaching values 
of $\sim$$7-10\,$$\Msun\,$yr$^{-1}$, whereas cold material has 
typical infall rates of $\sim$$3-5\,$$\Msun\,$yr$^{-1}$. At distances smaller 
than $\sim$$50\,$kpc the situation is reversed: the accretion rate of 
cold gas tends to be higher than that of hot gas. 
This behaviour is directly related to the larger cooling 
efficiency of cold material in comparison with gas at 
higher temperatures. In general -- and irrespectively of metallicity -- 
the cooling curves of the gas peak around $T\sim10^5\,$K, a temperature 
magnitude that is most commonly 
found in the inner galactic halo. As a result, the corresponding 
accretion rates of cold material in the central regions tend to increase.
\\

\item In our model, the MW$^{\rm c}$ and M31$^{\rm c}$ also accrete H\,{\sc i}
  gas reaching a peak of about $0.25\,$$\Msun\,$yr$^{-1}$ for a distance of $20\,$kpc, and decrease significantly 
for larger distances. Interestingly, most of the H\,{\sc i} accretion outside the discs takes place 
at distances of $\sim$$50-70\,$kpc at a rate of about $0.01-0.1\,$$\Msun\,$yr$^{-1}$ 
in agreement with HVC observations. 
The {\it integrated} H\,{\sc i} accretion rates within a sphere of $50\,$kpc radius 
(excluding the H\,{\sc i} galactic disc) 
give values of $0.34$ and $0.22\,$$\Msun\,$yr$^{-1}$ 
for MW$^{\rm c}$ and M31$^{\rm c}$ respectively, in line with recent observational estimates for the Milky Way. 
If we consider neutral material with typical HVC column densities we obtain 
similar results: e.g., at $N_{\rm HI}\geq 10^{19}\,$cm$^{-2}$ we get $0.28\,$$\Msun\,$yr$^{-1}$
for MW$^{\rm c}$ and $0.16\,$$\Msun\,$yr$^{-1}$ for M31$^{\rm c}$. In general, these accretion rate values 
are smaller than the SFRs of our simulated galaxies which are of the order of $1\,\Msun$yr$^{-1}$ 
(see Scannapieco et al., in preparation). However, most of the cold gas accretion in the simulation is not 
neutral but ionized, having typical values that are larger than the corresponding galactic SFRs.
\\

\item We also calculated ejection rates for the simulated MW$^{\rm c}$ and M31$^{\rm c}$. 
 At distances close to the virial radii of the galaxies, we found that  
 the gas ejection rate is dominated by hot
 material, reaching values of $\sim$$3-6\,$$\Msun\,$yr$^{-1}$, whereas the ejection of cold material
 is small in comparison ($\sim$$0.1\,$$\Msun\,$yr$^{-1}$). This can be understood by the fact
 that most of the ejected material is at high temperatures as a result of the supernova feedback
 process. At distances smaller than $\sim$$50\,$kpc the ejection rate of cold gas 
 ($\lesssim0.1-1\,$$\Msun\,$yr$^{-1}$) is lower than that of hot gas 
 ($\lesssim3-4\,$$\Msun\,$yr$^{-1}$), as expected. In both galaxies, the net infall 
 (accretion minus ejection) rate profiles show a clear transition scale, located at 
 a distance of about $100-120\,$kpc, above (below) which the hot material of the halo is being 
 mainly accreted (ejected) as a result of the interplay between supernova feedback and 
 gravity.  
 \\
\end{itemize}

We have calculated a number of properties of the simulated LG 
at larger scales, say $1-2$ Mpc, and investigated the gas distribution between
the MW$^{\rm c}$ and M31$^{\rm c}$ systems. In this respect, we found the following main results:

\begin{itemize}

\item The gas associated with the simulated LG, within a distance of 
 $\sim$$1\,$Mpc, is dominated by the hot component, with a hot
 gas fraction of $\sim$$80\%$. This means that a considerable fraction of
 the gas in the LG is most likely in a hot gas phase that is difficult 
 to detect observationally. At this scale, the resulting baryon fraction is $0.13$, i.e. 
 $\sim$$20\%$ below the universal value, as could be expected from cosmic variance.
 Within the individual MW$^{\rm c}$ and M31$^{\rm c}$ haloes, the baryon
 fraction is even lower (0.095 and  0.087 for the MW$^{\rm c}$ and M31$^{\rm c}$, respectively) 
 suggesting that galactic winds have played an important role in ejecting
 baryons in the form of gas away from the haloes (see Fig.~\ref{LG_gas_fractions}).   
 \\

\item We detected an excess of gas in the direction joining the MW$^{\rm c}$ and M31$^{\rm c}$ galaxies as a result 
 of the overlapping of their hot gaseous haloes. In comparison with other random directions, 
 this signal represents an $\sim$$20\sigma$ detection. 
 Our gas density estimate between the two galaxies is consistent 
 with the electron number densities derived by several authors, as is shown in 
 Fig.~\ref{density_profiles}. In particular, at distances close to the virial radius of 
 the MW$^{\rm c}$ we found an hydrogen number density of $n_{\rm H}\sim4\times10^{-5}\,$cm$^{-3}$, 
 which is close to the electron number density estimated by \citet{Blitz00} 
 for path-lengths $l\lesssim250\,$kpc. Interestingly, the existence of neutral gas clouds between the simulated 
 galaxies goes in line with the observations of \cite{Wolfe13} for the Andromeda/M33 system. 
 We found that, in our simulation, the excess originates at $z\sim 1$, when 
 the two galaxies start sharing the same filament and have overlapping gas distributions. 
 The gas bridge we detect at $z=0$ may explain the relatively strong O\,{\sc vi} absorption 
 observed towards the direction of Andromeda \citep[][]{Sembach03,Wakker03}. However, caution 
 must be taken with this interpretation since the observed line of sights are in 
 the same {\it general} direction of other prominent neutral gas structures in the Milky Way halo, 
 such as the HVC Complexes A, C, H and the Magellanic Stream. 
 \\

\end{itemize}

\section*{Acknowledgements} 
S.E.N. acknowledges support from the Deutsche Forschungsgemeinschaft under the 
grants MU1020 16-1 and NU 332/2-1.
C.S.  acknowledges support from the Leibniz Gemeinschaft through
grant SAW-2012-AIP-5 129. 
The simulation was performed on the Juropa supercomputer of the J\"ulich Supercomputing Centre (JSC).
We also thank the CLUES collaboration (www.clues-project.org) for providing the initial conditions 
for the constrained simulations used in this study.

\bibliographystyle{mn2emod}
\bibliography{paper}

\end{document}

%% file: mydefs.tex
\usepackage{xspace}

\def\Mpch{\mbox{$h^{-1}$\,Mpc }}
\def\Msunh{\mbox{$h^{-1}\,{\rm M}_\odot$}}
\def\Msun{\mbox{M$_\odot$}}